\documentclass[preprint]{elsarticle}

\usepackage{graphicx}
\usepackage{tikz}
\usetikzlibrary{trees}
\usetikzlibrary{calc}
\usepackage{soul}
\usepackage{booktabs}
\usepackage{amsmath}
\usepackage{longfbox}
\usepackage{multirow}
\usepackage{array}
\usepackage{natbib}
\usepackage[hidelinks]{hyperref}
\usepackage{cleveref}
\usepackage{enumitem}
\usepackage{longtable}
\usepackage{url}
\usepackage{forest}

\newlist{qwq}{itemize}{1}
\setlist[qwq]{label={}, nosep}
\newcommand{\indentbegin}{\begin{qwq} \item}
\newcommand{\indentend}{\end{qwq}}

\makeatletter
\@ifundefined{lhs2tex.lhs2tex.sty.read}%
  {\@namedef{lhs2tex.lhs2tex.sty.read}{}%
   \newcommand\SkipToFmtEnd{}%
   \newcommand\EndFmtInput{}%
   \long\def\SkipToFmtEnd#1\EndFmtInput{}%
  }\SkipToFmtEnd

\newcommand\ReadOnlyOnce[1]{\@ifundefined{#1}{\@namedef{#1}{}}\SkipToFmtEnd}
\usepackage{amstext}
\usepackage{amssymb}
\usepackage{stmaryrd}
\DeclareFontFamily{OT1}{cmtex}{}
\DeclareFontShape{OT1}{cmtex}{m}{n}
  {<5><6><7><8>cmtex8
   <9>cmtex9
   <10><10.95><12><14.4><17.28><20.74><24.88>cmtex10}{}
\DeclareFontShape{OT1}{cmtex}{m}{it}
  {<-> ssub * cmtt/m/it}{}

\DeclareFontShape{OT1}{cmtt}{bx}{n}
  {<5><6><7><8>cmtt8
   <9>cmbtt9
   <10><10.95><12><14.4><17.28><20.74><24.88>cmbtt10}{}
\DeclareFontShape{OT1}{cmtex}{bx}{n}
  {<-> ssub * cmtt/bx/n}{}

\newcommand{\Conid}[1]{\mathit{#1}}
\newcommand{\Varid}[1]{\mathit{#1}}
\newcommand{\anonymous}{\kern0.06em \vbox{\hrule\@width.5em}}

\newcommand{\bind}{\mathbin{>\!\!\!>\mkern-6.7mu=}}
\newcommand{\sequ}{\mathbin{>\!\!\!>}}

\usepackage{polytable}

\@ifundefined{mathindent}%
  {\newdimen\mathindent\mathindent\leftmargini}%
  {}%

\def\resethooks{%
  \global\let\SaveRestoreHook\empty
  \global\let\ColumnHook\empty}
\newcommand*{\savecolumns}[1][default]%
  {\g@addto@macro\SaveRestoreHook{\savecolumns[#1]}}
\newcommand*{\restorecolumns}[1][default]%
  {\g@addto@macro\SaveRestoreHook{\restorecolumns[#1]}}
\newcommand*{\aligncolumn}[2]%
  {\g@addto@macro\ColumnHook{\column{#1}{#2}}}

\resethooks

\newcommand{\onelinecommentchars}{\quad-{}- }
\newcommand{\commentbeginchars}{\enskip\{-}
\newcommand{\commentendchars}{-\}\enskip}

\newcommand{\visiblecomments}{%
  \let\onelinecomment=\onelinecommentchars
  \let\commentbegin=\commentbeginchars
  \let\commentend=\commentendchars}

\newcommand{\invisiblecomments}{%
  \let\onelinecomment=\empty
  \let\commentbegin=\empty
  \let\commentend=\empty}

\visiblecomments

\newlength{\blanklineskip}
\newcommand{\hsindent}[1]{\quad}
\let\hspre\empty
\let\hspost\empty

\EndFmtInput
\makeatother
\ReadOnlyOnce{polycode.fmt}%
\makeatletter

\newcommand{\hsnewpar}[1]%
  {{\parskip=0pt\parindent=0pt\par\vskip #1\noindent}}

\newcommand{\hscodestyle}{}

\newcommand{\sethscode}[1]%
  {\expandafter\let\expandafter\hscode\csname #1\endcsname
   \expandafter\let\expandafter\endhscode\csname end#1\endcsname}

  {\par\noindent
   \advance\leftskip\mathindent
   \hscodestyle
   \let\\=\@normalcr
   \let\hspre\(\let\hspost\)%
   \pboxed}%
  {\endpboxed\)%
   \par\noindent
   \ignorespacesafterend}

  {\hsnewpar\abovedisplayskip
   \advance\leftskip\mathindent
   \hscodestyle
   \let\hspre\(\let\hspost\)%
   \pboxed}%
  {\endpboxed%
   \hsnewpar\belowdisplayskip
   \ignorespacesafterend}

  {\hsnewpar\abovedisplayskip
   \advance\leftskip\mathindent
   \hscodestyle
   \let\\=\@normalcr
   \(\pboxed}%
  {\endpboxed\)%
   \hsnewpar\belowdisplayskip
   \ignorespacesafterend}

\newcommand{\plainhs}{\sethscode{plainhscode}}

\plainhs

  {\hsnewpar\abovedisplayskip
   \advance\leftskip\mathindent
   \hscodestyle
   \let\\=\@normalcr
   \(\parray}%
  {\endparray\)%
   \hsnewpar\belowdisplayskip
   \ignorespacesafterend}

  {\parray}{\endparray}

  {\(\parray}{\endparray\)}

\def\codeframewidth{\arrayrulewidth}
\RequirePackage{calc}

  {\parskip=\abovedisplayskip\par\noindent
   \hscodestyle
   \arrayrulewidth=\codeframewidth
   \tabular{@{}|p{\linewidth-2\arraycolsep-2\arrayrulewidth-2pt}|@{}}%
   \hline\framedhslinecorrect\\{-1.5ex}%
   \let\endoflinesave=\\
   \let\\=\@normalcr
   \(\pboxed}%
  {\endpboxed\)%
   \framedhslinecorrect\endoflinesave{.5ex}\hline
   \endtabular
   \parskip=\belowdisplayskip\par\noindent
   \ignorespacesafterend}

\newcommand{\framedhslinecorrect}[2]%
  {#1[#2]}

  {\(\def\column##1##2{}%
   \let\>\undefined\let\<\undefined\let\\\undefined
   \newcommand\>[1][]{}\newcommand\<[1][]{}\newcommand\\[1][]{}%
   \def\fromto##1##2##3{##3}%
   }{\) }%

  {\let\orighscode=\hscode
   \let\origendhscode=\endhscode
   \def\endhscode{\def\hscode{\endgroup\def\@currenvir{hscode}\\}\begingroup}
   \orighscode\def\hscode{\endgroup\def\@currenvir{hscode}}}%
  {\origendhscode
   \global\let\hscode=\orighscode
   \global\let\endhscode=\origendhscode}%

\makeatother
\EndFmtInput
\ReadOnlyOnce{forall.fmt}%
\makeatletter

\let\HaskellResetHook\empty
\newcommand*{\AtHaskellReset}[1]{%
  \g@addto@macro\HaskellResetHook{#1}}
\newcommand*{\HaskellReset}{\HaskellResetHook}

\newcommand\hsforall{\global\let\hsdot=\hsperiodonce}
\newcommand*\hsperiodonce[2]{#2\global\let\hsdot=\hscompose}
\newcommand*\hscompose[2]{#1}

\AtHaskellReset{\global\let\hsdot=\hscompose}

\HaskellReset

\makeatother
\EndFmtInput

\newcommand{\squote}[1][scare quotes]{`{#1}'}
\newcommand{\dquote}[1][double quotes]{``{#1}''}

\newcommand{\bigO}[1]{\ensuremath{\mathcal{O}({#1})}}

\newsavebox{\centerdotbox}
\newcommand{\morphicAtsqcdot}[2]{%
  \sbox{\centerdotbox}{$\centerdot$}%
  \ht\centerdotbox=.33333\ht\centerdotbox
  \vcenter{\box\centerdotbox}%
}

\DeclareSymbolFont{symbolsC}{U}{txsyc}{m}{n}
\SetSymbolFont{symbolsC}{bold}{U}{txsyc}{bx}{n}
\DeclareFontSubstitution{U}{txsyc}{m}{n}
\DeclareMathSymbol{\multimapboth}{\mathrel}{symbolsC}{"13}

\DeclareMathAlphabet{\mymathbb}{U}{bbold}{m}{n}

\newcommand{\nothingtodo}[1]{}

\newcounter{definition}
\newenvironment{definition}[1][]{\refstepcounter{definition}\par\medskip
  \noindent\textbf{Definition~\thedefinition~(#1).} \rmfamily \itshape}{\medskip}
\newcounter{property}
\newenvironment{property}[1][]{\refstepcounter{property}\par\medskip
  \noindent\textbf{Property~\theproperty~(#1).} \rmfamily \itshape}{}
\newcounter{specification}

\newcounter{theorem}
\newenvironment{theorem}[1][]{\refstepcounter{theorem}\par\bigskip
  \noindent\begin{tabular}{|p{\textwidth}|}\hline
  \textbf{Theorem~\thetheorem.} \rmfamily \itshape}{\\\hline\end{tabular}\\}
\newcounter{lemma}
\newenvironment{lemma}[1][]{\refstepcounter{lemma}\par\bigskip
  \noindent\begin{tabular}{|p{\textwidth}|}\hline
  \textbf{Lemma~\thelemma.} \rmfamily \itshape}{\\\hline\end{tabular}\\}
\newcounter{corollary}

\newenvironment{proof}{\paragraph{Proof:}}{\hfill$\square$}

\counterwithout*{equation}{section}            

\renewcommand\hscodestyle{%
   \setlength\leftskip{-.5cm}%
}
\renewcommand{\commentbegin}{\enskip\{-~}

\newcommand{\partialfrac}[2]{\frac{\partial #1}{\partial #2}}

\begin{document}




\title
      {Forward- or Reverse-Mode Automatic Differentiation: What's the Difference?}

\author[1]{Birthe van den Berg\corref{cor1}}
\ead{birthe.vandenberg@kuleuven.be}

\author[1]{Tom Schrijvers}
\ead{tom.schrijvers@kuleuven.be}

\author[2]{James McKinna}
\ead{j.mckinna@hw.ac.uk}

\author[3]{Alexander Vandenbroucke}
\ead{alexander.vandenbroucke@sc.com}

\affiliation[1]{organization={KU Leuven}, addressline={Celestijnenlaan 200A},city={Leuven},country={Belgium}}
\affiliation[2]{organization={Heriot-Watt University}, addressline={Earl Mountbatten Building G.52},city={Edinburgh EH14 4AS},country={United Kingdom}}
\affiliation[3]{organization={Standard Chartered}, addressline={1 Basinghall Avenue},city={London, EC2V 5DD},country={United Kingdom}}


\cortext[cor1]{Corresponding author}

\begin{abstract}

%
Automatic differentiation (AD) has been a topic of interest for researchers in
many disciplines, with increased popularity since its application to machine
learning and neural networks.
%
%
Although many researchers appreciate and know how to apply AD,
it remains a challenge to truly understand the underlying processes.
%
%
From an algebraic point of view, however, AD appears surprisingly natural:
it originates from the differentiation laws.
%
%
In this work we use Algebra of Programming techniques to reason about
different AD variants, leveraging Haskell to illustrate our observations.
%
%
Our findings stem from \emph{three fundamental algebraic abstractions}:
(1) the notion of semimodule,
(2) Nagata's construction of the \squote[idealization of a module], and
(3) Kronecker's delta function,
that together allow us to write a \emph{single-line abstract definition} of AD.
%
%
From this single-line definition, and by instantiating our algebraic
structures in various ways, we derive different AD variants,
that have the same \emph{extensional} behaviour, but different \emph{intensional} 
properties, mainly in terms of (asymptotic) computational complexity.
We show the different variants equivalent by means of \emph{Kronecker isomorphisms}, a further elaboration of our Haskell infrastructure which guarantees correctness by construction.
%
%
With this framework in place, this paper seeks to make AD variants
more comprehensible, taking an algebraic perspective on the matter.

\end{abstract}


\maketitle


\section{Introduction}

Many algorithms in artificial intelligence crucially depend not only on the
efficient large-scale evaluation of arithmetic expressions, but also on the
evaluation of their partial derivatives at specific points
\cite{DBLP:conf/osdi/AbadiBCCDDDGIIK16,JMLR:v18:17-468}. Symbolic and numerical
differentiation can be used for
these computations but suffer from performance and accuracy issues,
respectively. As a consequence, there is much interest in a third approach,
called automatic differentiation (AD) \cite{Wengert64}. AD efficiently and
accurately evaluates the derivatives of any computation composed of basic
arithmetic and exponential, logarithmic, and trigonometric functions.

The orthodox accounts of automatic differentiation, since its first
introduction by Wengert \cite{Wengert64}, emphasise it as a whole-program
transformation.
It is coupled with a shift from evaluation of the primitive
operations of the underlying language of numerical computation to those on the
so-called \emph{dual numbers} (due originally to Clifford \cite{Clifford73}, considerably
predating their subsequent application to AD).
The various AD modes (forward, reverse, mixed) then arise as further
optimisations of this basic picture in terms of the efficient organisation and
traversal of the underlying datastructures, 
typically exploiting mutability.

The vast literature on AD, which is impossible to survey here, covers
extensions of Wengert's original approach in all possible directions:
algorithmic improvements, language designs and extended expressivity, novel
applications and much more. The contribution of this paper is not to
extend this body of work even further, but rather to revisit the original algorithms
from an algebraic perspective.
Following Bird and de Moor's \dquote[Algebra of Programming]~\cite{AlgofProg}, we show how
algebra can be used to (re-)derive existing AD algorithms from a basic
executable specification, expressed in the purely functional programming
language Haskell.
The algebra involved consists of largely standard constructions
(with one notable exception, namely Nagata's construction described in Section~\ref{sec:nagata} below), which---because
they yield isomorphic mathematical objects---have no \emph{extensional}
consequences. Yet, they do establish the functional equivalence between the different
algorithmic variants.
At the algorithmic level the various (isomorphic) changes of representation
are precisely what allows us to exploit \emph{intensional} aspects,
specifically to change the computational complexity.


Specifically, this paper makes the following contributions.
\begin{itemize}
\item
  First, we show how symbolic differentiation and evaluation of symbolic expressions
  can both be seen algebraically as unique semiring homomorphisms from the free
  semiring. In terms of code, they are both instances of the same fold-like
  recursion scheme. This algebraic view enables us first to fuse the two homomorphisms
  to obtain basic forward mode AD, and then to recover symbolic differentiation
  as an instance of that (Section~\ref{sec:stage}).

\item
  Next, we abstract the AD algorithm in terms of three fundamental algebraic abstractions,
  whose specification and (equational) properties we realise via Haskell type classes:
  \emph{modules} over a semiring, a generalization
  of dual numbers which we dub \emph{Nagata numbers} (whose construction appears originally due to Nagata~\cite{Nagata1962}), and of Kronecker's \emph{delta} function.
  This abstract algorithm readily instantiates to a more efficient form of forward mode AD
  that computes the whole gradient vector of partial derivatives in one go (Section~\ref{sec:abstractad}).

\item
  Then, we show how other instances of the abstract algorithm can be obtained by
  replacing the representation of the gradient using an isomorphism of such structure,
  using Elliott's method of type class morphisms~\cite{Elliott2009-type-class-morphisms-TR}.
  This is demonstrated on a sparse vector representation (Section~\ref{sec:construction}).

\item
  Using the above methodology, we ultimately derive purely functional and imperative variants of
  reverse-mode AD: we successively improve the time complexity of scalar multiplication
  and gradient vector addition with representations based on homomorphisms, and switch to
  their equivalent mutable representation (Section~\ref{sec:rev}).

\item
  As an epilogue, we demonstrate that our approach readily admits common, useful
  extensions: additional primitive functions, direct computation with Nagata
  numbers to eliminate the overhead in building and evaluating symbolic expressions,
  higher derivatives, and \ensuremath{\mathbf{let}}-sharing (Section~\ref{sec:extensions}).

\item
  We provide an implementation\footnote{\href{https://github.com/birthevdb/Abstract-AD}{https://github.com/birthevdb/Abstract-AD}}
  of the approach and different AD versions.
\end{itemize}
Finally, the paper discusses related work (Section~\ref{sec:related}) and concludes (Section~\ref{sec:conclusion}).

We believe that this paper may be of interest to algebraists, (Haskell)
programmers and AD experts alike. While the paper does its best to be
accessible to these three groups, we acknowledge that this
requires working through unfamiliar concepts or unfamiliar
presentations of familiar concepts before the three strands come together.
\ref{app:functions} supports readers who are not familiar with Haskell, 
briefly introducing type classes and library functions that are used throughout this work.

\section{Setting the Stage}\label{sec:stage}

\noindent
This section provides a basic executable definition of automatic
differentiation (AD) based on a minimal setup.
For readers who want a gentle introduction to the methodology used here, 
we refer to ``Domain-Specific Languages of Mathematics'' \cite{DSLsofMathBook2022}.

\subsection{Symbolic Expressions and their Evaluation}


\noindent
In order to focus on the essential algorithmic mechanics, throughout most of
this paper we pare down to a minimal programming language in which
functions can be expressed: symbolic expressions \ensuremath{\Conid{Expr}\;\Varid{v}} that capture
polynomials over variables \ensuremath{\Varid{v}}, that vary independently\footnote{
In \Cref{sec:sharing-and-let} we extend the language of expressions with a \ensuremath{\mathbf{let}}-construct to permit 
variables that may be bound to complex subexpressions.}.

%

\indentbegin \begin{hscode}\SaveRestoreHook
\column{B}{@{}>{\hspre}l<{\hspost}@{}}%
\column{3}{@{}>{\hspre}l<{\hspost}@{}}%
\column{16}{@{}>{\hspre}c<{\hspost}@{}}%
\column{16E}{@{}l@{}}%
\column{19}{@{}>{\hspre}l<{\hspost}@{}}%
\column{28}{@{}>{\hspre}l<{\hspost}@{}}%
\column{36}{@{}>{\hspre}l<{\hspost}@{}}%
\column{45}{@{}>{\hspre}l<{\hspost}@{}}%
\column{52}{@{}>{\hspre}l<{\hspost}@{}}%
\column{E}{@{}>{\hspre}l<{\hspost}@{}}%
\>[3]{}\mathbf{data}\;\Conid{Expr}\;\Varid{v}{}\<[16]%
\>[16]{}\mathrel{=}{}\<[16E]%
\>[19]{}\Conid{Var}\;\Varid{v}\mid {}\<[28]%
\>[28]{}\Conid{Zero}\mid {}\<[36]%
\>[36]{}\Conid{One}{}\<[E]%
\\
\>[16]{}\mid {}\<[16E]%
\>[19]{}\Conid{Plus}\;(\Conid{Expr}\;\Varid{v})\;(\Conid{Expr}\;\Varid{v})\mid {}\<[45]%
\>[45]{}\Conid{Times}\;{}\<[52]%
\>[52]{}(\Conid{Expr}\;\Varid{v})\;(\Conid{Expr}\;\Varid{v}){}\<[E]%
\ColumnHook
\end{hscode}\resethooks
\indentend 
For example, for
univariate expressions we can define a type \ensuremath{\Conid{X}} with one variable.
The symbolic representation of \ensuremath{\Varid{x}\;\times\;(\Varid{x}\mathbin{+}\mathrm{1})} is captured in \ensuremath{\Varid{example}_{1}}.
\indentbegin \begin{hscode}\SaveRestoreHook
\column{B}{@{}>{\hspre}l<{\hspost}@{}}%
\column{3}{@{}>{\hspre}l<{\hspost}@{}}%
\column{E}{@{}>{\hspre}l<{\hspost}@{}}%
\>[3]{}\mathbf{data}\;\Conid{X}\mathrel{=}\Conid{X}{}\<[E]%
\\[\blanklineskip]%
\>[3]{}\Varid{example}_{1}\mathbin{::}\Conid{Expr}\;\Conid{X}{}\<[E]%
\\
\>[3]{}\Varid{example}_{1}\mathrel{=}\Conid{Times}\;(\Conid{Var}\;\Conid{X})\;(\Conid{Plus}\;(\Conid{Var}\;\Conid{X})\;\Conid{One}){}\<[E]%
\ColumnHook
\end{hscode}\resethooks
\indentend 
\noindent
We assign an overloaded meaning to symbolic expressions by mapping them to a
\emph{(commutative) semiring}. 
The `semi' in semiring refers to the fact that a semi-ring is a ring that does not need to have an 
additive inverse. We omit it here because AD does not require it. Nevertheless,
\Cref{sec:additional-primitives} shows that, if such an inverse exists, it can be accommodated.
The commutativity requirement we impose is not essential, but simplifies the 
presentation. \Cref{sec:non-comm} extends the approach 
to non-commutative semirings, which is necessary to support, for instance, matrices. 
The overloading
gives us the flexibility to interpret expressions in terms of both their plain
value and of different AD flavors.

\begin{definition}[Semiring]
A \emph{semiring} is a set, together with two
binary operators \ensuremath{\oplus} (``addition'') and \ensuremath{\otimes} (``multiplication'') and
two distinguished elements \ensuremath{\Varid{zero}} and \ensuremath{\Varid{one}}. These should obey the laws of
Figure~\ref{fig:semiring-laws}: \ensuremath{\oplus} and \ensuremath{\otimes} should both be
assocative and commutative and have neutral elements \ensuremath{\Varid{zero}} and \ensuremath{\Varid{one}}
respectively. Moreover, \ensuremath{\otimes} should distribute over \ensuremath{\oplus} and have \ensuremath{\Varid{zero}}
as annihilator.
\end{definition}

\begin{figure}
\begin{center}
\fbox{
\begin{minipage}[t]{0.55\textwidth}
\textbf{identity:}
\vspace{-3mm}\indentbegin \begin{hscode}\SaveRestoreHook
\column{B}{@{}>{\hspre}l<{\hspost}@{}}%
\column{3}{@{}>{\hspre}l<{\hspost}@{}}%
\column{9}{@{}>{\hspre}l<{\hspost}@{}}%
\column{18}{@{}>{\hspre}l<{\hspost}@{}}%
\column{21}{@{}>{\hspre}c<{\hspost}@{}}%
\column{21E}{@{}l@{}}%
\column{24}{@{}>{\hspre}l<{\hspost}@{}}%
\column{27}{@{}>{\hspre}c<{\hspost}@{}}%
\column{27E}{@{}l@{}}%
\column{30}{@{}>{\hspre}l<{\hspost}@{}}%
\column{33}{@{}>{\hspre}l<{\hspost}@{}}%
\column{42}{@{}>{\hspre}l<{\hspost}@{}}%
\column{E}{@{}>{\hspre}l<{\hspost}@{}}%
\>[3]{}\Varid{zero}{}\<[9]%
\>[9]{}\oplus{}\<[18]%
\>[18]{}\Varid{x}{}\<[21]%
\>[21]{}\mathrel{=}{}\<[21E]%
\>[24]{}\Varid{x}{}\<[27]%
\>[27]{}\mathrel{=}{}\<[27E]%
\>[30]{}\Varid{x}{}\<[33]%
\>[33]{}\oplus{}\<[42]%
\>[42]{}\Varid{zero}{}\<[E]%
\\
\>[3]{}\Varid{one}{}\<[9]%
\>[9]{}\otimes{}\<[18]%
\>[18]{}\Varid{x}{}\<[21]%
\>[21]{}\mathrel{=}{}\<[21E]%
\>[24]{}\Varid{x}{}\<[27]%
\>[27]{}\mathrel{=}{}\<[27E]%
\>[30]{}\Varid{x}{}\<[33]%
\>[33]{}\otimes{}\<[42]%
\>[42]{}\Varid{one}{}\<[E]%
\ColumnHook
\end{hscode}\resethooks
\indentend \textbf{associativity:}
\vspace{-3mm}\indentbegin \begin{hscode}\SaveRestoreHook
\column{B}{@{}>{\hspre}l<{\hspost}@{}}%
\column{3}{@{}>{\hspre}l<{\hspost}@{}}%
\column{6}{@{}>{\hspre}l<{\hspost}@{}}%
\column{15}{@{}>{\hspre}l<{\hspost}@{}}%
\column{19}{@{}>{\hspre}l<{\hspost}@{}}%
\column{28}{@{}>{\hspre}l<{\hspost}@{}}%
\column{32}{@{}>{\hspre}c<{\hspost}@{}}%
\column{32E}{@{}l@{}}%
\column{35}{@{}>{\hspre}l<{\hspost}@{}}%
\column{39}{@{}>{\hspre}l<{\hspost}@{}}%
\column{48}{@{}>{\hspre}l<{\hspost}@{}}%
\column{52}{@{}>{\hspre}l<{\hspost}@{}}%
\column{61}{@{}>{\hspre}l<{\hspost}@{}}%
\column{E}{@{}>{\hspre}l<{\hspost}@{}}%
\>[3]{}\Varid{x}{}\<[6]%
\>[6]{}\oplus{}\<[15]%
\>[15]{}(\Varid{y}{}\<[19]%
\>[19]{}\oplus{}\<[28]%
\>[28]{}\Varid{z}){}\<[32]%
\>[32]{}\mathrel{=}{}\<[32E]%
\>[35]{}(\Varid{x}{}\<[39]%
\>[39]{}\oplus{}\<[48]%
\>[48]{}\Varid{y}){}\<[52]%
\>[52]{}\oplus{}\<[61]%
\>[61]{}\Varid{z}{}\<[E]%
\\
\>[3]{}\Varid{x}{}\<[6]%
\>[6]{}\otimes{}\<[15]%
\>[15]{}(\Varid{y}{}\<[19]%
\>[19]{}\otimes{}\<[28]%
\>[28]{}\Varid{z}){}\<[32]%
\>[32]{}\mathrel{=}{}\<[32E]%
\>[35]{}(\Varid{x}{}\<[39]%
\>[39]{}\otimes{}\<[48]%
\>[48]{}\Varid{y}){}\<[52]%
\>[52]{}\otimes{}\<[61]%
\>[61]{}\Varid{z}{}\<[E]%
\ColumnHook
\end{hscode}\resethooks
\indentend \begin{minipage}{\textwidth}
\textbf{annihilator:}
\hspace{3mm}
\ensuremath{\Varid{zero}\otimes\Varid{x}\mathrel{=}\Varid{zero}\mathrel{=}\Varid{x}\otimes\Varid{zero}}
\end{minipage}
\end{minipage}%
\begin{minipage}[t]{0.45\textwidth}
\textbf{commutativity:}
\vspace{-3mm}\indentbegin \begin{hscode}\SaveRestoreHook
\column{B}{@{}>{\hspre}l<{\hspost}@{}}%
\column{3}{@{}>{\hspre}l<{\hspost}@{}}%
\column{6}{@{}>{\hspre}l<{\hspost}@{}}%
\column{15}{@{}>{\hspre}l<{\hspost}@{}}%
\column{18}{@{}>{\hspre}c<{\hspost}@{}}%
\column{18E}{@{}l@{}}%
\column{21}{@{}>{\hspre}l<{\hspost}@{}}%
\column{24}{@{}>{\hspre}l<{\hspost}@{}}%
\column{33}{@{}>{\hspre}l<{\hspost}@{}}%
\column{E}{@{}>{\hspre}l<{\hspost}@{}}%
\>[3]{}\Varid{x}{}\<[6]%
\>[6]{}\oplus{}\<[15]%
\>[15]{}\Varid{y}{}\<[18]%
\>[18]{}\mathrel{=}{}\<[18E]%
\>[21]{}\Varid{y}{}\<[24]%
\>[24]{}\oplus{}\<[33]%
\>[33]{}\Varid{x}{}\<[E]%
\\
\>[3]{}\Varid{x}{}\<[6]%
\>[6]{}\otimes{}\<[15]%
\>[15]{}\Varid{y}{}\<[18]%
\>[18]{}\mathrel{=}{}\<[18E]%
\>[21]{}\Varid{y}{}\<[24]%
\>[24]{}\otimes{}\<[33]%
\>[33]{}\Varid{x}{}\<[E]%
\ColumnHook
\end{hscode}\resethooks
\indentend \textbf{distributivity:}
\vspace{-3mm}\indentbegin \begin{hscode}\SaveRestoreHook
\column{B}{@{}>{\hspre}l<{\hspost}@{}}%
\column{3}{@{}>{\hspre}l<{\hspost}@{}}%
\column{7}{@{}>{\hspre}l<{\hspost}@{}}%
\column{17}{@{}>{\hspre}l<{\hspost}@{}}%
\column{29}{@{}>{\hspre}c<{\hspost}@{}}%
\column{29E}{@{}l@{}}%
\column{32}{@{}>{\hspre}l<{\hspost}@{}}%
\column{44}{@{}>{\hspre}l<{\hspost}@{}}%
\column{55}{@{}>{\hspre}l<{\hspost}@{}}%
\column{E}{@{}>{\hspre}l<{\hspost}@{}}%
\>[3]{}\Varid{x}\otimes(\Varid{y}{}\<[17]%
\>[17]{}\oplus\Varid{z}){}\<[29]%
\>[29]{}\mathrel{=}{}\<[29E]%
\>[32]{}(\Varid{x}\otimes{}\<[44]%
\>[44]{}\Varid{y})\oplus{}\<[55]%
\>[55]{}(\Varid{x}\otimes\Varid{z}){}\<[E]%
\\
\>[3]{}(\Varid{x}{}\<[7]%
\>[7]{}\oplus\Varid{y})\otimes\Varid{z}{}\<[29]%
\>[29]{}\mathrel{=}{}\<[29E]%
\>[32]{}(\Varid{x}\otimes{}\<[44]%
\>[44]{}\Varid{z})\oplus{}\<[55]%
\>[55]{}(\Varid{y}\otimes\Varid{z}){}\<[E]%
\ColumnHook
\end{hscode}\resethooks
\indentend \end{minipage}%
}
\end{center}
\caption{The commutative semiring laws.}\label{fig:semiring-laws}
\end{figure}

\noindent
In Haskell, we capture the algebraic concept of a semiring in the \ensuremath{\Conid{Semiring}} type class,
with instances for numeric types (e.g., \ensuremath{\Conid{Int}},
\ensuremath{\Conid{Float}}\footnote{If we ignore the limitations of floating point arithmetic.}) that
respect the laws.

\noindent
\begin{minipage}{0.5\textwidth}
\indentbegin \begin{hscode}\SaveRestoreHook
\column{B}{@{}>{\hspre}l<{\hspost}@{}}%
\column{3}{@{}>{\hspre}l<{\hspost}@{}}%
\column{5}{@{}>{\hspre}l<{\hspost}@{}}%
\column{12}{@{}>{\hspre}l<{\hspost}@{}}%
\column{E}{@{}>{\hspre}l<{\hspost}@{}}%
\>[3]{}\mathbf{class}\;\Conid{Semiring}\;\Varid{d}\;\mathbf{where}{}\<[E]%
\\
\>[3]{}\hsindent{2}{}\<[5]%
\>[5]{}\Varid{zero}{}\<[12]%
\>[12]{}\mathbin{::}\Varid{d}{}\<[E]%
\\
\>[3]{}\hsindent{2}{}\<[5]%
\>[5]{}\Varid{one}{}\<[12]%
\>[12]{}\mathbin{::}\Varid{d}{}\<[E]%
\\
\>[3]{}\hsindent{2}{}\<[5]%
\>[5]{}(\oplus){}\<[12]%
\>[12]{}\mathbin{::}\Varid{d}\to \Varid{d}\to \Varid{d}{}\<[E]%
\\
\>[3]{}\hsindent{2}{}\<[5]%
\>[5]{}(\otimes){}\<[12]%
\>[12]{}\mathbin{::}\Varid{d}\to \Varid{d}\to \Varid{d}{}\<[E]%
\ColumnHook
\end{hscode}\resethooks
\indentend \end{minipage} %
\begin{minipage}{0.5\textwidth}
\indentbegin \begin{hscode}\SaveRestoreHook
\column{B}{@{}>{\hspre}l<{\hspost}@{}}%
\column{3}{@{}>{\hspre}l<{\hspost}@{}}%
\column{5}{@{}>{\hspre}l<{\hspost}@{}}%
\column{12}{@{}>{\hspre}l<{\hspost}@{}}%
\column{E}{@{}>{\hspre}l<{\hspost}@{}}%
\>[3]{}\mathbf{instance}\;\Conid{Num}\;\Varid{a}\Rightarrow \Conid{Semiring}\;\Varid{a}\;\mathbf{where}{}\<[E]%
\\
\>[3]{}\hsindent{2}{}\<[5]%
\>[5]{}\Varid{zero}{}\<[12]%
\>[12]{}\mathrel{=}\mathrm{0}{}\<[E]%
\\
\>[3]{}\hsindent{2}{}\<[5]%
\>[5]{}\Varid{one}{}\<[12]%
\>[12]{}\mathrel{=}\mathrm{1}{}\<[E]%
\\
\>[3]{}\hsindent{2}{}\<[5]%
\>[5]{}(\oplus){}\<[12]%
\>[12]{}\mathrel{=}(\mathbin{+}){}\<[E]%
\\
\>[3]{}\hsindent{2}{}\<[5]%
\>[5]{}(\otimes){}\<[12]%
\>[12]{}\mathrel{=}(\mathbin{*}){}\<[E]%
\ColumnHook
\end{hscode}\resethooks
\indentend \end{minipage}


\noindent
There is one instance of semirings that is of special interest to us:
that for symbolic expressions \ensuremath{\Conid{Expr}\;\Varid{v}}, quotiented by the laws.\footnote{Quotienting by
the laws means that we should consider two (even structurally different) expressions equal
if the laws say so (e.g., \ensuremath{\Conid{Plus}\;\Conid{Zero}\;(\Conid{Var}\;\Conid{X})= =\Conid{Var}\;\Conid{X}}).}

\indentbegin \begin{hscode}\SaveRestoreHook
\column{B}{@{}>{\hspre}l<{\hspost}@{}}%
\column{3}{@{}>{\hspre}l<{\hspost}@{}}%
\column{5}{@{}>{\hspre}l<{\hspost}@{}}%
\column{12}{@{}>{\hspre}l<{\hspost}@{}}%
\column{E}{@{}>{\hspre}l<{\hspost}@{}}%
\>[3]{}\mathbf{instance}\;\Conid{Semiring}\;(\Conid{Expr}\;\Varid{v})\;\mathbf{where}{}\<[E]%
\\
\>[3]{}\hsindent{2}{}\<[5]%
\>[5]{}\Varid{zero}{}\<[12]%
\>[12]{}\mathrel{=}\Conid{Zero}{}\<[E]%
\\
\>[3]{}\hsindent{2}{}\<[5]%
\>[5]{}\Varid{one}{}\<[12]%
\>[12]{}\mathrel{=}\Conid{One}{}\<[E]%
\\
\>[3]{}\hsindent{2}{}\<[5]%
\>[5]{}(\oplus){}\<[12]%
\>[12]{}\mathrel{=}\Conid{Plus}{}\<[E]%
\\
\>[3]{}\hsindent{2}{}\<[5]%
\>[5]{}(\otimes){}\<[12]%
\>[12]{}\mathrel{=}\Conid{Times}{}\<[E]%
\ColumnHook
\end{hscode}\resethooks
\indentend What makes this instance special is that it is the \emph{free} semiring
instance, which gives rise to a fold-style \ensuremath{\Varid{eval}} function. This evaluator
proceeds by structural recursion over the symbolic expressions to map them to
their interpretation in a semiring \ensuremath{\Varid{d}}, given a mapping \ensuremath{\Varid{var}\mathbin{::}\Varid{v}\to \Varid{d}} for
variables.
\indentbegin \begin{hscode}\SaveRestoreHook
\column{B}{@{}>{\hspre}l<{\hspost}@{}}%
\column{3}{@{}>{\hspre}l<{\hspost}@{}}%
\column{20}{@{}>{\hspre}l<{\hspost}@{}}%
\column{28}{@{}>{\hspre}c<{\hspost}@{}}%
\column{28E}{@{}l@{}}%
\column{31}{@{}>{\hspre}l<{\hspost}@{}}%
\column{44}{@{}>{\hspre}l<{\hspost}@{}}%
\column{53}{@{}>{\hspre}l<{\hspost}@{}}%
\column{E}{@{}>{\hspre}l<{\hspost}@{}}%
\>[3]{}\Varid{eval}\mathbin{::}\Conid{Semiring}\;\Varid{d}\Rightarrow (\Varid{v}\to \Varid{d})\to \Conid{Expr}\;\Varid{v}\to \Varid{d}{}\<[E]%
\\
\>[3]{}\Varid{eval}\;\Varid{var}\;(\Conid{Var}\;\Varid{x}){}\<[28]%
\>[28]{}\mathrel{=}{}\<[28E]%
\>[31]{}\Varid{var}\;\Varid{x}{}\<[E]%
\\
\>[3]{}\Varid{eval}\;\Varid{var}\;\Conid{Zero}{}\<[28]%
\>[28]{}\mathrel{=}{}\<[28E]%
\>[31]{}\Varid{zero}{}\<[E]%
\\
\>[3]{}\Varid{eval}\;\Varid{var}\;\Conid{One}{}\<[28]%
\>[28]{}\mathrel{=}{}\<[28E]%
\>[31]{}\Varid{one}{}\<[E]%
\\
\>[3]{}\Varid{eval}\;\Varid{var}\;(\Conid{Plus}\;{}\<[20]%
\>[20]{}\Varid{e}_{1}\;\Varid{e}_{2}){}\<[28]%
\>[28]{}\mathrel{=}{}\<[28E]%
\>[31]{}\Varid{eval}\;\Varid{var}\;\Varid{e}_{1}{}\<[44]%
\>[44]{}\oplus{}\<[53]%
\>[53]{}\Varid{eval}\;\Varid{var}\;\Varid{e}_{2}{}\<[E]%
\\
\>[3]{}\Varid{eval}\;\Varid{var}\;(\Conid{Times}\;{}\<[20]%
\>[20]{}\Varid{e}_{1}\;\Varid{e}_{2}){}\<[28]%
\>[28]{}\mathrel{=}{}\<[28E]%
\>[31]{}\Varid{eval}\;\Varid{var}\;\Varid{e}_{1}{}\<[44]%
\>[44]{}\otimes{}\<[53]%
\>[53]{}\Varid{eval}\;\Varid{var}\;\Varid{e}_{2}{}\<[E]%
\ColumnHook
\end{hscode}\resethooks
\indentend This way we can evaluate \ensuremath{\Varid{example}_{1}} for \ensuremath{\Conid{X}\mathrel{=}\mathrm{5}} with the \ensuremath{\Conid{Int}} semiring.
\indentbegin \begin{hscode}\SaveRestoreHook
\column{B}{@{}>{\hspre}l<{\hspost}@{}}%
\column{3}{@{}>{\hspre}l<{\hspost}@{}}%
\column{E}{@{}>{\hspre}l<{\hspost}@{}}%
\>[3]{}\mathbin{>}\Varid{eval}\;(\lambda \Conid{X}\to \mathrm{5})\;\Varid{example}_{1}{}\<[E]%
\\
\>[3]{}\mathrm{30}{}\<[E]%
\ColumnHook
\end{hscode}\resethooks
\indentend 
\subsection{Homomorphisms}

\noindent
The notion of free semiring, the associated \ensuremath{\Varid{eval}} function and its properties
are special cases of generic results from the Algebra of
Programming~\cite{AlgofProg} which hold for any algebraic structure.
Another related notion, which will turn out to be central to this paper, is
that of \emph{homomorphism}. Informally, this is a function that preserves
algebraic structure. More formally, 

\begin{definition}[Semiring homomorphism]
A \emph{(semiring) homomorphism} between two semirings \ensuremath{\Conid{D}_{1}} and \ensuremath{\Conid{D}_{2}} is a function
\ensuremath{\Varid{h}\mathbin{::}\Conid{D}_{1}\to \Conid{D}_{2}} that preserves the semiring structure:\\
\begin{minipage}{0.5\textwidth}\indentbegin \begin{hscode}\SaveRestoreHook
\column{B}{@{}>{\hspre}l<{\hspost}@{}}%
\column{3}{@{}>{\hspre}l<{\hspost}@{}}%
\column{6}{@{}>{\hspre}l<{\hspost}@{}}%
\column{15}{@{}>{\hspre}c<{\hspost}@{}}%
\column{15E}{@{}l@{}}%
\column{18}{@{}>{\hspre}l<{\hspost}@{}}%
\column{E}{@{}>{\hspre}l<{\hspost}@{}}%
\>[3]{}\Varid{h}\;{}\<[6]%
\>[6]{}zero_{\text{\tiny D}_1}{}\<[15]%
\>[15]{}\mathrel{=}{}\<[15E]%
\>[18]{}zero_{\text{\tiny D}_2}{}\<[E]%
\\
\>[3]{}\Varid{h}\;{}\<[6]%
\>[6]{}one_{\text{\tiny D}_1}{}\<[15]%
\>[15]{}\mathrel{=}{}\<[15E]%
\>[18]{}one_{\text{\tiny D}_2}{}\<[E]%
\ColumnHook
\end{hscode}\resethooks
\indentend \end{minipage}%
\begin{minipage}{0.5\textwidth}\indentbegin \begin{hscode}\SaveRestoreHook
\column{B}{@{}>{\hspre}l<{\hspost}@{}}%
\column{3}{@{}>{\hspre}l<{\hspost}@{}}%
\column{6}{@{}>{\hspre}l<{\hspost}@{}}%
\column{10}{@{}>{\hspre}l<{\hspost}@{}}%
\column{22}{@{}>{\hspre}l<{\hspost}@{}}%
\column{27}{@{}>{\hspre}l<{\hspost}@{}}%
\column{34}{@{}>{\hspre}l<{\hspost}@{}}%
\column{46}{@{}>{\hspre}l<{\hspost}@{}}%
\column{E}{@{}>{\hspre}l<{\hspost}@{}}%
\>[3]{}\Varid{h}\;{}\<[6]%
\>[6]{}(\Varid{x}{}\<[10]%
\>[10]{}\oplus_{\text{\tiny D}_1}{}\<[22]%
\>[22]{}\Varid{y}){}\<[27]%
\>[27]{}\mathrel{=}\Varid{h}\;\Varid{x}{}\<[34]%
\>[34]{}\oplus_{\text{\tiny D}_2}{}\<[46]%
\>[46]{}\Varid{h}\;\Varid{y}{}\<[E]%
\\
\>[3]{}\Varid{h}\;{}\<[6]%
\>[6]{}(\Varid{x}{}\<[10]%
\>[10]{}\otimes_{\text{\tiny D}_1}{}\<[22]%
\>[22]{}\Varid{y}){}\<[27]%
\>[27]{}\mathrel{=}\Varid{h}\;\Varid{x}{}\<[34]%
\>[34]{}\otimes_{\text{\tiny D}_2}{}\<[46]%
\>[46]{}\Varid{h}\;\Varid{y}{}\<[E]%
\ColumnHook
\end{hscode}\resethooks
\indentend \end{minipage}
\vspace{-3mm}
\end{definition}

%
\noindent
Semiring homomorphisms compose: 
given homomorphisms \ensuremath{\Varid{h}_{1}\mathbin{::}\Conid{D}_{1}\to \Conid{D}_{2}} and \ensuremath{\Varid{h}_{2}\mathbin{::}\Conid{D}_{2}\to \Conid{D}_{3}},
their composition \ensuremath{\Varid{h}_{2}\hsdot{\circ }{.}\Varid{h}_{1}\mathbin{::}\Conid{D}_{1}\to \Conid{D}_{3}} is also a homomorphism; and the expected laws hold.
Furthermore, 
the identity function \ensuremath{\Varid{id}\mathbin{::}\Conid{D}\to \Conid{D}} is trivially a homomorphism.
%
%
%
The function \ensuremath{\Varid{eval}\;\Varid{var}} is a homomorphism for any choice of \ensuremath{\Varid{var}}.
In fact, it is the \emph{unique} homomorphism \ensuremath{\Varid{h}} such that \ensuremath{\Varid{h}\;(\Conid{Var}\;\Varid{v})\mathrel{=}\Varid{var}\;\Varid{v}}.
From this uniqueness property follow two useful properties:

\begin{property}[Fusion]\label{eq:eval:fusion}
The \emph{fusion property} allows us to incorporate any
homomorphism \ensuremath{\Varid{h}} into evaluation:
\ensuremath{\Varid{h}\hsdot{\circ }{.}\Varid{eval}\;\Varid{var}\mathrel{=}\Varid{eval}\;(\Varid{h}\hsdot{\circ }{.}\Varid{var})}.
\end{property}

\begin{property}[Reflection]\label{eq:eval:reflection}
The \emph{reflection property} observes that evaluation with \ensuremath{\Conid{Var}} is the identity:
\ensuremath{\Varid{eval}\;\Conid{Var}\mathrel{=}\Varid{id}}.
\end{property}

\subsection{Symbolic Differentiation and Dual Numbers}
\label{sec:dual}

\noindent
The \ensuremath{\Varid{eval}} homomorphism also allows us to do symbolic differentiation.
We capture its rules in the \ensuremath{\textit{derive}} function, using the \ensuremath{\Conid{Semiring}} instance
of \ensuremath{\Conid{Expr}\;\Varid{v}}, and exploiting the commutativity in the multiplication case.
\indentbegin \begin{hscode}\SaveRestoreHook
\column{B}{@{}>{\hspre}l<{\hspost}@{}}%
\column{3}{@{}>{\hspre}l<{\hspost}@{}}%
\column{19}{@{}>{\hspre}l<{\hspost}@{}}%
\column{27}{@{}>{\hspre}c<{\hspost}@{}}%
\column{27E}{@{}l@{}}%
\column{30}{@{}>{\hspre}l<{\hspost}@{}}%
\column{E}{@{}>{\hspre}l<{\hspost}@{}}%
\>[3]{}\textit{derive}\mathbin{::}\Conid{Eq}\;\Varid{v}\Rightarrow \Varid{v}\to \Conid{Expr}\;\Varid{v}\to \Conid{Expr}\;\Varid{v}{}\<[E]%
\\
\>[3]{}\textit{derive}\;\Varid{x}\;(\Conid{Var}\;\Varid{y}){}\<[27]%
\>[27]{}\mathrel{=}{}\<[27E]%
\>[30]{}\mathbf{if}\;\Varid{x}= =\Varid{y}\;\mathbf{then}\;\Varid{one}\;\mathbf{else}\;\Varid{zero}{}\<[E]%
\\
\>[3]{}\textit{derive}\;\Varid{x}\;\Conid{Zero}{}\<[27]%
\>[27]{}\mathrel{=}{}\<[27E]%
\>[30]{}\Varid{zero}{}\<[E]%
\\
\>[3]{}\textit{derive}\;\Varid{x}\;\Conid{One}{}\<[27]%
\>[27]{}\mathrel{=}{}\<[27E]%
\>[30]{}\Varid{zero}{}\<[E]%
\\
\>[3]{}\textit{derive}\;\Varid{x}\;(\Conid{Plus}\;{}\<[19]%
\>[19]{}\Varid{e}_{1}\;\Varid{e}_{2}){}\<[27]%
\>[27]{}\mathrel{=}{}\<[27E]%
\>[30]{}\textit{derive}\;\Varid{x}\;\Varid{e}_{1}\oplus\textit{derive}\;\Varid{x}\;\Varid{e}_{2}{}\<[E]%
\\
\>[3]{}\textit{derive}\;\Varid{x}\;(\Conid{Times}\;{}\<[19]%
\>[19]{}\Varid{e}_{1}\;\Varid{e}_{2}){}\<[27]%
\>[27]{}\mathrel{=}{}\<[27E]%
\>[30]{}(\colorbox{lightgray}{$\Varid{e}_{2}$}\otimes\textit{derive}\;\Varid{x}\;\Varid{e}_{1})\oplus(\colorbox{lightgray}{$\Varid{e}_{1}$}\otimes\textit{derive}\;\Varid{x}\;\Varid{e}_{2}){}\<[E]%
\ColumnHook
\end{hscode}\resethooks
\indentend 
These do not fit \ensuremath{\Varid{eval}}'s fold-style recursion scheme because, in the
case for \ensuremath{\Conid{Times}}, \ensuremath{\Varid{e}_{1}} and \ensuremath{\Varid{e}_{2}} appear outside of recursive calls (\ensuremath{\colorbox{lightgray}{$\Varid{gray}$}}).
Using the so-called ``banana-split'' property \cite{jfp/Hutton99,meijer1991,meijer1992},
%
we can turn this function into the
proper shape: we make it return a tuple \ensuremath{(\Conid{Expr}\;\Varid{v},\Conid{Expr}\;\Varid{v})} of the original
expression and the symbolic derivative.

\label{eq:deriv'}\indentbegin \begin{hscode}\SaveRestoreHook
\column{B}{@{}>{\hspre}l<{\hspost}@{}}%
\column{3}{@{}>{\hspre}l<{\hspost}@{}}%
\column{E}{@{}>{\hspre}l<{\hspost}@{}}%
\>[3]{}derive^\prime\;\Varid{x}\;\Varid{e}\mathrel{=}(\Varid{e},\textit{derive}\;\Varid{x}\;\Varid{e}){}\<[E]%
\ColumnHook
\end{hscode}\resethooks
\indentend 
\noindent
We thus obtain the following structurally recursive definition for \ensuremath{derive^\prime},
from which we may then recover \ensuremath{\textit{derive}} by projection on the second component:
\indentbegin \begin{hscode}\SaveRestoreHook
\column{B}{@{}>{\hspre}l<{\hspost}@{}}%
\column{3}{@{}>{\hspre}l<{\hspost}@{}}%
\column{20}{@{}>{\hspre}l<{\hspost}@{}}%
\column{28}{@{}>{\hspre}c<{\hspost}@{}}%
\column{28E}{@{}l@{}}%
\column{31}{@{}>{\hspre}l<{\hspost}@{}}%
\column{36}{@{}>{\hspre}l<{\hspost}@{}}%
\column{40}{@{}>{\hspre}l<{\hspost}@{}}%
\column{48}{@{}>{\hspre}c<{\hspost}@{}}%
\column{48E}{@{}l@{}}%
\column{51}{@{}>{\hspre}l<{\hspost}@{}}%
\column{E}{@{}>{\hspre}l<{\hspost}@{}}%
\>[3]{}derive^\prime\mathbin{::}\Conid{Eq}\;\Varid{v}\Rightarrow \Varid{v}\to \Conid{Expr}\;\Varid{v}\to (\Conid{Expr}\;\Varid{v},\Conid{Expr}\;\Varid{v}){}\<[E]%
\\
\>[3]{}derive^\prime\;\Varid{x}\;(\Conid{Var}\;\Varid{y}){}\<[28]%
\>[28]{}\mathrel{=}{}\<[28E]%
\>[31]{}(\Conid{Var}\;\Varid{y},{}\<[40]%
\>[40]{}\mathbf{if}\;\Varid{x}= =\Varid{y}\;\mathbf{then}\;\Varid{one}\;\mathbf{else}\;\Varid{zero}){}\<[E]%
\\
\>[3]{}derive^\prime\;\Varid{x}\;\Conid{Zero}{}\<[28]%
\>[28]{}\mathrel{=}{}\<[28E]%
\>[31]{}(\Varid{zero},{}\<[40]%
\>[40]{}\Varid{zero}){}\<[E]%
\\
\>[3]{}derive^\prime\;\Varid{x}\;\Conid{One}{}\<[28]%
\>[28]{}\mathrel{=}{}\<[28E]%
\>[31]{}(\Varid{one},{}\<[40]%
\>[40]{}\Varid{zero}){}\<[E]%
\\
\>[3]{}derive^\prime\;\Varid{x}\;(\Conid{Plus}\;{}\<[20]%
\>[20]{}\Varid{e}_{1}\;\Varid{e}_{2}){}\<[28]%
\>[28]{}\mathrel{=}{}\<[28E]%
\>[31]{}\mathbf{let}\;{}\<[36]%
\>[36]{}(\Varid{e}_{1}',\Varid{de}_{1}){}\<[48]%
\>[48]{}\mathrel{=}{}\<[48E]%
\>[51]{}derive^\prime\;\Varid{x}\;\Varid{e}_{1}{}\<[E]%
\\
\>[36]{}(\Varid{e}_{2}',\Varid{de}_{2}){}\<[48]%
\>[48]{}\mathrel{=}{}\<[48E]%
\>[51]{}derive^\prime\;\Varid{x}\;\Varid{e}_{2}{}\<[E]%
\\
\>[31]{}\mathbf{in}\;{}\<[36]%
\>[36]{}(\Varid{e}_{1}'\oplus\Varid{e}_{2}',\Varid{de}_{1}\oplus\Varid{de}_{2}){}\<[E]%
\\
\>[3]{}derive^\prime\;\Varid{x}\;(\Conid{Times}\;{}\<[20]%
\>[20]{}\Varid{e}_{1}\;\Varid{e}_{2}){}\<[28]%
\>[28]{}\mathrel{=}{}\<[28E]%
\>[31]{}\mathbf{let}\;{}\<[36]%
\>[36]{}(\Varid{e}_{1}',\Varid{de}_{1}){}\<[48]%
\>[48]{}\mathrel{=}{}\<[48E]%
\>[51]{}derive^\prime\;\Varid{x}\;\Varid{e}_{1}{}\<[E]%
\\
\>[36]{}(\Varid{e}_{2}',\Varid{de}_{2}){}\<[48]%
\>[48]{}\mathrel{=}{}\<[48E]%
\>[51]{}derive^\prime\;\Varid{x}\;\Varid{e}_{2}{}\<[E]%
\\
\>[31]{}\mathbf{in}\;{}\<[36]%
\>[36]{}(\Varid{e}_{1}\otimes\Varid{e}_{2}',(\Varid{e}_{2}'\otimes\Varid{de}_{1})\oplus(\Varid{e}_{1}'\otimes\Varid{de}_{2})){}\<[E]%
\ColumnHook
\end{hscode}\resethooks
\indentend 


\noindent
Before we write the functionality of \ensuremath{derive^\prime} in terms of \ensuremath{\Varid{eval}}, we generalize \ensuremath{(\Conid{Expr}\;\Varid{v},\Conid{Expr}\;\Varid{v})} to tuples whose components are drawn from an arbitrary semiring.
In the AD literature, these are known as~\emph{dual numbers},
originally due to Clifford~\citep{Clifford73}.

\noindent
\begin{minipage}{0.5\textwidth}
\indentbegin \begin{hscode}\SaveRestoreHook
\column{B}{@{}>{\hspre}l<{\hspost}@{}}%
\column{3}{@{}>{\hspre}l<{\hspost}@{}}%
\column{22}{@{}>{\hspre}l<{\hspost}@{}}%
\column{E}{@{}>{\hspre}l<{\hspost}@{}}%
\>[3]{}\mathbf{data}\;\Conid{Dual}\;\Varid{d}\mathrel{=}\Conid{D}\;\{\mskip1.5mu {}\<[22]%
\>[22]{}\Varid{pri}^D\mathbin{::}\Varid{d},{}\<[E]%
\\
\>[22]{}\Varid{tan}^D\mathbin{::}\Varid{d}\mskip1.5mu\}{}\<[E]%
\ColumnHook
\end{hscode}\resethooks
\indentend 
\end{minipage}%
\begin{minipage}{0.5\textwidth}
\indentbegin \begin{hscode}\SaveRestoreHook
\column{B}{@{}>{\hspre}l<{\hspost}@{}}%
\column{3}{@{}>{\hspre}l<{\hspost}@{}}%
\column{5}{@{}>{\hspre}l<{\hspost}@{}}%
\column{E}{@{}>{\hspre}l<{\hspost}@{}}%
\>[3]{}\mathbf{instance}\;\Conid{Functor}\;\Conid{Dual}\;\mathbf{where}{}\<[E]%
\\
\>[3]{}\hsindent{2}{}\<[5]%
\>[5]{}\Varid{fmap}\mathbin{::}(\Varid{a}\to \Varid{b})\to \Conid{Dual}\;\Varid{a}\to \Conid{Dual}\;\Varid{b}{}\<[E]%
\\
\>[3]{}\hsindent{2}{}\<[5]%
\>[5]{}\Varid{fmap}\;\Varid{h}\;(\Conid{D}\;\Varid{f}\;\Varid{df})\mathrel{=}\Conid{D}\;(\Varid{h}\;\Varid{f})\;(\Varid{h}\;\Varid{df}){}\<[E]%
\ColumnHook
\end{hscode}\resethooks
\indentend 
\end{minipage}
\noindent
Dual numbers are functorial; \ensuremath{\Varid{fmap}\;\Varid{f}} transforms both parameters using \ensuremath{\Varid{f}}.
The two components of a dual number \ensuremath{\Conid{D}\;\Varid{x}\;\Varid{dx}} are respectively known
as the \emph{primal} \ensuremath{\Varid{x}} and the \emph{tangent} \ensuremath{\Varid{dx}}.
%
For a semiring \ensuremath{\Varid{d}}, the type \ensuremath{\Conid{Dual}\;\Varid{d}} also admits a semiring structure, where
the definition of tangent precisely follows the corresponding cases of \ensuremath{\textit{derive}}.
\indentbegin \begin{hscode}\SaveRestoreHook
\column{B}{@{}>{\hspre}l<{\hspost}@{}}%
\column{3}{@{}>{\hspre}l<{\hspost}@{}}%
\column{5}{@{}>{\hspre}l<{\hspost}@{}}%
\column{15}{@{}>{\hspre}l<{\hspost}@{}}%
\column{24}{@{}>{\hspre}l<{\hspost}@{}}%
\column{34}{@{}>{\hspre}l<{\hspost}@{}}%
\column{39}{@{}>{\hspre}l<{\hspost}@{}}%
\column{50}{@{}>{\hspre}l<{\hspost}@{}}%
\column{54}{@{}>{\hspre}l<{\hspost}@{}}%
\column{E}{@{}>{\hspre}l<{\hspost}@{}}%
\>[3]{}\mathbf{instance}\;\Conid{Semiring}\;\Varid{d}\Rightarrow \Conid{Semiring}\;(\Conid{Dual}\;\Varid{d})\;\mathbf{where}{}\<[E]%
\\
\>[3]{}\hsindent{2}{}\<[5]%
\>[5]{}\Varid{zero}{}\<[34]%
\>[34]{}\mathrel{=}\Conid{D}\;{}\<[39]%
\>[39]{}\Varid{zero}\;{}\<[54]%
\>[54]{}\Varid{zero}{}\<[E]%
\\
\>[3]{}\hsindent{2}{}\<[5]%
\>[5]{}\Varid{one}{}\<[34]%
\>[34]{}\mathrel{=}\Conid{D}\;{}\<[39]%
\>[39]{}\Varid{one}\;{}\<[54]%
\>[54]{}\Varid{zero}{}\<[E]%
\\
\>[3]{}\hsindent{2}{}\<[5]%
\>[5]{}(\Conid{D}\;\Varid{f}\;\Varid{df}){}\<[15]%
\>[15]{}\oplus{}\<[24]%
\>[24]{}(\Conid{D}\;\Varid{g}\;\Varid{dg}){}\<[34]%
\>[34]{}\mathrel{=}\Conid{D}\;{}\<[39]%
\>[39]{}(\Varid{f}\oplus{}\<[50]%
\>[50]{}\Varid{g})\;{}\<[54]%
\>[54]{}(\Varid{df}\oplus\Varid{dg}){}\<[E]%
\\
\>[3]{}\hsindent{2}{}\<[5]%
\>[5]{}(\Conid{D}\;\Varid{f}\;\Varid{df}){}\<[15]%
\>[15]{}\otimes{}\<[24]%
\>[24]{}(\Conid{D}\;\Varid{g}\;\Varid{dg}){}\<[34]%
\>[34]{}\mathrel{=}\Conid{D}\;{}\<[39]%
\>[39]{}(\Varid{f}\otimes\Varid{g})\;{}\<[54]%
\>[54]{}((\Varid{g}\otimes\Varid{df})\oplus(\Varid{f}\otimes\Varid{dg})){}\<[E]%
\ColumnHook
\end{hscode}\resethooks
\indentend Now, symbolic differentiation may be seen as another instance of \ensuremath{\Varid{eval}}uation, but into
the semiring of dual numbers over symbolic expressions.
\indentbegin \begin{hscode}\SaveRestoreHook
\column{B}{@{}>{\hspre}l<{\hspost}@{}}%
\column{3}{@{}>{\hspre}l<{\hspost}@{}}%
\column{13}{@{}>{\hspre}l<{\hspost}@{}}%
\column{31}{@{}>{\hspre}l<{\hspost}@{}}%
\column{36}{@{}>{\hspre}l<{\hspost}@{}}%
\column{39}{@{}>{\hspre}l<{\hspost}@{}}%
\column{E}{@{}>{\hspre}l<{\hspost}@{}}%
\>[3]{}symbolic\mathbin{::}\Conid{Eq}\;\Varid{v}\Rightarrow \Varid{v}\to \Conid{Expr}\;\Varid{v}\to \Conid{Dual}\;(\Conid{Expr}\;\Varid{v}){}\<[E]%
\\
\>[3]{}symbolic\;\Varid{x}{}\<[13]%
\>[13]{}\mathrel{=}\Varid{eval}\;\Varid{gen}\;\mathbf{where}\;{}\<[31]%
\>[31]{}\Varid{gen}\;{}\<[36]%
\>[36]{}\Varid{y}{}\<[39]%
\>[39]{}\mathrel{=}\Conid{D}\;(\Conid{Var}\;\Varid{y})\;(\delta_{x}\;\Varid{y}){}\<[E]%
\\
\>[31]{}\delta_{x}\;{}\<[36]%
\>[36]{}\Varid{y}{}\<[39]%
\>[39]{}\mathrel{=}\mathbf{if}\;\Varid{x}= =\Varid{y}\;\mathbf{then}\;\Varid{one}\;\mathbf{else}\;\Varid{zero}{}\<[E]%
\ColumnHook
\end{hscode}\resethooks
\indentend 
%
%
%

\subsection{Classic Forward Automatic Differentiation}
\label{sec:classic}

\noindent
Automatic differentiation (AD) is the umbrella term for
algorithms that programmatically compute
both the value \emph{and} the derivative of an expression \emph{at a point}
\cite{pacmpl/KrawiecJKEEF22,higher-order,hoffmann2016hitchhiker}.
Thus, if we combine symbolic differentiation with subsequent evaluation into a semiring \ensuremath{\Varid{d}},
we obtain a specification for forward-mode AD: 


\label{spec:ad}\indentbegin \begin{hscode}\SaveRestoreHook
\column{B}{@{}>{\hspre}l<{\hspost}@{}}%
\column{3}{@{}>{\hspre}l<{\hspost}@{}}%
\column{E}{@{}>{\hspre}l<{\hspost}@{}}%
\>[3]{}\Varid{forwardAD}\;\Varid{var}\;\Varid{x}\mathrel{=}\Varid{fmap}\;(\Varid{eval}\;\Varid{var})\hsdot{\circ }{.}symbolic\;\Varid{x}{}\<[E]%
\ColumnHook
\end{hscode}\resethooks
\indentend 
%

\noindent
We can make this 
specification more efficient by exploiting \ensuremath{\Varid{eval}}'s
fusion property. 
\indentbegin \begin{hscode}\SaveRestoreHook
\column{B}{@{}>{\hspre}l<{\hspost}@{}}%
\column{3}{@{}>{\hspre}l<{\hspost}@{}}%
\column{6}{@{}>{\hspre}l<{\hspost}@{}}%
\column{40}{@{}>{\hspre}l<{\hspost}@{}}%
\column{42}{@{}>{\hspre}l<{\hspost}@{}}%
\column{47}{@{}>{\hspre}l<{\hspost}@{}}%
\column{49}{@{}>{\hspre}l<{\hspost}@{}}%
\column{E}{@{}>{\hspre}l<{\hspost}@{}}%
\>[6]{}\Varid{fmap}\;(\Varid{eval}\;\Varid{var})\hsdot{\circ }{.}symbolic\;\Varid{x}{}\<[E]%
\\
\>[3]{}\mathrel{=}\mbox{\commentbegin  definition of \ensuremath{symbolic}  \commentend}{}\<[E]%
\\
\>[3]{}\hsindent{3}{}\<[6]%
\>[6]{}\Varid{fmap}\;(\Varid{eval}\;\Varid{var})\hsdot{\circ }{.}\Varid{eval}\;\Varid{gen}\;\mathbf{where}\;{}\<[40]%
\>[40]{}\Varid{gen}\;\Varid{y}{}\<[47]%
\>[47]{}\mathrel{=}\Conid{D}\;(\Conid{Var}\;\Varid{y})\;(\delta_{x}\;\Varid{y}){}\<[E]%
\\
\>[40]{}\delta_{x}\;\Varid{y}{}\<[47]%
\>[47]{}\mathrel{=}\mathbf{if}\;\Varid{x}= =\Varid{y}\;\mathbf{then}\;\Varid{one}\;\mathbf{else}\;\Varid{zero}{}\<[E]%
\\
\>[3]{}\mathrel{=}\mbox{\commentbegin  \ensuremath{\Varid{eval}} fusion (Property~\ref{eq:eval:fusion})  \commentend}{}\<[E]%
\\
\>[3]{}\hsindent{3}{}\<[6]%
\>[6]{}\Varid{eval}\;(\Varid{fmap}\;(\Varid{eval}\;\Varid{var})\hsdot{\circ }{.}\Varid{gen})\;\mathbf{where}\;{}\<[42]%
\>[42]{}\Varid{gen}\;\Varid{y}{}\<[49]%
\>[49]{}\mathrel{=}\Conid{D}\;(\Conid{Var}\;\Varid{y})\;(\delta_{x}\;\Varid{y}){}\<[E]%
\\
\>[42]{}\delta_{x}\;\Varid{y}{}\<[49]%
\>[49]{}\mathrel{=}\mathbf{if}\;\Varid{x}= =\Varid{y}\;\mathbf{then}\;\Varid{one}\;\mathbf{else}\;\Varid{zero}{}\<[E]%
\ColumnHook
\end{hscode}\resethooks
\indentend After inlining \ensuremath{\Varid{fmap}\;(\Varid{eval}\;\Varid{var})} into \ensuremath{\Varid{gen}},
we get the classic definition of
forward-mode AD, which calls \ensuremath{\Varid{eval}} only once. Differences
compared with \ensuremath{symbolic} are highlighted in gray.
\indentbegin \begin{hscode}\SaveRestoreHook
\column{B}{@{}>{\hspre}l<{\hspost}@{}}%
\column{3}{@{}>{\hspre}l<{\hspost}@{}}%
\column{43}{@{}>{\hspre}l<{\hspost}@{}}%
\column{50}{@{}>{\hspre}l<{\hspost}@{}}%
\column{E}{@{}>{\hspre}l<{\hspost}@{}}%
\>[3]{}\Varid{forwardAD}\mathbin{::}(\Conid{Eq}\;\Varid{v},\Conid{Semiring}\;\Varid{d})\Rightarrow (\Varid{v}\to \Varid{d})\to \Varid{v}\to \Conid{Expr}\;\Varid{v}\to \Conid{Dual}\;\Varid{d}{}\<[E]%
\\
\>[3]{}\Varid{forwardAD}\;\colorbox{lightgray}{$\Varid{var}$}\;\Varid{x}\mathrel{=}\Varid{eval}\;\Varid{gen}\;\mathbf{where}\;{}\<[43]%
\>[43]{}\Varid{gen}\;\Varid{y}{}\<[50]%
\>[50]{}\mathrel{=}\Conid{D}\;(\colorbox{lightgray}{$\Varid{var}$}\;\Varid{y})\;(\delta_{x}\;\Varid{y}){}\<[E]%
\\
\>[43]{}\delta_{x}\;\Varid{y}{}\<[50]%
\>[50]{}\mathrel{=}\mathbf{if}\;\Varid{x}= =\Varid{y}\;\mathbf{then}\;\Varid{one}\;\mathbf{else}\;\Varid{zero}{}\<[E]%
\ColumnHook
\end{hscode}\resethooks
\indentend For example, the derivative of \ensuremath{\Varid{example}_{1}} at the point \ensuremath{\Conid{X}\mathrel{=}\mathrm{5}} is \ensuremath{\mathrm{11}}; the
process computing the result is depicted in Figure~\ref{fig:fmad}.
\indentbegin \begin{hscode}\SaveRestoreHook
\column{B}{@{}>{\hspre}l<{\hspost}@{}}%
\column{3}{@{}>{\hspre}l<{\hspost}@{}}%
\column{E}{@{}>{\hspre}l<{\hspost}@{}}%
\>[3]{}\mathbin{>}\Varid{forwardAD}\;(\lambda \Conid{X}\to \mathrm{5})\;\Conid{X}\;\Varid{example}_{1}{}\<[E]%
\\
\>[3]{}\Conid{D}\;\mathrm{30}\;\mathrm{11}{}\<[E]%
\ColumnHook
\end{hscode}\resethooks
\indentend 
\begin{figure}[t]
  \begin{center}
    \tikzset{every picture/.style={line width=0.75pt}}
    \tikzstyle{boo} = [circle]
    \tikzstyle{bag} = [rectangle,draw,inner sep=1pt, text centered,minimum width=1cm, minimum height=5mm]
    \tikzstyle{end} = [rectangle,draw,inner sep=1pt, text centered,minimum width=1cm, minimum height=5mm]
    \centering
    \begin{tikzpicture}[edge from parent/.style={latex-,draw},grow=left]
    \tikzstyle{level}=[level distance=4cm, sibling distance = 1cm]
    \node[boo] { }
        child {
            node[bag] {\ensuremath{\otimes}}
            child {
                node[end] {\ensuremath{\Varid{x}}}
                edge from parent
                node[above,xshift=1mm,yshift=1mm]  {\ensuremath{\Conid{D}\;\mathrm{5}\;\mathrm{1}}}
            }
            child {
                node[bag] {\ensuremath{\oplus}}
                child {
                  node[end] {\ensuremath{\Varid{x}}}
                  edge from parent
                  node[above,xshift=1mm,yshift=1mm]  {\ensuremath{\Conid{D}\;\mathrm{5}\;\mathrm{1}}}
                }
                child {
                  node[end] {\ensuremath{\Varid{one}}}
                  edge from parent
                  node[below,xshift=1mm,yshift=-1mm]  {\ensuremath{\Conid{D}\;\mathrm{1}\;\mathrm{0}}}
                }
                edge from parent
                node[below,xshift=1mm,yshift=-1mm]  {\ensuremath{\Conid{D}\;\mathrm{6}\;\mathrm{1}}}
            }
            edge from parent
            node[above]  {\ensuremath{\Conid{D}\;\mathrm{30}\;\mathrm{11}}}
        };
    \end{tikzpicture}
  \end{center}
\caption{Running forward-mode AD on \ensuremath{\Varid{example}_{1}} at \ensuremath{\Conid{X}\mathrel{=}\mathrm{5}}.}
\label{fig:fmad}
\end{figure}


\subsection{Symbolic AD as Forward AD}

\noindent
Observe that the only essential difference between symbolic and automatic
differentiation is that the former evaluates into the symbolic \ensuremath{\Conid{Expr}\;\Varid{v}} semiring, while
the latter uses a \squote[numeric] semiring \ensuremath{\Varid{d}}. In fact, we can also
have \ensuremath{\Varid{forwardAD}} use the symbolic semiring, interpreting variables with \ensuremath{\Conid{Var}}.
\indentbegin \begin{hscode}\SaveRestoreHook
\column{B}{@{}>{\hspre}l<{\hspost}@{}}%
\column{3}{@{}>{\hspre}l<{\hspost}@{}}%
\column{7}{@{}>{\hspre}l<{\hspost}@{}}%
\column{E}{@{}>{\hspre}l<{\hspost}@{}}%
\>[7]{}\Varid{forwardAD}\;\Conid{Var}{}\<[E]%
\\
\>[3]{}\mathrel{=}\mbox{\commentbegin  AD specification (Specification~\ref{spec:ad})  \commentend}{}\<[E]%
\\
\>[3]{}\hsindent{4}{}\<[7]%
\>[7]{}\Varid{fmap}\;(\Varid{eval}\;\Conid{Var})\hsdot{\circ }{.}symbolic{}\<[E]%
\\
\>[3]{}\mathrel{=}\mbox{\commentbegin  reflection property (Property~\ref{eq:eval:reflection})  \commentend}{}\<[E]%
\\
\>[3]{}\hsindent{4}{}\<[7]%
\>[7]{}\Varid{fmap}\;\Varid{id}\hsdot{\circ }{.}symbolic{}\<[E]%
\\
\>[3]{}\mathrel{=}\mbox{\commentbegin  functor law and \ensuremath{\Varid{id}} as unit of composition  \commentend}{}\<[E]%
\\
\>[3]{}\hsindent{4}{}\<[7]%
\>[7]{}symbolic{}\<[E]%
\ColumnHook
\end{hscode}\resethooks
\indentend In summary, \ensuremath{symbolic} is a special case of \ensuremath{\Varid{forwardAD}}:
\begin{equation*}
\ensuremath{symbolic} = \ensuremath{\Varid{forwardAD}\;\Conid{Var}}
\end{equation*}
This sets the stage for the remainder of this paper:
we use algebraic techniques to reason about the structure of differentiation algorithms
and  by exposing more abstract structure, one and the same abstract algorithm (\Cref{sec:abstractad})
subsumes various concrete AD flavors.


\section{Abstract Automatic Differentiation}\label{sec:abstractad}


\noindent
The central claim of this paper is that we can express the two previous
differentiation functions (symbolic differentiation and classic forward-mode
AD) as well as several more variants to come later, as the same one-line
generic program.
%
%
This section identifies three fundamental algebraic structures to accomplish
this:
\ensuremath{\Varid{d}}-modules \ensuremath{\Varid{e}}, generalized dual numbers \ensuremath{\Varid{d}\ltimes\Varid{e}}, and the Kronecker
delta function.
It also shows how to instantiate these to recover a version of
forward-mode AD that computes the gradient vector of all partial
derivatives \emph{at once}, rather than for a single variable at a time. 

\begin{figure}
\begin{center}
\fbox{
\begin{minipage}[t]{.5\textwidth}
\indentbegin \begin{hscode}\SaveRestoreHook
\column{B}{@{}>{\hspre}l<{\hspost}@{}}%
\column{3}{@{}>{\hspre}l<{\hspost}@{}}%
\column{6}{@{}>{\hspre}l<{\hspost}@{}}%
\column{16}{@{}>{\hspre}l<{\hspost}@{}}%
\column{35}{@{}>{\hspre}c<{\hspost}@{}}%
\column{35E}{@{}l@{}}%
\column{38}{@{}>{\hspre}l<{\hspost}@{}}%
\column{E}{@{}>{\hspre}l<{\hspost}@{}}%
\>[3]{}\Varid{e}{}\<[6]%
\>[6]{}\oplus_\text{\tiny E}{}\<[16]%
\>[16]{}zero_\text{\tiny E}{}\<[35]%
\>[35]{}\mathrel{=}{}\<[35E]%
\>[38]{}\Varid{e}{}\<[E]%
\\
\>[3]{}\Varid{e}{}\<[6]%
\>[6]{}\oplus_\text{\tiny E}{}\<[16]%
\>[16]{}(\Varid{e'}\oplus_\text{\tiny E}\Varid{e''}){}\<[35]%
\>[35]{}\mathrel{=}{}\<[35E]%
\>[38]{}(\Varid{e}\oplus_\text{\tiny E}\Varid{e'})\oplus_\text{\tiny E}\Varid{e''}{}\<[E]%
\\
\>[3]{}\Varid{e}{}\<[6]%
\>[6]{}\oplus_\text{\tiny E}{}\<[16]%
\>[16]{}\Varid{e'}{}\<[35]%
\>[35]{}\mathrel{=}{}\<[35E]%
\>[38]{}\Varid{e'}\oplus_\text{\tiny E}\Varid{e}{}\<[E]%
\ColumnHook
\end{hscode}\resethooks
\indentend 
\vspace{-3mm}\indentbegin \begin{hscode}\SaveRestoreHook
\column{B}{@{}>{\hspre}l<{\hspost}@{}}%
\column{3}{@{}>{\hspre}l<{\hspost}@{}}%
\column{6}{@{}>{\hspre}l<{\hspost}@{}}%
\column{14}{@{}>{\hspre}l<{\hspost}@{}}%
\column{31}{@{}>{\hspre}c<{\hspost}@{}}%
\column{31E}{@{}l@{}}%
\column{34}{@{}>{\hspre}l<{\hspost}@{}}%
\column{E}{@{}>{\hspre}l<{\hspost}@{}}%
\>[3]{}\Varid{d}{}\<[6]%
\>[6]{}\bullet{}\<[14]%
\>[14]{}zero_\text{\tiny E}{}\<[31]%
\>[31]{}\mathrel{=}{}\<[31E]%
\>[34]{}zero_\text{\tiny E}{}\<[E]%
\\
\>[3]{}\Varid{d}{}\<[6]%
\>[6]{}\bullet{}\<[14]%
\>[14]{}(\Varid{e}\oplus_\text{\tiny E}\Varid{e'}){}\<[31]%
\>[31]{}\mathrel{=}{}\<[31E]%
\>[34]{}(\Varid{d}\bullet\Varid{e})\oplus_\text{\tiny E}(\Varid{d}\bullet\Varid{e'}){}\<[E]%
\ColumnHook
\end{hscode}\resethooks
\indentend \vspace{-3mm}
\end{minipage}%
\begin{minipage}[t]{.5\textwidth}
\indentbegin \begin{hscode}\SaveRestoreHook
\column{B}{@{}>{\hspre}l<{\hspost}@{}}%
\column{3}{@{}>{\hspre}l<{\hspost}@{}}%
\column{21}{@{}>{\hspre}l<{\hspost}@{}}%
\column{29}{@{}>{\hspre}l<{\hspost}@{}}%
\column{32}{@{}>{\hspre}l<{\hspost}@{}}%
\column{35}{@{}>{\hspre}l<{\hspost}@{}}%
\column{E}{@{}>{\hspre}l<{\hspost}@{}}%
\>[3]{}zero_{\text{\tiny D}}{}\<[21]%
\>[21]{}\bullet{}\<[29]%
\>[29]{}\Varid{e}{}\<[32]%
\>[32]{}\mathrel{=}{}\<[35]%
\>[35]{}zero_\text{\tiny E}{}\<[E]%
\\
\>[3]{}(\Varid{d}\oplus_{\text{\tiny D}}\Varid{d'}){}\<[21]%
\>[21]{}\bullet{}\<[29]%
\>[29]{}\Varid{e}{}\<[32]%
\>[32]{}\mathrel{=}{}\<[35]%
\>[35]{}(\Varid{d}\bullet\Varid{e})\oplus_\text{\tiny E}(\Varid{d'}\bullet\Varid{e}){}\<[E]%
\\
\>[3]{}one_{\text{\tiny D}}{}\<[21]%
\>[21]{}\bullet{}\<[29]%
\>[29]{}\Varid{e}{}\<[32]%
\>[32]{}\mathrel{=}\Varid{e}{}\<[E]%
\\
\>[3]{}(\Varid{d}\otimes_{\text{\tiny D}}\Varid{d'}){}\<[21]%
\>[21]{}\bullet{}\<[29]%
\>[29]{}\Varid{e}{}\<[32]%
\>[32]{}\mathrel{=}\Varid{d}\bullet(\Varid{d'}\bullet\Varid{e}){}\<[E]%
\ColumnHook
\end{hscode}\resethooks
\indentend \end{minipage}
}
\end{center}
\vspace{-3mm}
\caption{The laws for a \ensuremath{\Conid{D}}-module \ensuremath{\Conid{E}} over a semiring \ensuremath{\Conid{D}}, where \ensuremath{\Conid{E}} is a commutative monoid.
We overload the monoidal operation \ensuremath{\oplus} for the corresponding \ensuremath{\Conid{Semiring}}, to emphasize its additive interpretation.}
\label{fig:module-laws}
\end{figure}
%

\subsection{Modules over Semirings}\label{sec:module}

\noindent
Our first abstraction generalizes the notion of gradient $\nabla e$
of an expression $e$.
\begin{equation*}
\nabla e = \left[
\begin{array}{c}
\partialfrac{e}{x_1} \\
\vdots \\
\partialfrac{e}{x_n}
\end{array}
\right]
\end{equation*}
By analogy with vector spaces (over a field) or modules (over a ring), we consider
semimodules over a semiring \ensuremath{\Varid{d}} \cite{Golan1999}, which we refer to as \emph{\ensuremath{\Varid{d}}-modules}.


\begin{definition}[Semimodule]
Given a semiring \ensuremath{\Conid{D}}, then a \ensuremath{\Conid{D}}-module \ensuremath{\Conid{E}} is defined as a
commutative monoid \ensuremath{\Conid{E}},
together with \emph{scalar multiplication} \ensuremath{(\bullet)\mathbin{::}\Conid{D}\to \Conid{E}\to \Conid{E}},
which distributes over the additive structure of \ensuremath{\Conid{E}}.
This means that \ensuremath{\Conid{E}} should have a commutative, associative binary operator
\ensuremath{\oplus} with neutral element \ensuremath{\Varid{zero}}.
Figure~\ref{fig:module-laws} gives \ensuremath{\Conid{E}}'s monoid laws and the laws that govern the interaction between
the monoid structure of \ensuremath{\Conid{E}} and the scalar multiplication.
\end{definition}

\noindent
In Haskell, we capture the concept of a semimodule in the type class 
\ensuremath{\Conid{Module}},\footnote{We use the functional dependency \ensuremath{\Varid{e}\to \Varid{d}} to anchor a \ensuremath{\Varid{d}}-module \ensuremath{\Varid{e}} to its underlying semiring \ensuremath{\Varid{d}} (\ref{app:functions}).}
on top of the \ensuremath{\Conid{Monoid}} type class.

\noindent
\begin{minipage}{0.45\textwidth}\indentbegin \begin{hscode}\SaveRestoreHook
\column{B}{@{}>{\hspre}l<{\hspost}@{}}%
\column{3}{@{}>{\hspre}l<{\hspost}@{}}%
\column{5}{@{}>{\hspre}l<{\hspost}@{}}%
\column{14}{@{}>{\hspre}l<{\hspost}@{}}%
\column{E}{@{}>{\hspre}l<{\hspost}@{}}%
\>[3]{}\mathbf{class}\;\Conid{Monoid}\;\Varid{e}\;\mathbf{where}{}\<[E]%
\\
\>[3]{}\hsindent{2}{}\<[5]%
\>[5]{}\Varid{zero}{}\<[14]%
\>[14]{}\mathbin{::}\Varid{e}{}\<[E]%
\\
\>[3]{}\hsindent{2}{}\<[5]%
\>[5]{}(\oplus){}\<[14]%
\>[14]{}\mathbin{::}\Varid{e}\to \Varid{e}\to \Varid{e}{}\<[E]%
\ColumnHook
\end{hscode}\resethooks
\indentend \end{minipage}%
\begin{minipage}{0.55\textwidth}
\indentbegin \begin{hscode}\SaveRestoreHook
\column{B}{@{}>{\hspre}l<{\hspost}@{}}%
\column{3}{@{}>{\hspre}l<{\hspost}@{}}%
\column{7}{@{}>{\hspre}l<{\hspost}@{}}%
\column{10}{@{}>{\hspre}l<{\hspost}@{}}%
\column{14}{@{}>{\hspre}l<{\hspost}@{}}%
\column{E}{@{}>{\hspre}l<{\hspost}@{}}%
\>[3]{}\mathbf{class}\;{}\<[10]%
\>[10]{}(\Conid{Semiring}\;\Varid{d},\Conid{Monoid}\;\Varid{e}){}\<[E]%
\\
\>[10]{}\Rightarrow {}\<[14]%
\>[14]{}\Conid{Module}\;\Varid{d}\;\Varid{e}\mid \Varid{e}\to \Varid{d}\;\mathbf{where}{}\<[E]%
\\
\>[3]{}\hsindent{4}{}\<[7]%
\>[7]{}(\bullet)\mathbin{::}\Varid{d}\to \Varid{e}\to \Varid{e}{}\<[E]%
\ColumnHook
\end{hscode}\resethooks
\indentend \end{minipage}

\noindent Every semiring \ensuremath{\Varid{d}} may be viewed trivially as a \ensuremath{\Varid{d}}-module, by taking \ensuremath{(\bullet)\mathrel{=}(\otimes)}.
A well-known example that is not itself a semiring is the vector space \ensuremath{\mathbb{R}^3}; it forms a
monoid with vector addition and the zero vector $\left[ \begin{array}{ccc} 0 & 0 & 0 \end{array} \right]$,
and an  \ensuremath{\mathbb{R}}-module with the usual scalar multiplication.

\subsection{Nagata Numbers}\label{sec:nagata}

\noindent
The second abstraction generalizes dual numbers of \Cref{sec:dual} 
to two components \ensuremath{\Varid{d}} and \ensuremath{\Varid{e}} that have \emph{distinct} types,
reflecting the idea that the value of an expression (its primal),
and that of its derivative (tangent), conceptually live in different spaces.
We exploit precisely the generalization that permits primal and tangent \emph{to vary
independently}.
This concept first appeared in the work of Nagata~\cite{Nagata1962} in the early 1950s, under the name \dquote[idealization of a module]. However, the application of Nagata's construction to the study of AD appears to be novel.
By analogy with dual numbers, we call the type \ensuremath{\Varid{d}\ltimes\Varid{e}} \emph{Nagata numbers}.

\noindent
\begin{minipage}{0.5\textwidth}
\indentbegin \begin{hscode}\SaveRestoreHook
\column{B}{@{}>{\hspre}l<{\hspost}@{}}%
\column{3}{@{}>{\hspre}l<{\hspost}@{}}%
\column{33}{@{}>{\hspre}l<{\hspost}@{}}%
\column{38}{@{}>{\hspre}l<{\hspost}@{}}%
\column{44}{@{}>{\hspre}c<{\hspost}@{}}%
\column{44E}{@{}l@{}}%
\column{E}{@{}>{\hspre}l<{\hspost}@{}}%
\>[3]{}\mathbf{data}\;\Varid{d}\ltimes\Varid{e}\mathrel{=}\Varid{N}\;\{\mskip1.5mu {}\<[33]%
\>[33]{}\Varid{pri}^{N}{}\<[38]%
\>[38]{}\mathbin{::}\Varid{d}{}\<[44]%
\>[44]{},{}\<[44E]%
\\
\>[33]{}\Varid{tan}^{N}{}\<[38]%
\>[38]{}\mathbin{::}\Varid{e}{}\<[44]%
\>[44]{}\mskip1.5mu\}{}\<[44E]%
\ColumnHook
\end{hscode}\resethooks
\indentend 
\end{minipage}%
\begin{minipage}{0.5\textwidth}
\indentbegin \begin{hscode}\SaveRestoreHook
\column{B}{@{}>{\hspre}l<{\hspost}@{}}%
\column{3}{@{}>{\hspre}l<{\hspost}@{}}%
\column{5}{@{}>{\hspre}l<{\hspost}@{}}%
\column{E}{@{}>{\hspre}l<{\hspost}@{}}%
\>[3]{}\mathbf{instance}\;\Conid{Functor}\;(\Varid{d}\ltimes\cdot )\;\mathbf{where}{}\<[E]%
\\
\>[3]{}\hsindent{2}{}\<[5]%
\>[5]{}\Varid{fmap}\;\Varid{h}\;(\Varid{N}\;\Varid{f}\;\Varid{df})\mathrel{=}\Varid{N}\;\Varid{f}\;(\Varid{h}\;\Varid{df}){}\<[E]%
\ColumnHook
\end{hscode}\resethooks
\indentend \end{minipage}
Observe that the type constructor \ensuremath{\Varid{d}\ltimes\cdot } is functorial in its second parameter
and a \emph{bi}functor in both parameters (\ref{app:hod}).

Nagata proved the following fundamental theorem.
\begin{theorem}\label{lem:nagata}
  Given a \ensuremath{\Varid{d}}-module \ensuremath{\Varid{e}}, then \ensuremath{\Varid{d}\ltimes\Varid{e}} admits a semiring structure.
\end{theorem}

\noindent
The Haskell type class instance below witnesses this theorem.

\noindent

\indentbegin \begin{hscode}\SaveRestoreHook
\column{B}{@{}>{\hspre}l<{\hspost}@{}}%
\column{3}{@{}>{\hspre}l<{\hspost}@{}}%
\column{5}{@{}>{\hspre}l<{\hspost}@{}}%
\column{23}{@{}>{\hspre}l<{\hspost}@{}}%
\column{34}{@{}>{\hspre}l<{\hspost}@{}}%
\column{40}{@{}>{\hspre}l<{\hspost}@{}}%
\column{51}{@{}>{\hspre}l<{\hspost}@{}}%
\column{55}{@{}>{\hspre}l<{\hspost}@{}}%
\column{E}{@{}>{\hspre}l<{\hspost}@{}}%
\>[3]{}\mathbf{instance}\;\Conid{Module}\;\Varid{d}\;\Varid{e}\Rightarrow \Conid{Semiring}\;(\Varid{d}\ltimes\Varid{e})\;\mathbf{where}{}\<[E]%
\\
\>[3]{}\hsindent{2}{}\<[5]%
\>[5]{}\Varid{zero}{}\<[34]%
\>[34]{}\mathrel{=}\Varid{N}\;{}\<[40]%
\>[40]{}\Varid{zero}\;{}\<[55]%
\>[55]{}\Varid{zero}{}\<[E]%
\\
\>[3]{}\hsindent{2}{}\<[5]%
\>[5]{}\Varid{one}{}\<[34]%
\>[34]{}\mathrel{=}\Varid{N}\;{}\<[40]%
\>[40]{}\Varid{one}\;{}\<[55]%
\>[55]{}\Varid{zero}{}\<[E]%
\\
\>[3]{}\hsindent{2}{}\<[5]%
\>[5]{}(\Varid{N}\;\Varid{f}\;\Varid{df})\oplus{}\<[23]%
\>[23]{}(\Varid{N}\;\Varid{g}\;\Varid{dg}){}\<[34]%
\>[34]{}\mathrel{=}\Varid{N}\;{}\<[40]%
\>[40]{}(\Varid{f}\oplus{}\<[51]%
\>[51]{}\Varid{g})\;{}\<[55]%
\>[55]{}(\Varid{df}\oplus\Varid{dg}){}\<[E]%
\\
\>[3]{}\hsindent{2}{}\<[5]%
\>[5]{}(\Varid{N}\;\Varid{f}\;\Varid{df})\otimes(\Varid{N}\;\Varid{g}\;\Varid{dg}){}\<[34]%
\>[34]{}\mathrel{=}\Varid{N}\;{}\<[40]%
\>[40]{}(\Varid{f}\otimes\Varid{g})\;{}\<[55]%
\>[55]{}((\Varid{f}\bullet\Varid{dg})\oplus(\Varid{g}\bullet\Varid{df})){}\<[E]%
\ColumnHook
\end{hscode}\resethooks
\indentend 
%

\subsection{Kronecker Delta}\label{sec:kronecker}

\noindent
The final piece of our jigsaw comes from the observation that 
the function \ensuremath{\delta_{x}\;\Varid{y}\mathrel{=}\mathbf{if}\;\Varid{x}= =\Varid{y}\;\mathbf{then}\;\Varid{one}\;\mathbf{else}\;\Varid{zero}} is
the Kronecker delta function.
\begin{equation*}
\delta_x(y) = \left\{
\begin{array}{l@{\hspace{5mm}}l}
1 & x = y \\
0 & \mathit{otherwise} 
\end{array}\right.
\end{equation*}
This gives rise to our third fundamental algebraic abstraction.
%
%
%
%
%
%
%
%
%
%
In Haskell, we capture a generalized notion of this concept in the \ensuremath{\Conid{Kronecker}}
type class.
\indentbegin \begin{hscode}\SaveRestoreHook
\column{B}{@{}>{\hspre}l<{\hspost}@{}}%
\column{3}{@{}>{\hspre}l<{\hspost}@{}}%
\column{5}{@{}>{\hspre}l<{\hspost}@{}}%
\column{E}{@{}>{\hspre}l<{\hspost}@{}}%
\>[3]{}\mathbf{class}\;\Conid{Module}\;\Varid{d}\;\Varid{e}\Rightarrow \Conid{Kronecker}\;\Varid{v}\;\Varid{d}\;\Varid{e}\;\mathbf{where}{}\<[E]%
\\
\>[3]{}\hsindent{2}{}\<[5]%
\>[5]{}\Varid{delta}\mathbin{::}\Varid{v}\to \Varid{e}{}\<[E]%
\ColumnHook
\end{hscode}\resethooks
\indentend 

\noindent
Notice that, at this point, no extra laws are imposed on \ensuremath{\Conid{Kronecker}\;\Varid{v}\;\Varid{d}\;\Varid{e}}; 
later, we will use isomorphism conditions (\Cref{defn:kronecker-iso}) to constrain different AD variants.
The idea is that \ensuremath{\Varid{delta}\;\Varid{x}} does not compute $\delta_x(y)$ for just a single
variable \ensuremath{\Varid{y}}, but for \emph{all} variables at once. Thus \ensuremath{\Varid{delta}\;\Varid{x}} is a value in a
\ensuremath{\Varid{d}}-module \ensuremath{\Varid{e}} where only the \ensuremath{\Varid{x}}-component is \ensuremath{\Varid{one}}, and all others are \ensuremath{\Varid{zero}}.
A helpful analogy may be to think of \ensuremath{\Varid{v}} as the type of basis vectors of \ensuremath{\Varid{e}}.
Then, \ensuremath{\Varid{delta}\;\Varid{x}} is the embedding of basis vector \ensuremath{\Varid{x}} into \ensuremath{\Varid{e}}.
Continuing Section~\ref{sec:module}'s example of the \ensuremath{\mathbb{R}}-module \ensuremath{\mathbb{R}^3}, there are usually three variables \ensuremath{\Varid{x},\Varid{y},\Varid{z}} such that
\begin{eqnarray*}
\ensuremath{\Varid{delta}\;\Varid{x}} & = & \left[
\begin{array}{ccc}
1 &
0 &
0
\end{array}
\right] \\
\ensuremath{\Varid{delta}\;\Varid{y}} & =  & \left[
\begin{array}{ccc}
0 &
1 &
0
\end{array}
\right] \\
\ensuremath{\Varid{delta}\;\Varid{z}} & = & \left[
\begin{array}{ccc}
0 & 
0 &
1
\end{array}
\right]
\end{eqnarray*}

In the AD literature, these basis vectors, where one entry is \ensuremath{\Varid{one}} 
and all the other are \ensuremath{\Varid{zero}}, are often referred to as \emph{one-hot vectors}.

\subsection{Abstract Automatic Differentiation}\label{sec:abstractad:approach}

\noindent
With the three algebraic structures in place, we present our abstract definition
of differentiation.
\indentbegin \begin{hscode}\SaveRestoreHook
\column{B}{@{}>{\hspre}l<{\hspost}@{}}%
\column{3}{@{}>{\hspre}l<{\hspost}@{}}%
\column{E}{@{}>{\hspre}l<{\hspost}@{}}%
\>[3]{}\Varid{abstractD}\mathbin{::}\Conid{Kronecker}\;\Varid{v}\;\Varid{d}\;\Varid{e}\Rightarrow (\Varid{v}\to \Varid{d})\to \Conid{Expr}\;\Varid{v}\to \Varid{d}\ltimes\Varid{e}{}\<[E]%
\\
\>[3]{}\Varid{abstractD}\;\Varid{var}\mathrel{=}\Varid{eval}\;\Varid{gen}\;\mathbf{where}\;\Varid{gen}\;\Varid{x}\mathrel{=}\Varid{N}\;(\Varid{var}\;\Varid{x})\;(\Varid{delta}\;\Varid{x}){}\<[E]%
\ColumnHook
\end{hscode}\resethooks
\indentend 
Next, we show how this abstract definition can be instantiated to recover
\ensuremath{\Varid{forwardAD}}. Later, we vary \ensuremath{\Varid{e}} to obtain more efficient versions.

\subsection{Base case: Forward-Mode AD as AD in the Dense Function Space}
\label{sec:forward-dense}

\noindent
We recover \ensuremath{\Varid{forwardAD}}---the classical forward-mode AD implementation given in
Section~\ref{sec:classic}---as an instance of \ensuremath{\Varid{abstractD}}, by using the standard function
space \ensuremath{\Varid{e}\mathrel{=}\Varid{v}\to \Varid{d}}. We call this space \emph{dense}\footnote{To avoid clutter, we present
\ensuremath{\Conid{Dense}} and later types as \ensuremath{\mathbf{type}} synonyms in the paper, but our implementation uses \ensuremath{\mathbf{newtype}}s for unambiguous type class resolution.}
to contrast with the \emph{sparse} function space coming up in Section~\ref{sec:sparse}, by analogy with dense/sparse polynomial representations in computer algebra~\cite{MCA3}.
\indentbegin \begin{hscode}\SaveRestoreHook
\column{B}{@{}>{\hspre}l<{\hspost}@{}}%
\column{3}{@{}>{\hspre}l<{\hspost}@{}}%
\column{E}{@{}>{\hspre}l<{\hspost}@{}}%
\>[3]{}\mathbf{type}\;\Conid{Dense}\;\Varid{v}\;\Varid{d}\mathrel{=}\Varid{v}\to \Varid{d}{}\<[E]%
\ColumnHook
\end{hscode}\resethooks
\indentend Informally, we can think of this structure as a functional representation of a
gradient vector, required by least square fitting, gradient descent and many other
applications. Just like a gradient vector, it contains one \ensuremath{\Varid{d}}-value per variable of type \ensuremath{\Varid{v}}.
The actions of its monoid and \ensuremath{\Varid{d}}-module instances are given pointwise,
using the \squote[{\ensuremath{\Varid{d}}-as-\ensuremath{\Varid{d}}-module construction}] of \Cref{sec:module}.
\indentbegin \begin{hscode}\SaveRestoreHook
\column{B}{@{}>{\hspre}l<{\hspost}@{}}%
\column{3}{@{}>{\hspre}l<{\hspost}@{}}%
\column{5}{@{}>{\hspre}l<{\hspost}@{}}%
\column{18}{@{}>{\hspre}c<{\hspost}@{}}%
\column{18E}{@{}l@{}}%
\column{21}{@{}>{\hspre}l<{\hspost}@{}}%
\column{45}{@{}>{\hspre}c<{\hspost}@{}}%
\column{45E}{@{}l@{}}%
\column{48}{@{}>{\hspre}l<{\hspost}@{}}%
\column{E}{@{}>{\hspre}l<{\hspost}@{}}%
\>[3]{}\mathbf{instance}\;\Conid{Semiring}\;\Varid{d}\Rightarrow \Conid{Monoid}\;(\Conid{Dense}\;\Varid{v}\;\Varid{d})\;\mathbf{where}{}\<[E]%
\\
\>[3]{}\hsindent{2}{}\<[5]%
\>[5]{}\Varid{zero}{}\<[18]%
\>[18]{}\mathrel{=}{}\<[18E]%
\>[21]{}\lambda \Varid{v}\to \Varid{zero}{}\<[E]%
\\
\>[3]{}\hsindent{2}{}\<[5]%
\>[5]{}\Varid{f}_{1}\oplus\Varid{f}_{2}{}\<[18]%
\>[18]{}\mathrel{=}{}\<[18E]%
\>[21]{}\lambda \Varid{v}\to \Varid{f}_{1}\;\Varid{v}\oplus\Varid{f}_{2}\;\Varid{v}{}\<[E]%
\\[\blanklineskip]%
\>[3]{}\mathbf{instance}\;\Conid{Semiring}\;\Varid{d}\Rightarrow \Conid{Module}\;\Varid{d}\;(\Conid{Dense}\;\Varid{v}\;\Varid{d})\;\mathbf{where}{}\<[E]%
\\
\>[3]{}\hsindent{2}{}\<[5]%
\>[5]{}\Varid{d}\bullet\Varid{f}{}\<[18]%
\>[18]{}\mathrel{=}{}\<[18E]%
\>[21]{}\lambda \Varid{v}\to \Varid{d}\bullet(\Varid{f}\;\Varid{v}){}\<[45]%
\>[45]{}\mathrel{=}{}\<[45E]%
\>[48]{}\lambda \Varid{v}\to \Varid{d}\otimes(\Varid{f}\;\Varid{v}){}\<[E]%
\ColumnHook
\end{hscode}\resethooks
\indentend The Kronecker delta function for dense functions is the textbook one.
\indentbegin \begin{hscode}\SaveRestoreHook
\column{B}{@{}>{\hspre}l<{\hspost}@{}}%
\column{3}{@{}>{\hspre}l<{\hspost}@{}}%
\column{5}{@{}>{\hspre}l<{\hspost}@{}}%
\column{18}{@{}>{\hspre}l<{\hspost}@{}}%
\column{E}{@{}>{\hspre}l<{\hspost}@{}}%
\>[3]{}\mathbf{instance}\;(\Conid{Eq}\;\Varid{v},\Conid{Semiring}\;\Varid{d})\Rightarrow \Conid{Kronecker}\;\Varid{v}\;\Varid{d}\;(\Conid{Dense}\;\Varid{v}\;\Varid{d})\;\mathbf{where}{}\<[E]%
\\
\>[3]{}\hsindent{2}{}\<[5]%
\>[5]{}\Varid{delta}\;\Varid{v}{}\<[18]%
\>[18]{}\mathrel{=}\lambda \Varid{w}\to \mathbf{if}\;\Varid{v}= =\Varid{w}\;\mathbf{then}\;\Varid{one}\;\mathbf{else}\;\Varid{zero}{}\<[E]%
\ColumnHook
\end{hscode}\resethooks
\indentend Using this \ensuremath{\Conid{Kronecker}} instance
and generalizing dual numbers into \mbox{\ensuremath{\Varid{d}\ltimes(\Conid{Dense}\;\Varid{v}\;\Varid{d})}},
we obtain the following definition for forward-mode AD.
\indentbegin \begin{hscode}\SaveRestoreHook
\column{B}{@{}>{\hspre}l<{\hspost}@{}}%
\column{3}{@{}>{\hspre}l<{\hspost}@{}}%
\column{E}{@{}>{\hspre}l<{\hspost}@{}}%
\>[3]{}\Varid{forwardAD}_{Dense}\mathbin{::}(\Conid{Eq}\;\Varid{v},\Conid{Semiring}\;\Varid{d})\Rightarrow (\Varid{v}\to \Varid{d})\to \Conid{Expr}\;\Varid{v}\to \Varid{d}\ltimes(\Conid{Dense}\;\Varid{v}\;\Varid{d}){}\<[E]%
\\
\>[3]{}\Varid{forwardAD}_{Dense}\mathrel{=}\Varid{abstractD}{}\<[E]%
\ColumnHook
\end{hscode}\resethooks
\indentend This is \ensuremath{\Varid{forwardAD}} with its parameters re-arranged and \ensuremath{\Conid{Dual}\;\Varid{d}} generalized:

\indentbegin \begin{hscode}\SaveRestoreHook
\column{B}{@{}>{\hspre}l<{\hspost}@{}}%
\column{3}{@{}>{\hspre}l<{\hspost}@{}}%
\column{9}{@{}>{\hspre}l<{\hspost}@{}}%
\column{30}{@{}>{\hspre}c<{\hspost}@{}}%
\column{30E}{@{}l@{}}%
\column{33}{@{}>{\hspre}l<{\hspost}@{}}%
\column{38}{@{}>{\hspre}l<{\hspost}@{}}%
\column{E}{@{}>{\hspre}l<{\hspost}@{}}%
\>[3]{}\Varid{pri}^D\;{}\<[9]%
\>[9]{}(\Varid{forwardAD}\;\Varid{var}\;\Varid{x}\;\Varid{e}){}\<[30]%
\>[30]{}\mathrel{=}{}\<[30E]%
\>[33]{}\Varid{pri}^{N}\;{}\<[38]%
\>[38]{}(\Varid{forwardAD}_{Dense}\;\Varid{var}\;\Varid{e}\;\Varid{x}){}\<[E]%
\\
\>[3]{}\Varid{tan}^D\;{}\<[9]%
\>[9]{}(\Varid{forwardAD}\;\Varid{var}\;\Varid{x}\;\Varid{e}){}\<[30]%
\>[30]{}\mathrel{=}{}\<[30E]%
\>[33]{}\Varid{tan}^{N}\;{}\<[38]%
\>[38]{}(\Varid{forwardAD}_{Dense}\;\Varid{var}\;\Varid{e}\;\Varid{x}){}\<[E]%
\ColumnHook
\end{hscode}\resethooks
\indentend 


\noindent
The essential difference between \ensuremath{\Varid{forwardAD}} and \ensuremath{\Varid{forwardAD}_{Dense}} is intensional.
In moving from computing a value in \ensuremath{\Varid{v}\to (\Varid{d}\ltimes\Varid{d})} to one in
\ensuremath{\Varid{d}\ltimes(\Varid{v}\to \Varid{d})}, we no longer compute both the primal and tangent for each
individual \ensuremath{\Varid{v}}, but instead only compute the primal \emph{once} for the whole
gradient. In other words, we share the computation of the primal among all partial
derivatives.
For instance, \ensuremath{\Varid{example}_{2}} represents expression \ensuremath{(\Varid{x}\;\times\;\Varid{y})\mathbin{+}\Varid{x}\mathbin{+}\mathrm{1}} in variables \ensuremath{\Varid{x}} and \ensuremath{\Varid{y}}
using our expression language.
\indentbegin \begin{hscode}\SaveRestoreHook
\column{B}{@{}>{\hspre}l<{\hspost}@{}}%
\column{3}{@{}>{\hspre}l<{\hspost}@{}}%
\column{E}{@{}>{\hspre}l<{\hspost}@{}}%
\>[3]{}\mathbf{data}\;\Varid{XY}\mathrel{=}\Varid{X}\mid \Varid{Y}{}\<[E]%
\\[\blanklineskip]%
\>[3]{}\Varid{example}_{2}\mathbin{::}\Conid{Expr}\;\Varid{XY}{}\<[E]%
\\
\>[3]{}\Varid{example}_{2}\mathrel{=}\Conid{Plus}\;(\Conid{Plus}\;(\Conid{Times}\;(\Conid{Var}\;\Varid{X})\;(\Conid{Var}\;\Varid{Y}))\;(\Conid{Var}\;\Varid{X}))\;\Conid{One}{}\<[E]%
\ColumnHook
\end{hscode}\resethooks
\indentend We compute its gradient at the point \ensuremath{(\mathrm{5},\mathrm{3})}:

\indentbegin \begin{hscode}\SaveRestoreHook
\column{B}{@{}>{\hspre}l<{\hspost}@{}}%
\column{3}{@{}>{\hspre}l<{\hspost}@{}}%
\column{6}{@{}>{\hspre}l<{\hspost}@{}}%
\column{11}{@{}>{\hspre}l<{\hspost}@{}}%
\column{E}{@{}>{\hspre}l<{\hspost}@{}}%
\>[3]{}\mathbin{>}{}\<[6]%
\>[6]{}\mathbf{let}\;{}\<[11]%
\>[11]{}\{\mskip1.5mu \Varid{var}\;\Varid{X}\mathrel{=}\mathrm{5};\Varid{var}\;\Varid{Y}\mathrel{=}\mathrm{3};\Varid{e}\mathrel{=}\Varid{tan}^{N}\;(\Varid{forwardAD}_{Dense}\;\Varid{var}\;\Varid{example}_{2})\mskip1.5mu\}{}\<[E]%
\\
\>[6]{}\mathbf{in}\;{}\<[11]%
\>[11]{}(\Varid{e}\;\Varid{X},\Varid{e}\;\Varid{Y}){}\<[E]%
\\
\>[3]{}(\mathrm{4},\mathrm{5}){}\<[E]%
\ColumnHook
\end{hscode}\resethooks
\indentend In contrast, when using \ensuremath{\Varid{forwardAD}}, we would have to call it twice:
\indentbegin \begin{hscode}\SaveRestoreHook
\column{B}{@{}>{\hspre}l<{\hspost}@{}}%
\column{3}{@{}>{\hspre}c<{\hspost}@{}}%
\column{3E}{@{}l@{}}%
\column{6}{@{}>{\hspre}l<{\hspost}@{}}%
\column{11}{@{}>{\hspre}l<{\hspost}@{}}%
\column{E}{@{}>{\hspre}l<{\hspost}@{}}%
\>[3]{}\mathbin{>}{}\<[3E]%
\>[6]{}\mathbf{let}\;{}\<[11]%
\>[11]{}\{\mskip1.5mu \Varid{var}\;\Varid{X}\mathrel{=}\mathrm{5};\Varid{var}\;\Varid{Y}\mathrel{=}\mathrm{3},\Varid{e}\mathrel{=}\Varid{example}_{2}\mskip1.5mu\}{}\<[E]%
\\
\>[6]{}\mathbf{in}\;{}\<[11]%
\>[11]{}(\Varid{tan}^D\;(\Varid{forwardAD}\;\Varid{var}\;\Varid{X}\;\Varid{e}),\Varid{tan}^D\;(\Varid{forwardAD}\;\Varid{var}\;\Varid{Y}\;\Varid{e})){}\<[E]%
\ColumnHook
\end{hscode}\resethooks
\indentend 
\section{AD Variants by Construction}\label{sec:construction}
\label{sec:forward}

\noindent
This section explains our methodology, which is
based on constructing \emph{Kronecker isomorphisms}, and demonstrates
it on the sparse variant of \ensuremath{\Varid{forwardAD}_{Dense}}.
This methodology will allow us to instantiate our abstract algorithm by varying
the tangent type to obtain different flavors of automatic differentiation.

\subsection{Relating AD Variants with Kronecker Isomorphisms}\label{sec:abstractad:approach}

\noindent
A key concern when devising new AD variants is their correctness.  We can
establish this by showing their functional equivalence to an established AD
baseline. We do so using \ensuremath{\Conid{Kronecker}} isomorphisms.

\begin{definition}[Kronecker Homomorphism]\label{defn:kronecker-homo}
Given a type of variables \ensuremath{\Conid{V}} and a semiring \ensuremath{\Conid{D}}, with types \ensuremath{\Conid{E}_{1}} and \ensuremath{\Conid{E}_{2}}
such that \ensuremath{\Conid{Kronecker}\;\Conid{V}\;\Conid{D}\;\Conid{E}_{1}} and \ensuremath{\Conid{Kronecker}\;\Conid{V}\;\Conid{D}\;\Conid{E}_{2}} hold (and thus
also \ensuremath{\Conid{Monoid}\;\Conid{E}_{1}}, \ensuremath{\Conid{Monoid}\;\Conid{E}_{2}}, \ensuremath{\Conid{Module}\;\Conid{D}\;\Conid{E}_{1}} and \ensuremath{\Conid{Module}\;\Conid{D}\;\Conid{E}_{2}}),
then a \emph{Kronecker homomorphism} is a function \ensuremath{\Varid{h}\mathbin{::}\Conid{E}_{1}\to \Conid{E}_{2}}
that preserves the \ensuremath{\Conid{Kronecker}\;\Conid{V}\;\Conid{D}} structure: \\
\noindent
\begin{minipage}{0.5\textwidth}\indentbegin \begin{hscode}\SaveRestoreHook
\column{B}{@{}>{\hspre}l<{\hspost}@{}}%
\column{3}{@{}>{\hspre}l<{\hspost}@{}}%
\column{6}{@{}>{\hspre}l<{\hspost}@{}}%
\column{26}{@{}>{\hspre}l<{\hspost}@{}}%
\column{E}{@{}>{\hspre}l<{\hspost}@{}}%
\>[3]{}\Varid{h}\;{}\<[6]%
\>[6]{}zero_{\text{\tiny E}_1}{}\<[26]%
\>[26]{}\mathrel{=}zero_{\text{\tiny E}_2}{}\<[E]%
\\
\>[3]{}\Varid{h}\;{}\<[6]%
\>[6]{}(\Varid{x}\oplus_{\text{\tiny E}_1}\Varid{y}){}\<[26]%
\>[26]{}\mathrel{=}(\Varid{h}\;\Varid{x})\oplus_{\text{\tiny E}_2}(\Varid{h}\;\Varid{y}){}\<[E]%
\ColumnHook
\end{hscode}\resethooks
\indentend \end{minipage}%
\begin{minipage}{0.5\textwidth}\indentbegin \begin{hscode}\SaveRestoreHook
\column{B}{@{}>{\hspre}l<{\hspost}@{}}%
\column{3}{@{}>{\hspre}l<{\hspost}@{}}%
\column{6}{@{}>{\hspre}l<{\hspost}@{}}%
\column{24}{@{}>{\hspre}l<{\hspost}@{}}%
\column{E}{@{}>{\hspre}l<{\hspost}@{}}%
\>[3]{}\Varid{h}\;{}\<[6]%
\>[6]{}(\Varid{d}\bullet_{\text{\tiny E}_1}\Varid{m}){}\<[24]%
\>[24]{}\mathrel{=}\Varid{d}\bullet_{\text{\tiny E}_2}(\Varid{h}\;\Varid{m}){}\<[E]%
\\
\>[3]{}\Varid{h}\;{}\<[6]%
\>[6]{}(delta_{\text{\tiny E}_1}\;\Varid{v}){}\<[24]%
\>[24]{}\mathrel{=}delta_{\text{\tiny E}_2}\;\Varid{v}{}\<[E]%
\ColumnHook
\end{hscode}\resethooks
\indentend \end{minipage}
\end{definition}
%
\begin{definition}[Kronecker Isomorphism]\label{defn:kronecker-iso}
A \emph{Kronecker isomorphism} is a pair of Kronecker homomorphisms \ensuremath{\Varid{h}\mathbin{::}\Conid{E}_{1}\to \Conid{E}_{2}} and \ensuremath{\Varid{h}^{-1}\mathbin{::}\Conid{E}_{2}\to \Conid{E}_{1}},
that are inverses:
\ensuremath{\Varid{h}\hsdot{\circ }{.}\Varid{h}^{-1}\mathrel{=}\Varid{id}} and \ensuremath{\Varid{h}^{-1}\hsdot{\circ }{.}\Varid{h}\mathrel{=}\Varid{id}}.
\end{definition}

\noindent
In the rest of the paper we use \ensuremath{\Conid{Dense}\;\Conid{V}\;\Conid{D}} as the baseline.  Hence, when
coming up with a new tangent type \ensuremath{\Conid{E}}, we also want to find a Kronecker
isomorphism \ensuremath{\Varid{h}\mathbin{::}\Conid{Dense}\;\Conid{V}\;\Conid{D}\to \Conid{E}} to establish its correctness.
This isomorphism is a function pair consisting of a representation function \ensuremath{\Varid{rep}}
and its inverse \ensuremath{\Varid{abs}} (abstraction), following the terminology of data representation~\cite{gill_hutton_2009,hoare1972}.
To streamline our approach, we capture this isomorphism
in a type class \ensuremath{\Conid{CorrectAD}}.
\indentbegin \begin{hscode}\SaveRestoreHook
\column{B}{@{}>{\hspre}l<{\hspost}@{}}%
\column{3}{@{}>{\hspre}l<{\hspost}@{}}%
\column{5}{@{}>{\hspre}l<{\hspost}@{}}%
\column{12}{@{}>{\hspre}l<{\hspost}@{}}%
\column{E}{@{}>{\hspre}l<{\hspost}@{}}%
\>[3]{}\mathbf{class}\;\Conid{Kronecker}\;\Varid{v}\;\Varid{d}\;\Varid{e}\Rightarrow \Conid{CorrectAD}\;\Varid{v}\;\Varid{d}\;\Varid{e}\;\mathbf{where}{}\<[E]%
\\
\>[3]{}\hsindent{2}{}\<[5]%
\>[5]{}\Varid{rep}{}\<[12]%
\>[12]{}\mathbin{::}\Conid{Dense}\;\Varid{v}\;\Varid{d}\to \Varid{e}{}\<[E]%
\\
\>[3]{}\hsindent{2}{}\<[5]%
\>[5]{}\Varid{abs}{}\<[12]%
\>[12]{}\mathbin{::}\Varid{e}\to \Conid{Dense}\;\Varid{v}\;\Varid{d}{}\<[E]%
\ColumnHook
\end{hscode}\resethooks
\indentend With the \emph{free} theorem~\cite{fpca/Wadler89} that follows from the
parametricity of \ensuremath{\Varid{abstractD}}~\cite{ifip/Reynolds83} we can then
relate the new instance of the algorithm to the baseline.

\begin{theorem}\label{thm:main}
Given a \ensuremath{\Conid{CorrectAD}\;\Conid{V}\;\Conid{D}\;\Conid{E}} instance,
we have that

\begin{minipage}{\textwidth}\indentbegin \begin{hscode}\SaveRestoreHook
\column{B}{@{}>{\hspre}l<{\hspost}@{}}%
\column{3}{@{}>{\hspre}l<{\hspost}@{}}%
\column{16}{@{}>{\hspre}l<{\hspost}@{}}%
\column{39}{@{}>{\hspre}c<{\hspost}@{}}%
\column{39E}{@{}l@{}}%
\column{44}{@{}>{\hspre}l<{\hspost}@{}}%
\column{56}{@{}>{\hspre}l<{\hspost}@{}}%
\column{E}{@{}>{\hspre}l<{\hspost}@{}}%
\>[3]{}\Varid{fmap}\;\Varid{abs}\;({}\<[16]%
\>[16]{}abstractD_{\text{\tiny E}}\;\Varid{var}\;\Varid{e}){}\<[39]%
\>[39]{}\mathrel{=}{}\<[39E]%
\>[44]{}abstractD_{\text{\tiny Dense}}\;\Varid{var}\;\Varid{e}{}\<[E]%
\\
\>[16]{}abstractD_{\text{\tiny E}}\;\Varid{var}\;\Varid{e}{}\<[39]%
\>[39]{}\mathrel{=}{}\<[39E]%
\>[44]{}\Varid{fmap}\;\Varid{rep}\;{}\<[56]%
\>[56]{}(abstractD_{\text{\tiny Dense}}\;\Varid{var}\;\Varid{e}){}\<[E]%
\ColumnHook
\end{hscode}\resethooks
\indentend \end{minipage}
\end{theorem}

\subsection{Kronecker Isomorphisms by Construction}

\noindent
Given the \ensuremath{\Conid{Kronecker}\;\Conid{V}\;\Conid{D}\;(\Conid{Dense}\;\Conid{V}\;\Conid{D})} structure as our starting point and a
newly desired representation \ensuremath{\Conid{E}}, we do not have to devise the \ensuremath{\Conid{Kronecker}\;\Conid{V}\;\Conid{D}\;\Conid{E}} structure and the Kronecker isomorphism out of the blue. Instead, following
the method outlined by Elliott~\cite{Elliott2009-type-class-morphisms-TR}, we can constructively derive both from
a plain invertible function \ensuremath{\Varid{h}\mathbin{::}\Conid{Dense}\;\Conid{V}\;\Conid{D}\to \Conid{E}}.
%
Indeed, we can create the instances for \ensuremath{\Conid{E}} in terms of the
instances of \ensuremath{\Conid{Dense}\;\Conid{V}\;\Conid{D}} and of \ensuremath{\Varid{h}}. \\
\noindent
\begin{minipage}{0.5\textwidth}\indentbegin \begin{hscode}\SaveRestoreHook
\column{B}{@{}>{\hspre}l<{\hspost}@{}}%
\column{3}{@{}>{\hspre}l<{\hspost}@{}}%
\column{5}{@{}>{\hspre}l<{\hspost}@{}}%
\column{13}{@{}>{\hspre}c<{\hspost}@{}}%
\column{13E}{@{}l@{}}%
\column{16}{@{}>{\hspre}l<{\hspost}@{}}%
\column{E}{@{}>{\hspre}l<{\hspost}@{}}%
\>[3]{}\mathbf{instance}\;\Conid{Monoid}\;\Conid{E}\;\mathbf{where}{}\<[E]%
\\
\>[3]{}\hsindent{2}{}\<[5]%
\>[5]{}\Varid{zero}{}\<[13]%
\>[13]{}\mathrel{=}{}\<[13E]%
\>[16]{}\Varid{h}\;zero_{\text{\tiny Dense}}{}\<[E]%
\\
\>[3]{}\hsindent{2}{}\<[5]%
\>[5]{}\Varid{x}\oplus\Varid{y}{}\<[13]%
\>[13]{}\mathrel{=}{}\<[13E]%
\>[16]{}\Varid{h}\;(\Varid{h}^{-1}\;\Varid{x}\oplus_{\text{\tiny Dense}}\Varid{h}^{-1}\;\Varid{y}){}\<[E]%
\\[\blanklineskip]%
\>[3]{}\mathbf{instance}\;\Conid{Module}\;\Conid{D}\;\Conid{E}\;\mathbf{where}{}\<[E]%
\\
\>[3]{}\hsindent{2}{}\<[5]%
\>[5]{}\Varid{d}\bullet\Varid{x}\mathrel{=}\Varid{h}\;(\Varid{d}\bullet_{\text{\tiny Dense}}\Varid{h}^{-1}\;\Varid{x}){}\<[E]%
\ColumnHook
\end{hscode}\resethooks
\indentend \end{minipage}%
\begin{minipage}{0.5\textwidth}\indentbegin \begin{hscode}\SaveRestoreHook
\column{B}{@{}>{\hspre}l<{\hspost}@{}}%
\column{3}{@{}>{\hspre}l<{\hspost}@{}}%
\column{5}{@{}>{\hspre}l<{\hspost}@{}}%
\column{12}{@{}>{\hspre}c<{\hspost}@{}}%
\column{12E}{@{}l@{}}%
\column{15}{@{}>{\hspre}l<{\hspost}@{}}%
\column{E}{@{}>{\hspre}l<{\hspost}@{}}%
\>[3]{}\mathbf{instance}\;\Conid{Kronecker}\;\Conid{V}\;\Conid{D}\;\Conid{E}\;\mathbf{where}{}\<[E]%
\\
\>[3]{}\hsindent{2}{}\<[5]%
\>[5]{}\Varid{delta}\;\Varid{v}\mathrel{=}\Varid{h}\;(delta_{\text{\tiny Dense}}\;\Varid{v}){}\<[E]%
\\[\blanklineskip]%
\>[3]{}\mathbf{instance}\;\Conid{CorrectAD}\;\Conid{V}\;\Conid{D}\;\Conid{E}\;\mathbf{where}{}\<[E]%
\\
\>[3]{}\hsindent{2}{}\<[5]%
\>[5]{}\Varid{rep}{}\<[12]%
\>[12]{}\mathrel{=}{}\<[12E]%
\>[15]{}\Varid{h}{}\<[E]%
\\
\>[3]{}\hsindent{2}{}\<[5]%
\>[5]{}\Varid{abs}{}\<[12]%
\>[12]{}\mathrel{=}{}\<[12E]%
\>[15]{}\Varid{h}^{-1}{}\<[E]%
\ColumnHook
\end{hscode}\resethooks
\indentend \end{minipage}
By construction, these instances both (1) satisfy all the necessary laws, and (2)
ensure that \ensuremath{\Varid{h}} is a Kronecker isomorphism.
For example, we can verify that \ensuremath{\Varid{x}\oplus\Varid{zero}\mathrel{=}\Varid{x}} and
that \ensuremath{\Varid{h}\;\Varid{x}\oplus\Varid{h}\;\Varid{y}\mathrel{=}\Varid{h}\;(\Varid{x}\oplus\Varid{y})}
as follows:

\noindent
\begin{minipage}[t]{0.6\textwidth}\indentbegin \begin{hscode}\SaveRestoreHook
\column{B}{@{}>{\hspre}l<{\hspost}@{}}%
\column{3}{@{}>{\hspre}l<{\hspost}@{}}%
\column{6}{@{}>{\hspre}l<{\hspost}@{}}%
\column{E}{@{}>{\hspre}l<{\hspost}@{}}%
\>[6]{}\Varid{x}\oplus\Varid{zero}{}\<[E]%
\\
\>[3]{}\mathrel{=}\mbox{\commentbegin  definitions of \ensuremath{\Varid{zero}} and \ensuremath{\oplus}  \commentend}{}\<[E]%
\\
\>[3]{}\hsindent{3}{}\<[6]%
\>[6]{}\Varid{h}\;(\Varid{h}^{-1}\;\Varid{x}\oplus\Varid{h}^{-1}\;(\Varid{h}\;\Varid{zero})){}\<[E]%
\\
\>[3]{}\mathrel{=}\mbox{\commentbegin  isomorphism  \commentend}{}\<[E]%
\\
\>[3]{}\hsindent{3}{}\<[6]%
\>[6]{}\Varid{h}\;(\Varid{h}^{-1}\;\Varid{x}\oplus\Varid{zero}){}\<[E]%
\\
\>[3]{}\mathrel{=}\mbox{\commentbegin  monoid identity law for \ensuremath{\Conid{Dense}\;\Varid{v}\;\Varid{d}}  \commentend}{}\<[E]%
\\
\>[3]{}\hsindent{3}{}\<[6]%
\>[6]{}\Varid{h}\;(\Varid{h}^{-1}\;\Varid{x}){}\<[E]%
\\
\>[3]{}\mathrel{=}\mbox{\commentbegin  isomorphism  \commentend}{}\<[E]%
\\
\>[3]{}\hsindent{3}{}\<[6]%
\>[6]{}\Varid{x}{}\<[E]%
\ColumnHook
\end{hscode}\resethooks
\indentend \end{minipage}%
\begin{minipage}[t]{0.4\textwidth}\indentbegin \begin{hscode}\SaveRestoreHook
\column{B}{@{}>{\hspre}l<{\hspost}@{}}%
\column{3}{@{}>{\hspre}l<{\hspost}@{}}%
\column{6}{@{}>{\hspre}l<{\hspost}@{}}%
\column{E}{@{}>{\hspre}l<{\hspost}@{}}%
\>[6]{}\Varid{h}\;\Varid{x}\oplus\Varid{h}\;\Varid{y}{}\<[E]%
\\
\>[3]{}\mathrel{=}\mbox{\commentbegin  definition of \ensuremath{\oplus}  \commentend}{}\<[E]%
\\
\>[3]{}\hsindent{3}{}\<[6]%
\>[6]{}\Varid{h}\;(\Varid{h}^{-1}\;(\Varid{h}\;\Varid{x})\oplus\Varid{h}^{-1}\;(\Varid{h}\;\Varid{y})){}\<[E]%
\\
\>[3]{}\mathrel{=}\mbox{\commentbegin  isomorphism  \commentend}{}\<[E]%
\\
\>[3]{}\hsindent{3}{}\<[6]%
\>[6]{}\Varid{h}\;(\Varid{x}\oplus\Varid{y}){}\<[E]%
\ColumnHook
\end{hscode}\resethooks
\indentend \end{minipage}

\noindent
Verifying the other laws and homomorphism properties proceeds similarly.

In short, each time we devise a new AD variant, we come up with a new
representation \ensuremath{\Conid{E}} and an invertible function \ensuremath{\Varid{h}\mathbin{::}\Conid{Dense}\;\Conid{V}\;\Conid{D}\to \Conid{E}}. 
The rest
of the infrastructure then follows mechanically, except for the fact that we
can simplify the definitions, sometimes exploiting further equations, for the sake
of optimizing the (algorithmic properties of the) representation.

The invertible function ensures, through the constructed Kronecker isomorphism,
that the extensional behaviour remains the same, and thus guarantees
correctness. At the same time, a well-chosen alternative representation may improve
intensional properties like asymptotic runtime.

\subsection{Sparse Maps}\label{sec:sparse}

\noindent
We demonstrate the above approach to switch from the dense function space
representation to that of sparse maps. This exploits the fact that many partial
derivatives of sub-expressions are zero. Indeed, if a variable \ensuremath{\Varid{x}} does not
occur in an expression \ensuremath{\Varid{e}}, the partial derivative $\partialfrac{e}{x}$ is
zero. Notably, in the base instance of dense functions, 
when the expression is just a variable,
all partial derivatives but one are by definition zero.
%
We avoid explicit representations of such zeros---as well as explicit
operations with them---by replacing the dense function space \ensuremath{\Varid{v}\to \Varid{d}} with the type of finite
key-value maps \ensuremath{\Conid{Map}\;\Varid{v}\;\Varid{d}}. We use Haskell's \ensuremath{\Conid{\Conid{Data}.Map}} library (\ref{app:functions})
instead of a specialized dictionary representation \cite{Shaikhha}.
\indentbegin \begin{hscode}\SaveRestoreHook
\column{B}{@{}>{\hspre}l<{\hspost}@{}}%
\column{3}{@{}>{\hspre}l<{\hspost}@{}}%
\column{E}{@{}>{\hspre}l<{\hspost}@{}}%
\>[3]{}\mathbf{type}\;\Conid{Sparse}\;\Varid{v}\;\Varid{d}\mathrel{=}\Conid{Map}\;\Varid{v}\;\Varid{d}{}\<[E]%
\ColumnHook
\end{hscode}\resethooks
\indentend The omitted values in the map are \ensuremath{\Varid{zero}}. This way we recover the \ensuremath{\Conid{Dense}}
representation from the sparse one by looking up the variables in the map and
defaulting to \ensuremath{\Varid{zero}} when a variable is not present.
\indentbegin \begin{hscode}\SaveRestoreHook
\column{B}{@{}>{\hspre}l<{\hspost}@{}}%
\column{3}{@{}>{\hspre}l<{\hspost}@{}}%
\column{25}{@{}>{\hspre}l<{\hspost}@{}}%
\column{54}{@{}>{\hspre}l<{\hspost}@{}}%
\column{E}{@{}>{\hspre}l<{\hspost}@{}}%
\>[3]{}abs_{\rightarrow}{}\<[25]%
\>[25]{}\mathbin{::}(\Conid{Ord}\;\Varid{v},\Conid{Semiring}\;\Varid{d})\Rightarrow \Conid{Sparse}\;\Varid{v}\;\Varid{d}\to \Conid{Dense}\;\Varid{v}\;\Varid{d}{}\<[E]%
\\
\>[3]{}abs_{\rightarrow}\;\Varid{m}{}\<[25]%
\>[25]{}\mathrel{=}\lambda \Varid{v}\to \Varid{findWithDefault}\;\Varid{zero}\;\Varid{v}\;\Varid{m}{}\<[E]%
\\[\blanklineskip]%
\>[3]{}rep_{\rightarrow}{}\<[25]%
\>[25]{}\mathbin{::}(\Conid{Ord}\;\Varid{v},\Conid{Bounded}\;\Varid{v},\Conid{Enum}\;\Varid{v},\Conid{Semiring}\;\Varid{d},\Conid{Eq}\;\Varid{d}){}\<[E]%
\\
\>[25]{}\Rightarrow \Conid{Dense}\;\Varid{v}\;\Varid{d}\to \Conid{Sparse}\;\Varid{v}\;\Varid{d}{}\<[E]%
\\
\>[3]{}rep_{\rightarrow}\;\Varid{dense}{}\<[25]%
\>[25]{}\mathrel{=}\Varid{fromList}\;[\mskip1.5mu (\Varid{v},\Varid{dense}\;\Varid{v})\mid {}\<[54]%
\>[54]{}\Varid{v}\leftarrow [\mskip1.5mu \Varid{minBound}\mathinner{\ldotp\ldotp}\Varid{maxBound}\mskip1.5mu],{}\<[E]%
\\
\>[54]{}\Varid{dense}\;\Varid{v}\not\equiv \Varid{zero}\mskip1.5mu]{}\<[E]%
\ColumnHook
\end{hscode}\resethooks
\indentend Observe that \ensuremath{abs_{\rightarrow}} and \ensuremath{rep_{\rightarrow}} are inverse functions under the constraints 
(modulo the equivalence of explicit and implicit zeroes).
%
%
%
%
%
With the new \ensuremath{\Conid{Sparse}\;\Varid{v}\;\Varid{d}} representation and the two inverse functions we can
derive the \ensuremath{\Conid{Monoid}}, \ensuremath{\Conid{Module}}, \ensuremath{\Conid{Kronecker}} and \ensuremath{\Conid{CorrectAD}} instances following the
approach outlined in Section~\ref{sec:abstractad:approach}.
%
This results in the all-zero map being empty. Addition of maps is performed key-wise, but
avoided if either or both maps have an implicit zero at a given (variable) key.
%
%

\indentbegin \begin{hscode}\SaveRestoreHook
\column{B}{@{}>{\hspre}l<{\hspost}@{}}%
\column{3}{@{}>{\hspre}l<{\hspost}@{}}%
\column{5}{@{}>{\hspre}l<{\hspost}@{}}%
\column{18}{@{}>{\hspre}l<{\hspost}@{}}%
\column{E}{@{}>{\hspre}l<{\hspost}@{}}%
\>[3]{}\mathbf{instance}\;(\Conid{Ord}\;\Varid{v},\Conid{Semiring}\;\Varid{d})\Rightarrow \Conid{Monoid}\;(\Conid{Sparse}\;\Varid{v}\;\Varid{d})\;\mathbf{where}{}\<[E]%
\\
\>[3]{}\hsindent{2}{}\<[5]%
\>[5]{}\Varid{zero}{}\<[18]%
\>[18]{}\mathrel{=}\Varid{empty}{}\<[E]%
\\
\>[3]{}\hsindent{2}{}\<[5]%
\>[5]{}(\oplus){}\<[18]%
\>[18]{}\mathrel{=}\Varid{unionWith}\;(\oplus){}\<[E]%
\ColumnHook
\end{hscode}\resethooks
\indentend 
Scalar multiplication is also key-wise. As \ensuremath{\Varid{d}\otimes\Varid{zero}\mathrel{=}\Varid{zero}} (\Cref{fig:semiring-laws}),
we can effortlessly preserve the implicit zero entries.
\indentbegin \begin{hscode}\SaveRestoreHook
\column{B}{@{}>{\hspre}l<{\hspost}@{}}%
\column{3}{@{}>{\hspre}l<{\hspost}@{}}%
\column{5}{@{}>{\hspre}l<{\hspost}@{}}%
\column{18}{@{}>{\hspre}l<{\hspost}@{}}%
\column{E}{@{}>{\hspre}l<{\hspost}@{}}%
\>[3]{}\mathbf{instance}\;(\Conid{Ord}\;\Varid{v},\Conid{Semiring}\;\Varid{d})\Rightarrow \Conid{Module}\;\Varid{d}\;(\Conid{Sparse}\;\Varid{v}\;\Varid{d})\;\mathbf{where}{}\<[E]%
\\
\>[3]{}\hsindent{2}{}\<[5]%
\>[5]{}\Varid{d}\bullet\Varid{m}{}\<[18]%
\>[18]{}\mathrel{=}\Varid{fmap}\;(\Varid{d}\otimes)\;\Varid{m}{}\<[E]%
\ColumnHook
\end{hscode}\resethooks
\indentend 
The only non-zero
entry for \ensuremath{\Varid{delta}\;\Varid{x}} is that for \ensuremath{\Varid{x}}.
\indentbegin \begin{hscode}\SaveRestoreHook
\column{B}{@{}>{\hspre}l<{\hspost}@{}}%
\column{3}{@{}>{\hspre}l<{\hspost}@{}}%
\column{5}{@{}>{\hspre}l<{\hspost}@{}}%
\column{18}{@{}>{\hspre}l<{\hspost}@{}}%
\column{E}{@{}>{\hspre}l<{\hspost}@{}}%
\>[3]{}\mathbf{instance}\;(\Conid{Ord}\;\Varid{v},\Conid{Semiring}\;\Varid{d})\Rightarrow \Conid{Kronecker}\;\Varid{v}\;\Varid{d}\;(\Conid{Sparse}\;\Varid{v}\;\Varid{d})\;\mathbf{where}{}\<[E]%
\\
\>[3]{}\hsindent{2}{}\<[5]%
\>[5]{}\Varid{delta}\;\Varid{x}{}\<[18]%
\>[18]{}\mathrel{=}\Varid{singleton}\;\Varid{x}\;\Varid{one}{}\<[E]%
\ColumnHook
\end{hscode}\resethooks
\indentend Moreover, and crucially in what follows, we observe the following relationship
between this \ensuremath{\Conid{Kronecker}} structure and its \ensuremath{\Conid{Module}} and \ensuremath{\Conid{Monoid}} instances:
\begin{align}
\ensuremath{\Varid{d}\bullet\Varid{delta}\;\Varid{v}} & = \ensuremath{\Varid{singleton}\;\Varid{v}\;\Varid{d}} \label{eq:singleton} \\
\ensuremath{\Varid{e}\oplus(\Varid{d}\bullet\Varid{delta}\;\Varid{v})} & = \ensuremath{\Varid{insertWith}\;(\oplus)\;\Varid{v}\;\Varid{d}\;\Varid{e}} \label{eq:insertWith}
\end{align}

\noindent
Lastly, the \ensuremath{\Conid{CorrectAD}} instance is straightforward.
\indentbegin \begin{hscode}\SaveRestoreHook
\column{B}{@{}>{\hspre}l<{\hspost}@{}}%
\column{3}{@{}>{\hspre}l<{\hspost}@{}}%
\column{5}{@{}>{\hspre}l<{\hspost}@{}}%
\column{7}{@{}>{\hspre}l<{\hspost}@{}}%
\column{14}{@{}>{\hspre}l<{\hspost}@{}}%
\column{E}{@{}>{\hspre}l<{\hspost}@{}}%
\>[3]{}\mathbf{instance}\;(\Conid{Ord}\;\Varid{v},\Conid{Bounded}\;\Varid{v},\Conid{Enum}\;\Varid{v},\Conid{Semiring}\;\Varid{d},\Conid{Eq}\;\Varid{d})\Rightarrow {}\<[E]%
\\
\>[3]{}\hsindent{2}{}\<[5]%
\>[5]{}\Conid{CorrectAD}\;\Varid{v}\;\Varid{d}\;(\Conid{Sparse}\;\Varid{v}\;\Varid{d})\;\mathbf{where}{}\<[E]%
\\
\>[5]{}\hsindent{2}{}\<[7]%
\>[7]{}\Varid{rep}{}\<[14]%
\>[14]{}\mathrel{=}rep_{\rightarrow}{}\<[E]%
\\
\>[5]{}\hsindent{2}{}\<[7]%
\>[7]{}\Varid{abs}{}\<[14]%
\>[14]{}\mathrel{=}abs_{\rightarrow}{}\<[E]%
\ColumnHook
\end{hscode}\resethooks
\indentend 

\noindent
In summary, all that the sparse gradient computation requires,
in comparison to the dense computation, is the switch from
\ensuremath{\Varid{d}\ltimes(\Conid{Dense}\;\Varid{v}\;\Varid{d})} to \ensuremath{\Varid{d}\ltimes(\Conid{Sparse}\;\Varid{v}\;\Varid{d})}.
\indentbegin \begin{hscode}\SaveRestoreHook
\column{B}{@{}>{\hspre}l<{\hspost}@{}}%
\column{3}{@{}>{\hspre}l<{\hspost}@{}}%
\column{E}{@{}>{\hspre}l<{\hspost}@{}}%
\>[3]{}\Varid{forwardAD}_{Sparse}\mathbin{::}(\Conid{Ord}\;\Varid{v},\Conid{Semiring}\;\Varid{d})\Rightarrow (\Varid{v}\to \Varid{d})\to \Conid{Expr}\;\Varid{v}\to \Varid{d}\ltimes\colorbox{lightgray}{$\Conid{Sparse}\;\Varid{v}\;\Varid{d}$}{}\<[E]%
\\
\>[3]{}\Varid{forwardAD}_{Sparse}\mathrel{=}\Varid{abstractD}{}\<[E]%
\ColumnHook
\end{hscode}\resethooks
\indentend Theorem~\ref{thm:main} now holds by construction.
%
%
%
%
We can verify this correspondence by revisiting \ensuremath{\Varid{example}_{2}} at the point \ensuremath{(\mathrm{5},\mathrm{3})}.
\indentbegin \begin{hscode}\SaveRestoreHook
\column{B}{@{}>{\hspre}l<{\hspost}@{}}%
\column{3}{@{}>{\hspre}l<{\hspost}@{}}%
\column{E}{@{}>{\hspre}l<{\hspost}@{}}%
\>[3]{}\mathbin{>}\mathbf{let}\;\{\mskip1.5mu \Varid{var}\;\Varid{X}\mathrel{=}\mathrm{5};\Varid{var}\;\Varid{Y}\mathrel{=}\mathrm{3}\mskip1.5mu\}\;\mathbf{in}\;\Varid{tan}^{N}\;(\Varid{forwardAD}_{Sparse}\;\Varid{var}\;\Varid{example}_{2}){}\<[E]%
\\
\>[3]{}\{\mskip1.5mu \Varid{X}\mapsto\mathrm{4};\Varid{Y}\mapsto\mathrm{5}\mskip1.5mu\}{}\<[E]%
\ColumnHook
\end{hscode}\resethooks
\indentend 


\section{Reverse-Mode Automatic Differentiation}\label{sec:rev}

\noindent
We now develop the appropriate \ensuremath{\Varid{d}}-module structure that exhibits
reverse-mode AD as an instance of \ensuremath{\Varid{abstractD}} by refining
the above sparse representation.
Motivated by the quest for improved efficiency, we optimize the time
complexity of the essential functions for
abstract differentiation with the underlying Nagata numbers semiring:
scalar multiplication \ensuremath{(\bullet)}, addition \ensuremath{(\oplus)} and Kronecker's \ensuremath{\Varid{delta}}.

\subsection{Accumulating Multiplications}

\noindent
We start with another function-based representation.
Instead of the \ensuremath{\Varid{d}}-module \ensuremath{\Varid{e}}, we consider the space of \emph{\ensuremath{\Varid{d}}-module
homomorphisms} from \ensuremath{\Varid{d}}
to \ensuremath{\Varid{e}}. It is a
standard result in commutative algebra~\cite[{p.18}]{atiyah-macdonald:CommAlg}\footnote{For the case of commutative \emph{rings} and their modules; the argument for semirings and modules is identical, but simpler, because there are fewer equations to check. } that these two structures are isomorphic, considered as
\ensuremath{\Varid{d}}-modules.
In terms of the \emph{underlying} Haskell types, this amounts to taking the plain function space.
\indentbegin \begin{hscode}\SaveRestoreHook
\column{B}{@{}>{\hspre}l<{\hspost}@{}}%
\column{3}{@{}>{\hspre}l<{\hspost}@{}}%
\column{E}{@{}>{\hspre}l<{\hspost}@{}}%
\>[3]{}\mathbf{type}\;\Varid{d}\multimap\Varid{e}\mathrel{=}\Varid{d}\to \Varid{e}{}\<[E]%
\ColumnHook
\end{hscode}\resethooks
\indentend But we intend this type \ensuremath{\Varid{d}\multimap\Varid{e}} to represent only the \ensuremath{\Varid{d}}-module  homomorphisms.
This is the case for \emph{linear} functions from \ensuremath{\Varid{d}} to
\ensuremath{\Varid{e}}.  A function \ensuremath{\Varid{f}\mathbin{::}\Varid{d}\to \Varid{e}} is linear\footnote{Preservation of addition follows from the fact that
the result of \ensuremath{\Varid{f}} is a semimodule.} 
iff it has the multiplicative \emph{homogeneity} property:
\begin{eqnarray*}
\ensuremath{\Varid{f}\;(\Varid{x}\otimes\Varid{y})} & = & \ensuremath{\Varid{x}\bullet\Varid{f}\;\Varid{y}}
\end{eqnarray*}
As the Haskell type \ensuremath{\Varid{d}\to \Varid{e}} also includes non-linear functions, we
have to make disciplined use of \ensuremath{\Varid{d}\multimap\Varid{e}}, being careful to avoid the non-linear ones.
Intuitively, we can think of \ensuremath{\Varid{d}\multimap\Varid{e}} as augmenting \ensuremath{\Varid{e}} with an accumulator for
a scalar multiplier of type \ensuremath{\Varid{d}}.  By supplying the scalar \ensuremath{\Varid{one}}, we recover the original
module.

\noindent
\begin{minipage}{0.5\textwidth}\indentbegin \begin{hscode}\SaveRestoreHook
\column{B}{@{}>{\hspre}l<{\hspost}@{}}%
\column{3}{@{}>{\hspre}l<{\hspost}@{}}%
\column{E}{@{}>{\hspre}l<{\hspost}@{}}%
\>[3]{}\Varid{rep}_{\multimap}\mathbin{::}\Conid{Module}\;\Varid{d}\;\Varid{e}\Rightarrow \Varid{e}\to (\Varid{d}\multimap\Varid{e}){}\<[E]%
\\
\>[3]{}\Varid{rep}_{\multimap}\;\Varid{e}\mathrel{=}\lambda \Varid{d}\to \Varid{d}\bullet\Varid{e}{}\<[E]%
\ColumnHook
\end{hscode}\resethooks
\indentend \end{minipage}%
\begin{minipage}{0.5\textwidth}\indentbegin \begin{hscode}\SaveRestoreHook
\column{B}{@{}>{\hspre}l<{\hspost}@{}}%
\column{3}{@{}>{\hspre}l<{\hspost}@{}}%
\column{E}{@{}>{\hspre}l<{\hspost}@{}}%
\>[3]{}\Varid{abs}_{\multimap}\mathbin{::}\Conid{Module}\;\Varid{d}\;\Varid{e}\Rightarrow (\Varid{d}\multimap\Varid{e})\to \Varid{e}{}\<[E]%
\\
\>[3]{}\Varid{abs}_{\multimap}\;\Varid{f}\mathrel{=}\Varid{f}\;\Varid{one}{}\<[E]%
\ColumnHook
\end{hscode}\resethooks
\indentend \end{minipage}


\noindent
In particular, as \ensuremath{\Varid{one}\bullet\Varid{e}\mathrel{=}\Varid{e}} (\Cref{fig:module-laws}), we have that \ensuremath{\Varid{abs}_{\multimap}\hsdot{\circ }{.}\Varid{rep}_{\multimap}\mathrel{=}\Varid{id}}.
Moreover, for linear functions also \ensuremath{\Varid{rep}_{\multimap}\hsdot{\circ }{.}\Varid{abs}_{\multimap}\mathrel{=}\Varid{id}}.
Now we derive \ensuremath{\Conid{Monoid}}, \ensuremath{\Conid{Module}} and \ensuremath{\Conid{Kronecker}} instances that
make \ensuremath{\Varid{rep}_{\multimap}} a homomorphism by construction.
It follows that \ensuremath{\Varid{d}\multimap\Varid{e}} inherits its monoidal structure in a pointwise
fashion from \ensuremath{\Varid{e}}. For \ensuremath{\Varid{e}\mathrel{=}\Conid{Dense}\;\Varid{v}\;\Varid{d}},
the time complexity for additive operations remains linear in the number of variables.
\indentbegin \begin{hscode}\SaveRestoreHook
\column{B}{@{}>{\hspre}l<{\hspost}@{}}%
\column{3}{@{}>{\hspre}l<{\hspost}@{}}%
\column{5}{@{}>{\hspre}l<{\hspost}@{}}%
\column{21}{@{}>{\hspre}l<{\hspost}@{}}%
\column{E}{@{}>{\hspre}l<{\hspost}@{}}%
\>[3]{}\mathbf{instance}\;\Conid{Monoid}\;\Varid{e}\Rightarrow \Conid{Monoid}\;(\Varid{d}\multimap\Varid{e})\;\mathbf{where}{}\<[E]%
\\
\>[3]{}\hsindent{2}{}\<[5]%
\>[5]{}\Varid{zero}{}\<[21]%
\>[21]{}\mathrel{=}\lambda \Varid{d}\to \Varid{zero}{}\<[E]%
\\
\>[3]{}\hsindent{2}{}\<[5]%
\>[5]{}\Varid{f}\oplus\Varid{g}{}\<[21]%
\>[21]{}\mathrel{=}\lambda \Varid{d}\to \Varid{f}\;\Varid{d}\oplus\Varid{g}\;\Varid{d}{}\<[E]%
\ColumnHook
\end{hscode}\resethooks
\indentend Furthermore, its \ensuremath{\Varid{d}}-module structure uses
the function parameter as an accumulator for the scalar multiplier.
This affords a \emph{``strength reduction''}:
the expensive scalar multiplication \ensuremath{(\bullet)} is replaced by the
cheaper semiring multiplication \ensuremath{(\otimes)}.

\indentbegin \begin{hscode}\SaveRestoreHook
\column{B}{@{}>{\hspre}l<{\hspost}@{}}%
\column{3}{@{}>{\hspre}l<{\hspost}@{}}%
\column{5}{@{}>{\hspre}l<{\hspost}@{}}%
\column{E}{@{}>{\hspre}l<{\hspost}@{}}%
\>[3]{}\mathbf{instance}\;\Conid{Module}\;\Varid{d}\;\Varid{e}\Rightarrow \Conid{Module}\;\Varid{d}\;(\Varid{d}\multimap\Varid{e})\;\mathbf{where}{}\<[E]%
\\
\>[3]{}\hsindent{2}{}\<[5]%
\>[5]{}\Varid{d'}\bullet\Varid{f}\mathrel{=}\lambda \Varid{d}\to \Varid{f}\;(\Varid{d'}\otimes\Varid{d}){}\<[E]%
\ColumnHook
\end{hscode}\resethooks
\indentend This reduces the time complexity of scalar multiplication
from \bigO{V} (where \ensuremath{\Conid{V}} denotes the number of variables with non-zero values in the given sparse map)
for the \ensuremath{\Conid{Sparse}\;\Varid{v}\;\Varid{d}} representation
to \bigO{1}
for that of \ensuremath{\Varid{d}\multimap\Conid{Sparse}\;\Varid{v}\;\Varid{d}}.
%
%
%
%
%
%
However, the general \ensuremath{\Conid{Kronecker}} instance of \ensuremath{\Varid{d}\multimap\Varid{e}} still contains the expensive \ensuremath{(\bullet)} operator.
\indentbegin \begin{hscode}\SaveRestoreHook
\column{B}{@{}>{\hspre}l<{\hspost}@{}}%
\column{3}{@{}>{\hspre}l<{\hspost}@{}}%
\column{5}{@{}>{\hspre}l<{\hspost}@{}}%
\column{E}{@{}>{\hspre}l<{\hspost}@{}}%
\>[3]{}\mathbf{instance}\;\Conid{Kronecker}\;\Varid{v}\;\Varid{d}\;\Varid{e}\Rightarrow \Conid{Kronecker}\;\Varid{v}\;\Varid{d}\;(\Varid{d}\multimap\Varid{e})\;\mathbf{where}{}\<[E]%
\\
\>[3]{}\hsindent{2}{}\<[5]%
\>[5]{}\Varid{delta}\;\Varid{v}\mathrel{=}\lambda \Varid{d}\to \Varid{d}\bullet\Varid{delta}\;\Varid{v}{}\<[E]%
\ColumnHook
\end{hscode}\resethooks
\indentend 
Fortunately, we only require a specialized instance for sparse maps, which operates in constant time, as a consequence of
equation $(\ref{eq:singleton})$ for \ensuremath{\Varid{d}\bullet\Varid{delta}\;\Varid{v}}.


\indentbegin \begin{hscode}\SaveRestoreHook
\column{B}{@{}>{\hspre}l<{\hspost}@{}}%
\column{3}{@{}>{\hspre}l<{\hspost}@{}}%
\column{5}{@{}>{\hspre}l<{\hspost}@{}}%
\column{E}{@{}>{\hspre}l<{\hspost}@{}}%
\>[3]{}\mathbf{instance}\;(\Conid{Ord}\;\Varid{v},\Conid{Semiring}\;\Varid{d})\Rightarrow \Conid{Kronecker}\;\Varid{v}\;\Varid{d}\;(\Varid{d}\multimap\Conid{Sparse}\;\Varid{v}\;\Varid{d})\;\mathbf{where}{}\<[E]%
\\
\>[3]{}\hsindent{2}{}\<[5]%
\>[5]{}\Varid{delta}\mathrel{=}\Varid{singleton}{}\<[E]%
\ColumnHook
\end{hscode}\resethooks
\indentend Finally, the \ensuremath{\Conid{CorrectAD}} instance for \ensuremath{\Varid{d}\multimap\Varid{e}} is defined as the composition of \ensuremath{\Varid{rep}_{\multimap}}/\ensuremath{\Varid{abs}_{\multimap}} with \ensuremath{\Varid{rep}}/\ensuremath{\Varid{abs}} of \ensuremath{\Varid{e}}.
\indentbegin \begin{hscode}\SaveRestoreHook
\column{B}{@{}>{\hspre}l<{\hspost}@{}}%
\column{3}{@{}>{\hspre}l<{\hspost}@{}}%
\column{7}{@{}>{\hspre}l<{\hspost}@{}}%
\column{14}{@{}>{\hspre}l<{\hspost}@{}}%
\column{E}{@{}>{\hspre}l<{\hspost}@{}}%
\>[3]{}\mathbf{instance}\;\Conid{CorrectAD}\;\Varid{v}\;\Varid{d}\;\Varid{e}\Rightarrow \Conid{CorrectAD}\;\Varid{v}\;\Varid{d}\;(\Varid{d}\multimap\Varid{e})\;\mathbf{where}{}\<[E]%
\\
\>[3]{}\hsindent{4}{}\<[7]%
\>[7]{}\Varid{rep}{}\<[14]%
\>[14]{}\mathrel{=}\Varid{rep}_{\multimap}\hsdot{\circ }{.}\Varid{rep}{}\<[E]%
\\
\>[3]{}\hsindent{4}{}\<[7]%
\>[7]{}\Varid{abs}{}\<[14]%
\>[14]{}\mathrel{=}\Varid{abs}\hsdot{\circ }{.}\Varid{abs}_{\multimap}{}\<[E]%
\ColumnHook
\end{hscode}\resethooks
\indentend In summary, by using \ensuremath{\Varid{d}\ltimes(\Varid{d}\multimap\Conid{Sparse}\;\Varid{v}\;\Varid{d})} we obtain reverse-mode AD. 
\indentbegin \begin{hscode}\SaveRestoreHook
\column{B}{@{}>{\hspre}l<{\hspost}@{}}%
\column{3}{@{}>{\hspre}l<{\hspost}@{}}%
\column{E}{@{}>{\hspre}l<{\hspost}@{}}%
\>[3]{}\Varid{reverseAD}\mathbin{::}(\Conid{Ord}\;\Varid{v},\Conid{Semiring}\;\Varid{d})\Rightarrow (\Varid{v}\to \Varid{d})\to \Conid{Expr}\;\Varid{v}\to \Varid{d}\ltimes(\colorbox{lightgray}{$\Varid{d}\multimap\Conid{Sparse}\;\Varid{v}\;\Varid{d}$}){}\<[E]%
\\
\>[3]{}\Varid{reverseAD}\mathrel{=}\Varid{abstractD}{}\<[E]%
\ColumnHook
\end{hscode}\resethooks
\indentend 

\noindent
%
For instance, consider the more extensive example expression
$x \times ((x + 1) \times (x + x))$, encoded in Haskell as \ensuremath{\Varid{example}_{3}}.
\indentbegin \begin{hscode}\SaveRestoreHook
\column{B}{@{}>{\hspre}l<{\hspost}@{}}%
\column{3}{@{}>{\hspre}l<{\hspost}@{}}%
\column{E}{@{}>{\hspre}l<{\hspost}@{}}%
\>[3]{}\Varid{example}_{3}\mathbin{::}\Conid{Expr}\;\Conid{X}{}\<[E]%
\\
\>[3]{}\Varid{example}_{3}\mathrel{=}\Conid{Times}\;(\Conid{Var}\;\Conid{X})\;(\Conid{Times}\;(\Conid{Plus}\;(\Conid{Var}\;\Conid{X})\;\Conid{One})\;(\Conid{Plus}\;(\Conid{Var}\;\Conid{X})\;(\Conid{Var}\;\Conid{X}))){}\<[E]%
\ColumnHook
\end{hscode}\resethooks
\indentend 
Running reverse-mode AD on this expression,
the derivative at \ensuremath{\Conid{X}\mathrel{=}\mathrm{5}} is 170.
\indentbegin \begin{hscode}\SaveRestoreHook
\column{B}{@{}>{\hspre}l<{\hspost}@{}}%
\column{3}{@{}>{\hspre}l<{\hspost}@{}}%
\column{E}{@{}>{\hspre}l<{\hspost}@{}}%
\>[3]{}\mathbin{>}(\Varid{abs}_{\multimap}\hsdot{\circ }{.}\Varid{tan}^{N})\;(\Varid{reverseAD}\;(\lambda \Conid{X}\to \mathrm{5})\;\Varid{example}_{3}){}\<[E]%
\\
\>[3]{}\{\mskip1.5mu \Conid{X}\mapsto\mathrm{170}\mskip1.5mu\}{}\<[E]%
\ColumnHook
\end{hscode}\resethooks
\indentend 
\newcommand\DoubleLine[9][3pt]{%
  \path(#2)--(#3)coordinate[at start](h1)coordinate[at end](h2);
  \draw[#4]($(h1)!#1!90:(h2)$)-- node[#8]{#6}($(h2)!#1!-90:(h1)$);
  \draw[#5]($(h1)!#1!-90:(h2)$)--node[#9]{#7}($(h2)!#1!90:(h1)$);
}

\newcommand\BackwardForward[4]{%
  \DoubleLine{#1}{#2}{backwardarc}{forwardarc}{#3}{#4}{backward}{forward}
}

\newcommand\Forward[4]{%
  \path(#1)--(#2)coordinate[at start](h1)coordinate[at end](h2);
  \draw[forwardarc]($(h1)!0pt!-90:(h2)$)--node[forward]{#3}($(h2)!0pt!90:(h1)$);
  \draw[forwardarc]($(h1)!0pt!-90:(h2)$)--node[forward2]{#4}($(h2)!0pt!90:(h1)$);
}

\newcommand\factorbox[2]{%
  \path (#1.south)+(2pt,-8pt) node [factor] {#2} +(0,0);
}

\begin{figure}
\textbf{Forward Pass:}
\vspace{-0.5cm}
\begin{center}
\begin{tikzpicture}[scale=0.8,font = \small]
  \tikzstyle{level}=[level distance=4cm, sibling distance = 1cm]
  \tikzstyle{internal} = [rectangle,draw,outer sep = 3pt,inner sep=1pt, text centered,minimum width=1cm, minimum height=5mm]
  \tikzstyle{leaf} = [circle,outer sep = 4pt]
  \tikzstyle{forward} = [below,black, font = \small]
  \tikzstyle{forward2} = [above,black, font = \small]
  \tikzstyle{forwardarc} = [draw,->,black]
  \tikzstyle{backward} = [above,outer sep = 2pt,red]
  \tikzstyle{backwardarc} = [draw,<-,red]
  \tikzstyle{factor} = [draw,rectangle,rounded corners = 3]

  \node [leaf] (x1) at (0,0)   {\ensuremath{\Varid{x}}};
  \node [leaf] (x2) at (0,-0.5)  {\ensuremath{\Varid{x}}};
  \node [leaf] (c1) at (0,-2.5)  {one};
  \node [leaf] (x3) at (0,-3)  {\ensuremath{\Varid{x}}};
  \node [leaf] (x4) at (0,-5)  {\ensuremath{\Varid{x}}};
  \node [internal] (plus1) at (4,-1.5) {\ensuremath{\oplus}};
  \node [internal] (plus2) at (4,-4) {\ensuremath{\oplus}};
  \node [internal] (times1) at (8,-2.75) {\ensuremath{\otimes}};
  \node [internal] (times2) at (12,-1.375) {\ensuremath{\otimes}};
  \node [] (top) at (15,-1.375) {};

  \Forward{times2}{top}{\ensuremath{\Varid{c}_{\otimes_2}}}{\ensuremath{\mathrm{300}}}
  \Forward{x1}{times2}{\ensuremath{\Varid{c}_{x_1}}}{\ensuremath{\mathrm{5}}}
  \Forward{x2}{plus1}{\ensuremath{\Varid{c}_{x_2}}}{\ensuremath{\mathrm{5}}}
  \Forward{c1}{plus1}{\ensuremath{\Varid{c}_{1}}}{\ensuremath{\mathrm{1}}}
  \Forward{x3}{plus2}{\ensuremath{\Varid{c}_{x_3}}}{\ensuremath{\mathrm{5}}}
  \Forward{x4}{plus2}{\ensuremath{\Varid{c}_{x_4}}}{\ensuremath{\mathrm{5}}}
  \Forward{plus1}{times1}{\ensuremath{\Varid{c}_{\oplus_2}}}{\ensuremath{\mathrm{6}}}
  \Forward{plus2}{times1}{\ensuremath{\Varid{c}_{\oplus_1}}}{\ensuremath{\mathrm{10}}}
  \Forward{times1}{times2}{\ensuremath{\Varid{c}_{\otimes_1}}}{\ensuremath{\mathrm{60}}}

  \path[forwardarc] (12,0.5) -- (14,0.5) node [forward] {\ensuremath{\Varid{eval}}};

\end{tikzpicture}
\end{center}

\vspace{-0.25cm}
\textbf{Backward Pass:}
\vspace{-2.75cm}
\begin{center}
\begin{tikzpicture}[scale=0.8,font = \small]
  \tikzstyle{internal} = [rectangle,draw,outer sep = 3pt,inner sep=1pt, text centered,minimum width=1cm, minimum height=5mm]
  \tikzstyle{leaf} = [circle,outer sep = 4pt]
  \tikzstyle{forward} = [above,outer sep = 2pt,blue, font = \small]
  \tikzstyle{forwardarc} = [draw,->,blue]
  \tikzstyle{legendforward} = [draw,<-,blue]
  \tikzstyle{backward} = [below,outer sep = 2pt,red]
  \tikzstyle{backwardarc} = [draw,<-,red]
  \tikzstyle{legendbackward} = [draw,->,red]
  \tikzstyle{factor} = [draw,rectangle,rounded corners = 3]

  \node [leaf] (x1) at (15,0)  {\ensuremath{\Varid{c}_{x_1}}};
  \node [leaf](x2) at (15,-1)  {\ensuremath{\Varid{c}_{x_2}}};
  \node [leaf] (c1) at (15,-4)  {\ensuremath{\Varid{c}_{1}}};
  \node [leaf] (x3) at (15,-5)  {\ensuremath{\Varid{c}_{x_3}}};
  \node [leaf] (x4) at (15,-8) {\ensuremath{\Varid{c}_{x_4}}};
  \node [internal] (plus1) at (11,-2.5) {\ensuremath{\Varid{c}_{\oplus_1}}};
  \node [internal] (plus2) at (11,-6.5) {\ensuremath{\Varid{c}_{\oplus_2}}};
  \node [internal] (times1) at (7,-4.5) {\ensuremath{\Varid{c}_{\otimes_1}}};
  \node [internal] (times2) at (3,-3.75) {\ensuremath{\Varid{c}_{\otimes_2}}};
  \node [] (top) at (0,-3.75) {};

  \BackwardForward{times2}{top}{$1$}{$170$}
  \BackwardForward{x1}{times2}{$60$}{$60$}
  \BackwardForward{x2}{plus1}{$50$}{$50$}
  \BackwardForward{c1}{plus1}{$50$}{$0$}
  \BackwardForward{x3}{plus2}{$30$}{$30$}
  \BackwardForward{x4}{plus2}{$30$}{$30$}
  \BackwardForward{plus1}{times1}{$50$}{$50$}
  \BackwardForward{plus2}{times1}{$30$}{$60$}
  \BackwardForward{times1}{times2}{$5$}{$110$}

  \path[legendforward] (0,-6) -- (2,-6) node [forward] {evaluation result};
  \path[legendbackward] (0,-6.5) -- (2,-6.5) node [backward] {argument};

\end{tikzpicture}
\end{center}

\vspace{-1cm}
\textbf{Constructed Function Closures:}
\vspace{-0.5cm}
\begin{center}
\begin{tabular}{llll}
\toprule
Derivative & Closure & Argument ($d$) & Result\\
\midrule
\ensuremath{\Varid{c}_{\otimes_2}} & \ensuremath{\lambda \Varid{d}\to \Varid{c}_{x_1}\;(\Varid{d}\otimes\mathrm{60})\oplus\Varid{c}_{\otimes_1}\;(\Varid{d}\otimes\mathrm{5})} & 1 & 170\\
\ensuremath{\Varid{c}_{\otimes_1}} & \ensuremath{\lambda \Varid{d}\to \Varid{c}_{\oplus_1}\;(\Varid{d}\otimes\mathrm{10})\oplus\Varid{c}_{\oplus_2}\;(\Varid{d}\otimes\mathrm{6})} & 5 & 110\\
\ensuremath{\Varid{c}_{\oplus_1}} & \ensuremath{\lambda \Varid{d}\to \Varid{c}_{x_2}\;\Varid{d}\oplus\Varid{c}_{1}\;\Varid{d}} & 50 & 50\\
\ensuremath{\Varid{c}_{\oplus_2}} & \ensuremath{\lambda \Varid{d}\to \Varid{c}_{x_3}\;\Varid{d}\oplus\Varid{c}_{x_4}\;\Varid{d}} & 30 & 60\\
\ensuremath{\Varid{c}_{1}} & \ensuremath{\lambda \Varid{d}\to \{\mskip1.5mu \mskip1.5mu\}} & 50 & 0\\
\ensuremath{\Varid{c}_{x_i}} & \ensuremath{\lambda \Varid{d}\to \{\mskip1.5mu \Varid{x}\mapsto\Varid{d}\mskip1.5mu\}} & 60, 50, 30, 30 & 60, 50, 30, 30\\
\bottomrule
\end{tabular}
\end{center}

\caption{Reverse-mode AD of $x\times(x+1)\times(x+x)$, for $x = 5$.}
\label{fig:reverse-mode}
\end{figure}

\noindent
Figure~\ref{fig:reverse-mode} shows that this computation effectively works in multiple passes, which
is a key characteristic of reverse-mode AD.
During the forward (bottom-up) pass (shown by the black arrows) the algorithm
computes Nagata numbers, which contain the value of each sub-expression, and
the dependencies between the nodes as function closures \ensuremath{\Varid{c}\mathbin{::}\Varid{d}\multimap\Conid{Sparse}\;\Varid{v}\;\Varid{d}}, e.g., \ensuremath{\Varid{N}\;\mathrm{300}\;\Varid{c}_{\otimes_2}} for the final outcome.
In the backward (top-down) pass (shown by the red left-to-right arrows), we invoke the
function closure \ensuremath{\Varid{c}_{\otimes_2}} on argument \ensuremath{\Varid{one}} (using \ensuremath{\Varid{abs}_{\multimap}}).  It propagates the
scalar accumulator downwards.
At each downward step, the accumulator changes to reflect how much the overall derivative
changes given that the subexpression changes by one.
Finally, the results at the leaves are added (shown by
the blue right-to-left arrows) to yield the overall derivative. The conventional imperative
algorithms typically accomplish this with imperative
updates. We derive that form in Section~\ref{sec:rev:array}, but first we address the
linear cost of addition.

\subsection{Accumulating Additions}

\noindent
We can optimize the additive structure of \ensuremath{\Varid{d}}-modules
in much the same way as we have optimized scalar multiplications:
by considering a representation of \ensuremath{\Varid{e}\to \Varid{e}} functions that have the \emph{additive}
homogeneity property:
\begin{equation*}
\ensuremath{\Varid{f}\;(\Varid{x}\oplus\Varid{y})\mathrel{=}\Varid{x}\oplus\Varid{f}\;\Varid{y}}
\end{equation*}
This representation is also known as the
Cayley representation~\citep{DBLP:conf/mpc/Hinze12,DBLP:journals/pacmpl/BoisseauG18} of \ensuremath{\Varid{e}}, 
the most well-known example of which is difference lists~\citep{clark1977firstorder} or
Hughes lists \cite{DBLP:journals/ipl/Hughes86}.
Again, the Haskell type \ensuremath{\Varid{e}\to \Varid{e}} includes functions that
do not have the homogeneity property; those are not included in the \ensuremath{\Conid{Cayley}\;\Varid{e}} type.
%

\noindent
\begin{minipage}[b]{0.5\textwidth}\indentbegin \begin{hscode}\SaveRestoreHook
\column{B}{@{}>{\hspre}l<{\hspost}@{}}%
\column{3}{@{}>{\hspre}l<{\hspost}@{}}%
\column{E}{@{}>{\hspre}l<{\hspost}@{}}%
\>[3]{}\mathbf{type}\;\Conid{Cayley}\;\Varid{e}\mathrel{=}\Varid{e}\to \Varid{e}{}\<[E]%
\\[\blanklineskip]%
\>[3]{}\Varid{rep}_{\diamond}\mathbin{::}\Conid{Monoid}\;\Varid{e}\Rightarrow \Varid{e}\to \Conid{Cayley}\;\Varid{e}{}\<[E]%
\\
\>[3]{}\Varid{rep}_{\diamond}\;\Varid{e}\mathrel{=}\lambda \Varid{e'}\to \Varid{e'}\oplus\Varid{e}{}\<[E]%
\ColumnHook
\end{hscode}\resethooks
\indentend \end{minipage}%
\begin{minipage}[b]{0.5\textwidth}\indentbegin \begin{hscode}\SaveRestoreHook
\column{B}{@{}>{\hspre}l<{\hspost}@{}}%
\column{3}{@{}>{\hspre}l<{\hspost}@{}}%
\column{E}{@{}>{\hspre}l<{\hspost}@{}}%
\>[3]{}\Varid{abs}_{\diamond}\mathbin{::}\Conid{Monoid}\;\Varid{e}\Rightarrow \Conid{Cayley}\;\Varid{e}\to \Varid{e}{}\<[E]%
\\
\>[3]{}\Varid{abs}_{\diamond}\;\Varid{f}\mathrel{=}\Varid{f}\;\Varid{zero}{}\<[E]%
\ColumnHook
\end{hscode}\resethooks
\indentend \end{minipage}

\noindent
This representation typically implies another ``strength reduction'', namely
from \ensuremath{\Varid{e}}'s binary operator \ensuremath{(\oplus)}
to function composition \ensuremath{(\hsdot{\circ }{.})}.
We thereby reduce the time complexity of addition from \bigO{V}
for \ensuremath{\Conid{Sparse}\;\Varid{v}\;\Varid{d}}
to \bigO{1}
for \ensuremath{\Conid{Cayley}\;(\Conid{Sparse}\;\Varid{v}\;\Varid{d})}.

\indentbegin \begin{hscode}\SaveRestoreHook
\column{B}{@{}>{\hspre}l<{\hspost}@{}}%
\column{3}{@{}>{\hspre}l<{\hspost}@{}}%
\column{5}{@{}>{\hspre}l<{\hspost}@{}}%
\column{20}{@{}>{\hspre}l<{\hspost}@{}}%
\column{E}{@{}>{\hspre}l<{\hspost}@{}}%
\>[3]{}\mathbf{instance}\;\Conid{Monoid}\;(\Conid{Cayley}\;\Varid{e})\;\mathbf{where}{}\<[E]%
\\
\>[3]{}\hsindent{2}{}\<[5]%
\>[5]{}\Varid{zero}{}\<[20]%
\>[20]{}\mathrel{=}\Varid{id}{}\<[E]%
\\
\>[3]{}\hsindent{2}{}\<[5]%
\>[5]{}\Varid{f}\oplus\Varid{g}{}\<[20]%
\>[20]{}\mathrel{=}\Varid{g}\hsdot{\circ }{.}\Varid{f}{}\<[E]%
\ColumnHook
\end{hscode}\resethooks
\indentend %
%
%
%

\noindent
We derive the \ensuremath{\Conid{Module}} and \ensuremath{\Conid{Kronecker}} instances following our standard recipe.
However, we cannot eliminate the expensive operations from the \ensuremath{\Conid{Cayley}\;\Varid{e}}
instances for \ensuremath{\Conid{Module}} and \ensuremath{\Conid{Kronecker}}.
Indeed, the \ensuremath{\Conid{Module}} instance
requires converting from \ensuremath{\Conid{Cayley}\;\Varid{e}} to \ensuremath{\Varid{e}} and back, where notably
\ensuremath{\Varid{rep}_{\diamond}} re-introduces \ensuremath{\Varid{e}}'s expensive \bigO{V}\ addition operation \ensuremath{(\oplus)},
which we have just managed to eliminate from the \ensuremath{\Conid{Monoid}} instance.
\indentbegin \begin{hscode}\SaveRestoreHook
\column{B}{@{}>{\hspre}l<{\hspost}@{}}%
\column{3}{@{}>{\hspre}l<{\hspost}@{}}%
\column{5}{@{}>{\hspre}l<{\hspost}@{}}%
\column{E}{@{}>{\hspre}l<{\hspost}@{}}%
\>[3]{}\mathbf{instance}\;\Conid{Module}\;\Varid{d}\;\Varid{e}\Rightarrow \Conid{Module}\;\Varid{d}\;(\Conid{Cayley}\;\Varid{e})\;\mathbf{where}{}\<[E]%
\\
\>[3]{}\hsindent{2}{}\<[5]%
\>[5]{}\Varid{d}\bullet\Varid{f}\mathrel{=}\Varid{rep}_{\diamond}\;(\Varid{d}\bullet\Varid{abs}_{\diamond}\;\Varid{f}){}\<[E]%
\ColumnHook
\end{hscode}\resethooks
\indentend The \ensuremath{\Conid{Kronecker}} instance derived via \ensuremath{\Varid{rep}_{\diamond}} from \ensuremath{\Varid{e}}
similarly features \ensuremath{\Varid{e}}'s addition operation \ensuremath{(\oplus)}.
\indentbegin \begin{hscode}\SaveRestoreHook
\column{B}{@{}>{\hspre}l<{\hspost}@{}}%
\column{3}{@{}>{\hspre}l<{\hspost}@{}}%
\column{5}{@{}>{\hspre}l<{\hspost}@{}}%
\column{20}{@{}>{\hspre}l<{\hspost}@{}}%
\column{E}{@{}>{\hspre}l<{\hspost}@{}}%
\>[3]{}\mathbf{instance}\;\Conid{Kronecker}\;\Varid{v}\;\Varid{d}\;\Varid{e}\Rightarrow \Conid{Kronecker}\;\Varid{v}\;\Varid{d}\;(\Conid{Cayley}\;\Varid{e})\;\mathbf{where}{}\<[E]%
\\
\>[3]{}\hsindent{2}{}\<[5]%
\>[5]{}\Varid{delta}\;\Varid{v}\mathrel{=}{}\<[20]%
\>[20]{}\lambda \Varid{e}\to \Varid{e}\oplus\Varid{delta}\;\Varid{v}{}\<[E]%
\ColumnHook
\end{hscode}\resethooks
\indentend Fortunately, we can again define a specialized \ensuremath{\Conid{Kronecker}} instance for the type
\ensuremath{\Varid{e}\mathrel{=}\Varid{d}\multimap\Conid{Cayley}\;(\Conid{Sparse}\;\Varid{v}\;\Varid{d})}, which 
avoids appeal to the expensive \ensuremath{\oplus} operation. Indeed, the \ensuremath{(\bullet)} operation defined on
\ensuremath{\Varid{e}}
need not make use of
that defined
on \ensuremath{\Conid{Cayley}\;(\Conid{Sparse}\;\Varid{v}\;\Varid{d})}.
For the specialized \ensuremath{\Varid{delta}}, we instead calculate:


\indentbegin \begin{hscode}\SaveRestoreHook
\column{B}{@{}>{\hspre}l<{\hspost}@{}}%
\column{3}{@{}>{\hspre}l<{\hspost}@{}}%
\column{12}{@{}>{\hspre}c<{\hspost}@{}}%
\column{12E}{@{}l@{}}%
\column{15}{@{}>{\hspre}l<{\hspost}@{}}%
\column{E}{@{}>{\hspre}l<{\hspost}@{}}%
\>[3]{}\Varid{delta}\;\Varid{v}{}\<[12]%
\>[12]{}\mathrel{=}{}\<[12E]%
\>[15]{}\mbox{\commentbegin  definition of \ensuremath{\Varid{delta}} for \ensuremath{\Varid{d}\multimap}  \commentend}{}\<[E]%
\\
\>[15]{}\lambda \Varid{d}\to \Varid{d}\bullet\Varid{delta}\;\Varid{v}{}\<[E]%
\\
\>[12]{}\mathrel{=}{}\<[12E]%
\>[15]{}\mbox{\commentbegin  definition of \ensuremath{(\bullet)} and \ensuremath{\Varid{delta}} for \ensuremath{\Conid{Cayley}}  \commentend}{}\<[E]%
\\
\>[15]{}\lambda \Varid{d}\to \Varid{rep}_{\diamond}\;(\Varid{d}\bullet\Varid{abs}_{\diamond}\;(\lambda \Varid{e}\to \Varid{e}\oplus\Varid{delta}\;\Varid{v})){}\<[E]%
\\
\>[12]{}\mathrel{=}{}\<[12E]%
\>[15]{}\mbox{\commentbegin  definition of \ensuremath{\Varid{rep}_{\diamond}} and \ensuremath{\Varid{abs}_{\diamond}}  \commentend}{}\<[E]%
\\
\>[15]{}\lambda \Varid{d}\to \lambda \Varid{e}\to \Varid{e}\oplus(\Varid{d}\bullet(\lambda \Varid{e}\to \Varid{e}\oplus\Varid{delta}\;\Varid{v})\;\Varid{zero}){}\<[E]%
\\
\>[12]{}\mathrel{=}{}\<[12E]%
\>[15]{}\mbox{\commentbegin  \ensuremath{\beta}-reduction and unit of monoid (\Cref{fig:module-laws})  \commentend}{}\<[E]%
\\
\>[15]{}\lambda \Varid{d}\to \lambda \Varid{e}\to \Varid{e}\oplus(\Varid{d}\bullet\Varid{delta}\;\Varid{v}){}\<[E]%
\\
\>[12]{}\mathrel{=}{}\<[12E]%
\>[15]{}\mbox{\commentbegin  equation (\ref{eq:insertWith})  \commentend}{}\<[E]%
\\
\>[15]{}\lambda \Varid{d}\to \lambda \Varid{e}\to \Varid{insertWith}\;(\oplus)\;\Varid{v}\;\Varid{d}\;\Varid{e}{}\<[E]%
\\
\>[12]{}\mathrel{=}{}\<[12E]%
\>[15]{}\mbox{\commentbegin  \ensuremath{\eta}-reduction  \commentend}{}\<[E]%
\\
\>[15]{}\lambda \Varid{d}\to \Varid{insertWith}\;(\oplus)\;\Varid{v}\;\Varid{d}{}\<[E]%
\ColumnHook
\end{hscode}\resethooks
\indentend Hence, we obtain an implementation of \ensuremath{\Varid{delta}} that takes only \bigO{\log V}\ time.
\indentbegin \begin{hscode}\SaveRestoreHook
\column{B}{@{}>{\hspre}l<{\hspost}@{}}%
\column{3}{@{}>{\hspre}l<{\hspost}@{}}%
\column{5}{@{}>{\hspre}l<{\hspost}@{}}%
\column{12}{@{}>{\hspre}l<{\hspost}@{}}%
\column{E}{@{}>{\hspre}l<{\hspost}@{}}%
\>[3]{}\mathbf{instance}\;(\Conid{Ord}\;\Varid{v},\Conid{Semiring}\;\Varid{d})\Rightarrow \Conid{Kronecker}\;\Varid{v}\;\Varid{d}\;(\Varid{d}\multimap\Conid{Cayley}\;(\Conid{Sparse}\;\Varid{v}\;\Varid{d})){}\<[E]%
\\
\>[3]{}\hsindent{2}{}\<[5]%
\>[5]{}\mathbf{where}\;{}\<[12]%
\>[12]{}\Varid{delta}\mathrel{=}\Varid{insertWith}\;(\oplus){}\<[E]%
\ColumnHook
\end{hscode}\resethooks
\indentend Moreover, the \ensuremath{\Conid{Cayley}\;\Varid{e}} representation preserves the correctness of \ensuremath{\Varid{e}}.
\indentbegin \begin{hscode}\SaveRestoreHook
\column{B}{@{}>{\hspre}l<{\hspost}@{}}%
\column{3}{@{}>{\hspre}l<{\hspost}@{}}%
\column{5}{@{}>{\hspre}l<{\hspost}@{}}%
\column{11}{@{}>{\hspre}l<{\hspost}@{}}%
\column{E}{@{}>{\hspre}l<{\hspost}@{}}%
\>[3]{}\mathbf{instance}\;\Conid{CorrectAD}\;\Varid{v}\;\Varid{d}\;\Varid{e}\Rightarrow \Conid{CorrectAD}\;\Varid{v}\;\Varid{d}\;(\Conid{Cayley}\;\Varid{e})\;\mathbf{where}{}\<[E]%
\\
\>[3]{}\hsindent{2}{}\<[5]%
\>[5]{}\Varid{rep}\mathrel{=}\Varid{rep}_{\diamond}\hsdot{\circ }{.}\Varid{rep}{}\<[E]%
\\
\>[3]{}\hsindent{2}{}\<[5]%
\>[5]{}\Varid{abs}{}\<[11]%
\>[11]{}\mathrel{=}\Varid{abs}\hsdot{\circ }{.}\Varid{abs}_{\diamond}{}\<[E]%
\ColumnHook
\end{hscode}\resethooks
\indentend %
%
%
In summary, with type \ensuremath{\Varid{d}\multimap\Conid{Cayley}\;(\Conid{Sparse}\;\Varid{v}\;\Varid{d})}, 
both \ensuremath{(\bullet)} and \ensuremath{(\oplus)} take
\bigO{1}\ and \ensuremath{\Varid{delta}} takes \bigO{\log V}.
As a consequence, the time complexity of \ensuremath{\Varid{abstractD}} for a program with \ensuremath{\Conid{N}} nodes and \ensuremath{\Conid{V}} variables is
now \bigO{N \times \log V}.
\indentbegin \begin{hscode}\SaveRestoreHook
\column{B}{@{}>{\hspre}l<{\hspost}@{}}%
\column{3}{@{}>{\hspre}l<{\hspost}@{}}%
\column{17}{@{}>{\hspre}l<{\hspost}@{}}%
\column{E}{@{}>{\hspre}l<{\hspost}@{}}%
\>[3]{}\Varid{reverseAD}_{Cayley}{}\<[17]%
\>[17]{}\mathbin{::}(\Conid{Ord}\;\Varid{v},\Conid{Semiring}\;\Varid{d}){}\<[E]%
\\
\>[17]{}\Rightarrow (\Varid{v}\to \Varid{d})\to \Conid{Expr}\;\Varid{v}\to \Varid{d}\ltimes(\colorbox{lightgray}{$\Varid{d}\multimap\Conid{Cayley}\;(\Conid{Sparse}\;\Varid{v}\;\Varid{d})$}){}\<[E]%
\\
\>[3]{}\Varid{reverseAD}_{Cayley}{}\<[17]%
\>[17]{}\mathrel{=}\Varid{abstractD}{}\<[E]%
\ColumnHook
\end{hscode}\resethooks
\indentend 
%
%
%
%


\noindent
For example, as before, the derivative of \ensuremath{\Varid{example}_{3}} at \ensuremath{\Conid{X}\mathrel{=}\mathrm{5}} is \ensuremath{\mathrm{170}}.
\indentbegin \begin{hscode}\SaveRestoreHook
\column{B}{@{}>{\hspre}l<{\hspost}@{}}%
\column{3}{@{}>{\hspre}l<{\hspost}@{}}%
\column{4}{@{}>{\hspre}l<{\hspost}@{}}%
\column{E}{@{}>{\hspre}l<{\hspost}@{}}%
\>[3]{}\mathbin{>}(\Varid{abs}_{\diamond}\hsdot{\circ }{.}\Varid{abs}_{\multimap}\hsdot{\circ }{.}\Varid{tan}^{N})\;(\Varid{reverseAD}_{Cayley}\;(\lambda \Conid{X}\to \mathrm{5})\;\Varid{example}_{3}){}\<[E]%
\\
\>[3]{}\hsindent{1}{}\<[4]%
\>[4]{}\{\mskip1.5mu \Conid{X}\mapsto\mathrm{170}\mskip1.5mu\}{}\<[E]%
\ColumnHook
\end{hscode}\resethooks
\indentend 
Notice that we can see values of type \ensuremath{\Conid{Cayley}\;(\Conid{Sparse}\;\Varid{v}\;\Varid{d})} as a functional
model for computations that modify a state of type \ensuremath{\Conid{Sparse}\;\Varid{v}\;\Varid{d}}, i.e., immutable maps.

\subsection{Array-based Reverse Mode}\label{sec:rev:array}

\noindent
Performance-wise, we can do even better by switching from immutable maps to
mutable arrays (from \ensuremath{\Conid{\Conid{Data}.\Conid{Array}.ST}}) (\ref{app:functions}). Using these to represent the state,
instead of the purely functional implementation above, we can perform in-place
updates.  Furthermore, the array can be provided by an implicit environment
(\ensuremath{\Conid{\Conid{Control}.\Conid{Monad}.Reader}}). Hence, instead of type \ensuremath{\Conid{Cayley}\;(\Conid{Sparse}\;\Varid{v}\;\Varid{d})}, we can
work with the type \ensuremath{\Conid{ReaderT}\;(\Conid{STArray}\;\Varid{s}\;\Varid{v}\;\Varid{d})\;(\Conid{ST}\;\Varid{s})\;()}. It is convenient
to encapsulate this rather wordy type in the type synonym \ensuremath{\Conid{STCayley}\;\Varid{v}\;\Varid{d}}.
\indentbegin \begin{hscode}\SaveRestoreHook
\column{B}{@{}>{\hspre}l<{\hspost}@{}}%
\column{3}{@{}>{\hspre}l<{\hspost}@{}}%
\column{E}{@{}>{\hspre}l<{\hspost}@{}}%
\>[3]{}\mathbf{type}\;\Conid{STCayley}\;\Varid{v}\;\Varid{d}\mathrel{=}\forall \Varid{s}\hsforall \hsdot{\circ }{.}\Conid{ReaderT}\;(\Conid{STArray}\;\Varid{s}\;\Varid{v}\;\Varid{d})\;(\Conid{ST}\;\Varid{s})\;(){}\<[E]%
\ColumnHook
\end{hscode}\resethooks
\indentend Observe that \ensuremath{\Conid{STCayley}\;\Varid{v}\;\Varid{d}} polymorphically quantifies over the state thread
parameter \ensuremath{\Varid{s}}, which is necessary if we want to run the computation
using Haskell's built-in primitive
\ensuremath{\Varid{runST}\mathbin{::}(\forall \Varid{s}\hsforall \hsdot{\circ }{.}\Conid{ST}\;\Varid{s}\;\Varid{a})\to \Varid{a}}.

The isomorphism between \ensuremath{\Conid{Cayley}\;(\Conid{Sparse}\;\Varid{v}\;\Varid{d})} and \ensuremath{\Conid{STCayley}\;\Varid{v}\;\Varid{d}} essentially converts
between the array and map representations.

\indentbegin \begin{hscode}\SaveRestoreHook
\column{B}{@{}>{\hspre}l<{\hspost}@{}}%
\column{3}{@{}>{\hspre}l<{\hspost}@{}}%
\column{13}{@{}>{\hspre}l<{\hspost}@{}}%
\column{24}{@{}>{\hspre}l<{\hspost}@{}}%
\column{E}{@{}>{\hspre}l<{\hspost}@{}}%
\>[3]{}\Varid{rep}_{>\!\!\!>}\mathbin{::}(\Conid{Ix}\;\Varid{v},\Conid{Semiring}\;\Varid{d})\Rightarrow \Conid{Cayley}\;(\Conid{Sparse}\;\Varid{v}\;\Varid{d})\to \Conid{STCayley}\;\Varid{v}\;\Varid{d}{}\<[E]%
\\
\>[3]{}\Varid{rep}_{>\!\!\!>}\;\Varid{p}\mathrel{=}\Varid{foldMap}\;(\lambda (\Varid{v},\Varid{d})\to \Varid{modifyAt}\;(\oplus\Varid{d})\;\Varid{v})\;(\Varid{toList}\;(\Varid{p}\;\Varid{zero})){}\<[E]%
\\[\blanklineskip]%
\>[3]{}\Varid{abs}_{>\!\!\!>}\mathbin{::}(\Conid{Ix}\;\Varid{v},\Conid{Bounded}\;\Varid{v},\Conid{Semiring}\;\Varid{d})\Rightarrow \Conid{STCayley}\;\Varid{v}\;\Varid{d}\to \Conid{Cayley}\;(\Conid{Sparse}\;\Varid{v}\;\Varid{d}){}\<[E]%
\\
\>[3]{}\Varid{abs}_{>\!\!\!>}\;\Varid{q}\mathrel{=}{}\<[13]%
\>[13]{}\Varid{runST}\;(\mathbf{do}\;{}\<[24]%
\>[24]{}\Varid{arr}\leftarrow \Varid{newArray}\;(\Varid{minBound},\Varid{maxBound})\;\Varid{zero}{}\<[E]%
\\
\>[24]{}\Varid{runReaderT}\;\Varid{q}\;\Varid{arr}{}\<[E]%
\\
\>[24]{}\Varid{l}\leftarrow \Varid{getAssocs}\;\Varid{arr}{}\<[E]%
\\
\>[24]{}\Varid{return}\mathbin{\$}\Varid{foldMap}\;(\lambda (\Varid{v},\Varid{d})\to \Varid{insertWith}\;(\oplus)\;\Varid{v}\;\Varid{d})\;\Varid{l}){}\<[E]%
\ColumnHook
\end{hscode}\resethooks
\indentend The \ensuremath{\Varid{rep}_{>\!\!\!>}} function uses a helper function \ensuremath{\Varid{modifyAt}} to edit the value at a
specific position of the array, which is the counterpart of \ensuremath{\Varid{insertWith}} on maps.
\indentbegin \begin{hscode}\SaveRestoreHook
\column{B}{@{}>{\hspre}l<{\hspost}@{}}%
\column{3}{@{}>{\hspre}l<{\hspost}@{}}%
\column{E}{@{}>{\hspre}l<{\hspost}@{}}%
\>[3]{}\Varid{modifyAt}\mathbin{::}\Conid{Ix}\;\Varid{v}\Rightarrow (\Varid{d}\to \Varid{d})\to \Varid{v}\to \Conid{STCayley}\;\Varid{v}\;\Varid{d}{}\<[E]%
\\
\>[3]{}\Varid{modifyAt}\;\Varid{f}\;\Varid{v}\mathrel{=}\mathbf{do}\;\Varid{arr}\leftarrow \Varid{ask};\Varid{x}\leftarrow \Varid{readArray}\;\Varid{arr}\;\Varid{v};\Varid{writeArray}\;\Varid{arr}\;\Varid{v}\;(\Varid{f}\;\Varid{x}){}\<[E]%
\ColumnHook
\end{hscode}\resethooks
\indentend 

\noindent
Using the methodology of Section~\ref{sec:abstractad:approach} the \ensuremath{\Conid{Monoid}}
instance works out to be the generic instance that is available to any type \ensuremath{\Varid{m}\;()} where \ensuremath{\Varid{m}} is a monad.
\indentbegin \begin{hscode}\SaveRestoreHook
\column{B}{@{}>{\hspre}l<{\hspost}@{}}%
\column{3}{@{}>{\hspre}l<{\hspost}@{}}%
\column{5}{@{}>{\hspre}l<{\hspost}@{}}%
\column{13}{@{}>{\hspre}l<{\hspost}@{}}%
\column{E}{@{}>{\hspre}l<{\hspost}@{}}%
\>[3]{}\mathbf{instance}\;\Conid{Monoid}\;(\Conid{STCayley}\;\Varid{v}\;\Varid{d})\;\mathbf{where}{}\<[E]%
\\
\>[3]{}\hsindent{2}{}\<[5]%
\>[5]{}\Varid{zero}{}\<[13]%
\>[13]{}\mathrel{=}\Varid{return}\;(){}\<[E]%
\\
\>[3]{}\hsindent{2}{}\<[5]%
\>[5]{}\Varid{p}\oplus\Varid{q}{}\<[13]%
\>[13]{}\mathrel{=}\Varid{p}\sequ \Varid{q}{}\<[E]%
\ColumnHook
\end{hscode}\resethooks
\indentend As we work with \ensuremath{\Varid{d}\multimap\Conid{STCayley}\;\Varid{v}\;\Varid{d}}, the \ensuremath{\Conid{Module}} structure of \ensuremath{\Conid{STCayley}\;\Varid{v}\;\Varid{d}}
is not used. For that reason, we skip immediately to the \ensuremath{\Conid{Kronecker}} instance where
the purpose of using mutable arrays is to reduce the \bigO{\log V}\
time complexity of \ensuremath{\Varid{insertWith}} in \ensuremath{\Varid{delta}} to the \bigO{1}\
time complexity of \ensuremath{\Varid{modifyAt}}.

\indentbegin \begin{hscode}\SaveRestoreHook
\column{B}{@{}>{\hspre}l<{\hspost}@{}}%
\column{3}{@{}>{\hspre}l<{\hspost}@{}}%
\column{5}{@{}>{\hspre}l<{\hspost}@{}}%
\column{E}{@{}>{\hspre}l<{\hspost}@{}}%
\>[3]{}\mathbf{instance}\;(\Conid{Ix}\;\Varid{v},\Conid{Semiring}\;\Varid{d})\Rightarrow \Conid{Kronecker}\;\Varid{v}\;\Varid{d}\;(\Conid{STCayley}\;\Varid{v}\;\Varid{d})\;\mathbf{where}{}\<[E]%
\\
\>[3]{}\hsindent{2}{}\<[5]%
\>[5]{}\Varid{delta}\;\Varid{v}\mathrel{=}\Varid{modifyAt}\;(\oplus\Varid{one})\;\Varid{v}{}\<[E]%
\ColumnHook
\end{hscode}\resethooks
\indentend In the \ensuremath{\Varid{d}\multimap\Conid{STCayley}\;\Varid{v}\;\Varid{d}} combination this becomes:

\indentbegin \begin{hscode}\SaveRestoreHook
\column{B}{@{}>{\hspre}l<{\hspost}@{}}%
\column{3}{@{}>{\hspre}l<{\hspost}@{}}%
\column{5}{@{}>{\hspre}l<{\hspost}@{}}%
\column{E}{@{}>{\hspre}l<{\hspost}@{}}%
\>[3]{}\mathbf{instance}\;(\Conid{Ix}\;\Varid{v},\Conid{Semiring}\;\Varid{d})\Rightarrow \Conid{Kronecker}\;\Varid{v}\;\Varid{d}\;(\Varid{d}\multimap\Conid{STCayley}\;\Varid{v}\;\Varid{d})\;\mathbf{where}{}\<[E]%
\\
\>[3]{}\hsindent{2}{}\<[5]%
\>[5]{}\Varid{delta}\;\Varid{v}\mathrel{=}\lambda \Varid{d}\to \Varid{modifyAt}\;(\oplus\Varid{d})\;\Varid{v}{}\<[E]%
\ColumnHook
\end{hscode}\resethooks
\indentend 
The correctness of \ensuremath{\Conid{STCayley}\;\Varid{v}\;\Varid{d}} is established in terms of that of \ensuremath{\Conid{Cayley}\;(\Conid{Sparse}\;\Varid{v}\;\Varid{d})}.
\indentbegin \begin{hscode}\SaveRestoreHook
\column{B}{@{}>{\hspre}l<{\hspost}@{}}%
\column{3}{@{}>{\hspre}l<{\hspost}@{}}%
\column{5}{@{}>{\hspre}l<{\hspost}@{}}%
\column{7}{@{}>{\hspre}l<{\hspost}@{}}%
\column{13}{@{}>{\hspre}l<{\hspost}@{}}%
\column{33}{@{}>{\hspre}l<{\hspost}@{}}%
\column{E}{@{}>{\hspre}l<{\hspost}@{}}%
\>[3]{}\mathbf{instance}\;(\Conid{Ix}\;\Varid{v},\Conid{Ord}\;\Varid{v},\Conid{Bounded}\;\Varid{v},\Conid{Enum}\;\Varid{v},\Conid{Semiring}\;\Varid{d},\Conid{Eq}\;\Varid{d})\Rightarrow {}\<[E]%
\\
\>[3]{}\hsindent{2}{}\<[5]%
\>[5]{}\Conid{CorrectAD}\;\Varid{v}\;\Varid{d}\;(\Conid{STCayley}\;\Varid{v}\;\Varid{d})\;{}\<[33]%
\>[33]{}\mathbf{where}{}\<[E]%
\\
\>[5]{}\hsindent{2}{}\<[7]%
\>[7]{}\Varid{rep}\mathrel{=}\Varid{rep}_{>\!\!\!>}\hsdot{\circ }{.}\Varid{rep}{}\<[E]%
\\
\>[5]{}\hsindent{2}{}\<[7]%
\>[7]{}\Varid{abs}{}\<[13]%
\>[13]{}\mathrel{=}\Varid{abs}\hsdot{\circ }{.}\Varid{abs}_{>\!\!\!>}{}\<[E]%
\ColumnHook
\end{hscode}\resethooks
\indentend Finally, we get the resulting monadic definition for reverse-mode AD.

\indentbegin \begin{hscode}\SaveRestoreHook
\column{B}{@{}>{\hspre}l<{\hspost}@{}}%
\column{3}{@{}>{\hspre}l<{\hspost}@{}}%
\column{24}{@{}>{\hspre}l<{\hspost}@{}}%
\column{E}{@{}>{\hspre}l<{\hspost}@{}}%
\>[3]{}\Varid{reverseAD}_{m}{}\<[24]%
\>[24]{}\mathbin{::}(\Conid{Ix}\;\Varid{v},\Conid{Semiring}\;\Varid{d}){}\<[E]%
\\
\>[24]{}\Rightarrow (\Varid{v}\to \Varid{d})\to \Conid{Expr}\;\Varid{v}\to \Varid{d}\ltimes(\colorbox{lightgray}{$\Varid{d}\multimap\Conid{STCayley}\;\Varid{v}\;\Varid{d}$}){}\<[E]%
\\
\>[3]{}\Varid{reverseAD}_{m}{}\<[24]%
\>[24]{}\mathrel{=}\Varid{abstractD}{}\<[E]%
\ColumnHook
\end{hscode}\resethooks
\indentend 
\noindent
We use this efficient version of reverse-mode AD on \ensuremath{\Varid{example}_{3}} as follows:
\indentbegin \begin{hscode}\SaveRestoreHook
\column{B}{@{}>{\hspre}l<{\hspost}@{}}%
\column{3}{@{}>{\hspre}l<{\hspost}@{}}%
\column{E}{@{}>{\hspre}l<{\hspost}@{}}%
\>[3]{}\mathbin{>}(\Varid{abs}_{\diamond}\hsdot{\circ }{.}\Varid{abs}_{>\!\!\!>}\hsdot{\circ }{.}\Varid{abs}_{\multimap}\hsdot{\circ }{.}\Varid{tan}^{N})\;(\Varid{reverseAD}_{m}\;(\lambda \Conid{X}\to \mathrm{5})\;\Varid{example}_{3}){}\<[E]%
\\
\>[3]{}\{\mskip1.5mu \Conid{X}\mapsto\mathrm{170}\mskip1.5mu\}{}\<[E]%
\ColumnHook
\end{hscode}\resethooks
\indentend 
%
%
%


\subsection{Summary}

\begin{figure}[h]
\centering
\begin{tabular}{@{}l l l@{}}
\bf Differentiation & \bf Semiring      & \bf Time       \\
\bf Mode            & \bf Type          & \bf Complexity \\
\hline
Symbolic & \ensuremath{(\Conid{Expr}\;\Varid{v})\ltimes(\Conid{Expr}\;\Varid{v})} & \bigO{N^2} \\
Forward Mode & \ensuremath{\Varid{d}\ltimes(\Conid{Dense}\;\Varid{v}\;\Varid{d})} & \bigO{N \times V} \\
Forward Mode & \ensuremath{\Varid{d}\ltimes(\Conid{Sparse}\;\Varid{v}\;\Varid{d})} & \bigO{N \times V} \\
Reverse Mode & \ensuremath{\Varid{d}\ltimes(\Varid{d}\multimap\Conid{Sparse}\;\Varid{v}\;\Varid{d})} & \bigO{N \times V} \\
Reverse Mode & \ensuremath{\Varid{d}\ltimes(\Varid{d}\multimap\Conid{Cayley}\;(\Conid{Sparse}\;\Varid{v}\;\Varid{d}))} & \bigO{N \times \log V} \\
Reverse Mode & \ensuremath{\Varid{d}\ltimes(\Varid{d}\multimap\Conid{STCayley}\;\Varid{v}\;\Varid{d})} & \bigO{N+V}
\end{tabular}
\caption{The successive refinements of the semiring type, with
  their corresponding differentiation modes and time complexity, for an expression of $N$ nodes in $V$ variables.}
\label{fig:overview}
\end{figure}

\noindent
Figure~\ref{fig:overview} summarizes the successive differentiation algorithms we
have constructed by instantiating \ensuremath{\Varid{abstractD}} with
semiring types \ensuremath{\Varid{d}\ltimes\Varid{e}} for different choices of tangent type \ensuremath{\Varid{e}}.
It also shows the worst-case time complexity of each algorithm.
Notice that, when $V$ is large, i.e.,
there are many variables, reverse-mode AD, and the last two variants in particular,
is the most interesting. However, for few variables, forward-mode AD
is preferable as it only takes a single pass.

Finally, in Figure~\ref{fig:inlined} we collapse the layers of abstraction to
show what the one-but-last entry of Figure~\ref{fig:overview}---our most
efficient purely functional reverse-mode AD algorithm---looks like after
aggressive inlining. For the sake of simplicity we have also replaced the custom
algebraic datatype for Nagata numbers with plain Haskell tuples.

\begin{figure}[h]
\indentbegin \begin{hscode}\SaveRestoreHook
\column{B}{@{}>{\hspre}l<{\hspost}@{}}%
\column{3}{@{}>{\hspre}l<{\hspost}@{}}%
\column{5}{@{}>{\hspre}l<{\hspost}@{}}%
\column{9}{@{}>{\hspre}l<{\hspost}@{}}%
\column{21}{@{}>{\hspre}l<{\hspost}@{}}%
\column{29}{@{}>{\hspre}c<{\hspost}@{}}%
\column{29E}{@{}l@{}}%
\column{32}{@{}>{\hspre}l<{\hspost}@{}}%
\column{37}{@{}>{\hspre}l<{\hspost}@{}}%
\column{41}{@{}>{\hspre}l<{\hspost}@{}}%
\column{46}{@{}>{\hspre}c<{\hspost}@{}}%
\column{46E}{@{}l@{}}%
\column{48}{@{}>{\hspre}l<{\hspost}@{}}%
\column{49}{@{}>{\hspre}l<{\hspost}@{}}%
\column{52}{@{}>{\hspre}l<{\hspost}@{}}%
\column{E}{@{}>{\hspre}l<{\hspost}@{}}%
\>[3]{}\Varid{autodiff}\mathbin{::}(\Conid{Ord}\;\Varid{v},\Conid{Semiring}\;\Varid{d})\Rightarrow (\Varid{v}\to \Varid{d})\to \Conid{Expr}\;\Varid{v}\to \Conid{Map}\;\Varid{v}\;\Varid{d}{}\<[E]%
\\
\>[3]{}\Varid{autodiff}\;\Varid{var}\;\Varid{e}\mathrel{=}\Varid{snd}\;(\Varid{go}\;\Varid{var}\;\Varid{e})\;\Varid{one}\;\Varid{zero}\;\mathbf{where}{}\<[E]%
\\
\>[3]{}\hsindent{2}{}\<[5]%
\>[5]{}\Varid{go}{}\<[9]%
\>[9]{}\mathbin{::}(\Conid{Ord}\;\Varid{v},\Conid{Semiring}\;\Varid{d}){}\<[E]%
\\
\>[9]{}\Rightarrow (\Varid{v}\to \Varid{d})\to \Conid{Expr}\;\Varid{v}\to (\Varid{d},\Varid{d}\to \Conid{Map}\;\Varid{v}\;\Varid{d}\to \Conid{Map}\;\Varid{v}\;\Varid{d}){}\<[E]%
\\
\>[3]{}\hsindent{2}{}\<[5]%
\>[5]{}\Varid{go}\;{}\<[9]%
\>[9]{}\Varid{var}\;(\Conid{Var}\;\Varid{v}){}\<[29]%
\>[29]{}\mathrel{=}{}\<[29E]%
\>[32]{}(\Varid{var}\;\Varid{v},{}\<[41]%
\>[41]{}\lambda \Varid{d}\to \lambda \Varid{m}\to \Varid{insertWith}\;(\oplus)\;\Varid{v}\;\Varid{d}\;\Varid{m}){}\<[E]%
\\
\>[3]{}\hsindent{2}{}\<[5]%
\>[5]{}\Varid{go}\;{}\<[9]%
\>[9]{}\Varid{var}\;\Conid{Zero}{}\<[29]%
\>[29]{}\mathrel{=}{}\<[29E]%
\>[32]{}(\Varid{zero},{}\<[41]%
\>[41]{}\lambda \Varid{d}\to \lambda \Varid{m}\to \Varid{m}){}\<[E]%
\\
\>[3]{}\hsindent{2}{}\<[5]%
\>[5]{}\Varid{go}\;{}\<[9]%
\>[9]{}\Varid{var}\;\Conid{One}{}\<[29]%
\>[29]{}\mathrel{=}{}\<[29E]%
\>[32]{}(\Varid{one},{}\<[41]%
\>[41]{}\lambda \Varid{d}\to \lambda \Varid{m}\to \Varid{m}){}\<[E]%
\\
\>[3]{}\hsindent{2}{}\<[5]%
\>[5]{}\Varid{go}\;{}\<[9]%
\>[9]{}\Varid{var}\;(\Conid{Plus}\;{}\<[21]%
\>[21]{}\Varid{e}_{1}\;\Varid{e}_{2}){}\<[29]%
\>[29]{}\mathrel{=}{}\<[29E]%
\>[32]{}\mathbf{let}\;{}\<[37]%
\>[37]{}(\Varid{f},\Varid{df}){}\<[46]%
\>[46]{}\mathrel{=}{}\<[46E]%
\>[49]{}\Varid{go}\;\Varid{var}\;\Varid{e}_{1}{}\<[E]%
\\
\>[37]{}(\Varid{g},\Varid{dg}){}\<[46]%
\>[46]{}\mathrel{=}{}\<[46E]%
\>[49]{}\Varid{go}\;\Varid{var}\;\Varid{e}_{2}{}\<[E]%
\\
\>[32]{}\mathbf{in}\;{}\<[37]%
\>[37]{}(\Varid{f}\oplus{}\<[48]%
\>[48]{}\Varid{g},{}\<[52]%
\>[52]{}\lambda \Varid{d}\to \lambda \Varid{m}\to \Varid{df}\;\Varid{d}\;(\Varid{dg}\;\Varid{d}\;\Varid{m})){}\<[E]%
\\
\>[3]{}\hsindent{2}{}\<[5]%
\>[5]{}\Varid{go}\;{}\<[9]%
\>[9]{}\Varid{var}\;(\Conid{Times}\;{}\<[21]%
\>[21]{}\Varid{e}_{1}\;\Varid{e}_{2}){}\<[29]%
\>[29]{}\mathrel{=}{}\<[29E]%
\>[32]{}\mathbf{let}\;{}\<[37]%
\>[37]{}(\Varid{f},\Varid{df}){}\<[46]%
\>[46]{}\mathrel{=}{}\<[46E]%
\>[49]{}\Varid{go}\;\Varid{var}\;\Varid{e}_{1}{}\<[E]%
\\
\>[37]{}(\Varid{g},\Varid{dg}){}\<[46]%
\>[46]{}\mathrel{=}{}\<[46E]%
\>[49]{}\Varid{go}\;\Varid{var}\;\Varid{e}_{2}{}\<[E]%
\\
\>[32]{}\mathbf{in}\;{}\<[37]%
\>[37]{}(\Varid{f}\otimes\Varid{g},{}\<[52]%
\>[52]{}\lambda \Varid{d}\to \lambda \Varid{m}\to \Varid{df}\;(\Varid{d}\otimes\Varid{g})\;(\Varid{dg}\;(\Varid{d}\otimes\Varid{f})\;\Varid{m})){}\<[E]%
\ColumnHook
\end{hscode}\resethooks
\indentend \caption{Fully inlined AD code based on the \ensuremath{\Varid{d}\ltimes(\Varid{d}\multimap\Conid{Cayley}\;(\Conid{Sparse}\;\Varid{v}\;\Varid{d}))} semiring.}
\label{fig:inlined}
\end{figure}



\section{Extensions}\label{sec:extensions}

\noindent
Our setup allows for various extensions that make AD more useful.
Notice that, in these extensions, we often revert to the more naive \ensuremath{\Varid{forwardAD}_{Dense}}
with the \ensuremath{\Conid{Dense}} representation for the sake of brevity.

\subsection{Additional Primitives}\label{sec:additional-primitives}

\noindent
We have so far focused only on the minimal \ensuremath{\Conid{Semiring}} interface. Yet, it is
well-known that dual numbers (and, by extension, Nagata numbers) easily adapt to other primitives, such as trigonometric,
exponential and logarithmic functions.
%
We model such functions as subclasses of the \ensuremath{\Conid{Semiring}} type class.
For instance, we can extend the approach from semirings
to rings by creating a \ensuremath{\Conid{Ring}} subclass with a \ensuremath{\Varid{negate}} method
for additive inverses.\footnote{This deviates from the Haskell Prelude where \ensuremath{\Varid{negate}} is a method in the \ensuremath{\Conid{Num}} type class.}

\noindent
\begin{tabular}{@{}l@{\hspace{5mm}}l@{}}
\begin{minipage}[b]{0.33\textwidth}
\indentbegin \begin{hscode}\SaveRestoreHook
\column{B}{@{}>{\hspre}l<{\hspost}@{}}%
\column{3}{@{}>{\hspre}l<{\hspost}@{}}%
\column{5}{@{}>{\hspre}l<{\hspost}@{}}%
\column{E}{@{}>{\hspre}l<{\hspost}@{}}%
\>[3]{}\mathbf{class}\;\Conid{Semiring}\;\Varid{d}\Rightarrow \Conid{Ring}\;\Varid{d}{}\<[E]%
\\
\>[3]{}\hsindent{2}{}\<[5]%
\>[5]{}\mathbf{where}\;\Varid{negate}\mathbin{::}\Varid{d}\to \Varid{d}{}\<[E]%
\ColumnHook
\end{hscode}\resethooks
\indentend \end{minipage}
&
\begin{minipage}[b]{0.67\textwidth}
\indentbegin \begin{hscode}\SaveRestoreHook
\column{B}{@{}>{\hspre}l<{\hspost}@{}}%
\column{3}{@{}>{\hspre}l<{\hspost}@{}}%
\column{5}{@{}>{\hspre}l<{\hspost}@{}}%
\column{12}{@{}>{\hspre}l<{\hspost}@{}}%
\column{13}{@{}>{\hspre}l<{\hspost}@{}}%
\column{30}{@{}>{\hspre}c<{\hspost}@{}}%
\column{30E}{@{}l@{}}%
\column{33}{@{}>{\hspre}l<{\hspost}@{}}%
\column{36}{@{}>{\hspre}l<{\hspost}@{}}%
\column{E}{@{}>{\hspre}l<{\hspost}@{}}%
\>[3]{}\mathbf{instance}\;{}\<[13]%
\>[13]{}(\Conid{Module}\;\Varid{d}\;\Varid{e},\Conid{Ring}\;\Varid{d}){}\<[36]%
\>[36]{}\Rightarrow \Conid{Ring}\;(\Varid{d}\ltimes\Varid{e}){}\<[E]%
\\
\>[3]{}\hsindent{2}{}\<[5]%
\>[5]{}\mathbf{where}\;{}\<[12]%
\>[12]{}\Varid{negate}\;(\Varid{N}\;\Varid{f}\;\Varid{df}){}\<[30]%
\>[30]{}\mathrel{=}{}\<[30E]%
\>[33]{}\Varid{N}\;(\mathbin{-}\Varid{f})\;(\mathbin{-}\Varid{one}\bullet\Varid{df}){}\<[E]%
\ColumnHook
\end{hscode}\resethooks
\indentend \end{minipage}
\end{tabular}

\noindent
Here, we use the fact that Haskell allows us to write \ensuremath{\mathbin{-}\Varid{x}} for \ensuremath{\Varid{negate}\;\Varid{x}}.

Likewise, we collect trigonometric functions in \ensuremath{\Conid{Trig}}, whose Nagata number instance
relies on the well-known chain rule.
In general, to enrich \ensuremath{\Varid{d}} (and hence \ensuremath{\Conid{Expr}\;\Varid{v}}) with differentiable primitives requires us to know both the operation (\ensuremath{\Varid{sin}}), \emph{and} its derivative (\ensuremath{\Varid{cos}}). Here, in order to be able to negate \ensuremath{\Varid{sin}} as the derivative of \ensuremath{\Varid{cos}}, \ensuremath{\Varid{d}} must be a \ensuremath{\Conid{Ring}}.


\noindent
\begin{minipage}[t]{0.29\textwidth}
\indentbegin \begin{hscode}\SaveRestoreHook
\column{B}{@{}>{\hspre}l<{\hspost}@{}}%
\column{3}{@{}>{\hspre}l<{\hspost}@{}}%
\column{5}{@{}>{\hspre}l<{\hspost}@{}}%
\column{10}{@{}>{\hspre}l<{\hspost}@{}}%
\column{E}{@{}>{\hspre}l<{\hspost}@{}}%
\>[3]{}\mathbf{class}\;\Conid{Trig}\;\Varid{d}\;\mathbf{where}{}\<[E]%
\\
\>[3]{}\hsindent{2}{}\<[5]%
\>[5]{}\Varid{sin}{}\<[10]%
\>[10]{}\mathbin{::}\Varid{d}\to \Varid{d}{}\<[E]%
\\
\>[3]{}\hsindent{2}{}\<[5]%
\>[5]{}\Varid{cos}{}\<[10]%
\>[10]{}\mathbin{::}\Varid{d}\to \Varid{d}{}\<[E]%
\ColumnHook
\end{hscode}\resethooks
\indentend \end{minipage}%
\begin{minipage}[t]{0.7\textwidth}
\indentbegin \begin{hscode}\SaveRestoreHook
\column{B}{@{}>{\hspre}l<{\hspost}@{}}%
\column{3}{@{}>{\hspre}l<{\hspost}@{}}%
\column{5}{@{}>{\hspre}l<{\hspost}@{}}%
\column{12}{@{}>{\hspre}l<{\hspost}@{}}%
\column{13}{@{}>{\hspre}l<{\hspost}@{}}%
\column{17}{@{}>{\hspre}l<{\hspost}@{}}%
\column{28}{@{}>{\hspre}c<{\hspost}@{}}%
\column{28E}{@{}l@{}}%
\column{31}{@{}>{\hspre}l<{\hspost}@{}}%
\column{44}{@{}>{\hspre}l<{\hspost}@{}}%
\column{46}{@{}>{\hspre}l<{\hspost}@{}}%
\column{53}{@{}>{\hspre}l<{\hspost}@{}}%
\column{E}{@{}>{\hspre}l<{\hspost}@{}}%
\>[3]{}\mathbf{instance}\;{}\<[13]%
\>[13]{}(\Conid{Module}\;\Varid{d}\;\Varid{e},\Conid{Trig}\;\Varid{d},\Conid{Ring}\;\Varid{d}){}\<[44]%
\>[44]{}\Rightarrow \Conid{Trig}\;(\Varid{d}\ltimes\Varid{e}){}\<[E]%
\\
\>[3]{}\hsindent{2}{}\<[5]%
\>[5]{}\mathbf{where}\;{}\<[12]%
\>[12]{}\Varid{sin}\;{}\<[17]%
\>[17]{}(\Varid{N}\;\Varid{f}\;\Varid{df}){}\<[28]%
\>[28]{}\mathrel{=}{}\<[28E]%
\>[31]{}\Varid{N}\;(\Varid{sin}\;\Varid{f})\;({}\<[46]%
\>[46]{}\Varid{cos}\;\Varid{f}{}\<[53]%
\>[53]{}\bullet\Varid{df}){}\<[E]%
\\
\>[12]{}\Varid{cos}\;{}\<[17]%
\>[17]{}(\Varid{N}\;\Varid{f}\;\Varid{df}){}\<[28]%
\>[28]{}\mathrel{=}{}\<[28E]%
\>[31]{}\Varid{N}\;(\Varid{cos}\;\Varid{f})\;(\mathbin{-}{}\<[46]%
\>[46]{}\Varid{sin}\;\Varid{f}{}\<[53]%
\>[53]{}\bullet\Varid{df}){}\<[E]%
\ColumnHook
\end{hscode}\resethooks
\indentend \end{minipage}%

\noindent
Observe that, while \ensuremath{\Varid{d}} must now also have a \ensuremath{\Conid{Trig}} instance, no additional
demands are made of \ensuremath{\Varid{e}}. Hence, all the AD variations we have covered, which
are only based on variations of \ensuremath{\Varid{e}}, also support additional
primitive functions.

\subsection{Non-Commutative Semirings}
\label{sec:non-comm}

We can generalize our approach from commutative semirings to semirings with
non-commutative multiplication \ensuremath{(\otimes)}, such as (square) matrices over
semi-rings and the semiring of formal languages over an alphabet. 

The key change in our approach is that rather than starting from the following
rule for symbolic differentation, which exploits the commutativity:
\indentbegin \begin{hscode}\SaveRestoreHook
\column{B}{@{}>{\hspre}l<{\hspost}@{}}%
\column{3}{@{}>{\hspre}l<{\hspost}@{}}%
\column{19}{@{}>{\hspre}l<{\hspost}@{}}%
\column{27}{@{}>{\hspre}c<{\hspost}@{}}%
\column{27E}{@{}l@{}}%
\column{30}{@{}>{\hspre}l<{\hspost}@{}}%
\column{E}{@{}>{\hspre}l<{\hspost}@{}}%
\>[3]{}\textit{derive}\;\Varid{x}\;(\Conid{Times}\;{}\<[19]%
\>[19]{}\Varid{e}_{1}\;\Varid{e}_{2}){}\<[27]%
\>[27]{}\mathrel{=}{}\<[27E]%
\>[30]{}(\colorbox{lightgray}{$\Varid{e}_{2}\otimes$}\;\textit{derive}\;\Varid{x}\;\Varid{e}_{1})\oplus(\colorbox{lightgray}{$\Varid{e}_{1}\otimes$}\;\textit{derive}\;\Varid{x}\;\Varid{e}_{2}){}\<[E]%
\ColumnHook
\end{hscode}\resethooks
\indentend where \ensuremath{\Varid{e}_{1}} and \ensuremath{\Varid{e}_{2}}'s derivatives are \emph{both scaled on the left}, we instead use the more conventional
rule:
\indentbegin \begin{hscode}\SaveRestoreHook
\column{B}{@{}>{\hspre}l<{\hspost}@{}}%
\column{3}{@{}>{\hspre}l<{\hspost}@{}}%
\column{19}{@{}>{\hspre}l<{\hspost}@{}}%
\column{27}{@{}>{\hspre}c<{\hspost}@{}}%
\column{27E}{@{}l@{}}%
\column{30}{@{}>{\hspre}l<{\hspost}@{}}%
\column{43}{@{}>{\hspre}l<{\hspost}@{}}%
\column{E}{@{}>{\hspre}l<{\hspost}@{}}%
\>[3]{}\textit{derive}\;\Varid{x}\;(\Conid{Times}\;{}\<[19]%
\>[19]{}\Varid{e}_{1}\;\Varid{e}_{2}){}\<[27]%
\>[27]{}\mathrel{=}{}\<[27E]%
\>[30]{}(\textit{derive}\;\Varid{x}\;\Varid{e}_{1}\;{}\<[43]%
\>[43]{}\colorbox{lightgray}{$\otimes\Varid{e}_{2}$})\oplus(\colorbox{lightgray}{$\Varid{e}_{1}\otimes$}\;\textit{derive}\;\Varid{x}\;\Varid{e}_{2}){}\<[E]%
\ColumnHook
\end{hscode}\resethooks
\indentend which scales \ensuremath{\textit{derive}\;\Varid{x}\;\Varid{e}_{2}} on the left and \ensuremath{\textit{derive}\;\Varid{x}\;\Varid{e}_{1}} on the right.

More abstractly, we can model this
with \emph{\ensuremath{\Varid{d}}-\ensuremath{\Varid{d}}-bimodules} that feature distinct left-scaling
and right-scaling operators.
\indentbegin \begin{hscode}\SaveRestoreHook
\column{B}{@{}>{\hspre}l<{\hspost}@{}}%
\column{3}{@{}>{\hspre}l<{\hspost}@{}}%
\column{5}{@{}>{\hspre}l<{\hspost}@{}}%
\column{12}{@{}>{\hspre}c<{\hspost}@{}}%
\column{12E}{@{}l@{}}%
\column{16}{@{}>{\hspre}l<{\hspost}@{}}%
\column{22}{@{}>{\hspre}l<{\hspost}@{}}%
\column{28}{@{}>{\hspre}l<{\hspost}@{}}%
\column{E}{@{}>{\hspre}l<{\hspost}@{}}%
\>[3]{}\mathbf{class}\;\Conid{Semiring}\;\Varid{d}\Rightarrow \Conid{Bimodule}\;\Varid{d}\;\Varid{e}\mid \Varid{e}\to \Varid{d}\;\mathbf{where}{}\<[E]%
\\
\>[3]{}\hsindent{2}{}\<[5]%
\>[5]{}(\overset{\shortrightarrow}{\bullet}){}\<[12]%
\>[12]{}\mathbin{::}{}\<[12E]%
\>[16]{}\Varid{d}\to {}\<[22]%
\>[22]{}\Varid{e}\to {}\<[28]%
\>[28]{}\Varid{e}{}\<[E]%
\\
\>[3]{}\hsindent{2}{}\<[5]%
\>[5]{}(\overset{\shortleftarrow}{\bullet}){}\<[12]%
\>[12]{}\mathbin{::}{}\<[12E]%
\>[16]{}\Varid{e}\to {}\<[22]%
\>[22]{}\Varid{d}\to {}\<[28]%
\>[28]{}\Varid{e}{}\<[E]%
\ColumnHook
\end{hscode}\resethooks
\indentend Note that, for every \ensuremath{\Varid{d}} whose multiplication is commutative, every \ensuremath{\Conid{Module}\;\Varid{d}\;\Varid{e}} instance gives rise to a \ensuremath{\Conid{Bimodule}\;\Varid{d}\;\Varid{e}} instance, by taking the two scaling operations to be the same underlying \ensuremath{(\bullet)} operation. If multiplication in \ensuremath{\Varid{d}} is not commutative, nevertheless the distinguished case \ensuremath{\Varid{e}\mathrel{=}\Varid{d}} yields a \ensuremath{\Conid{Bimodule}} structure with \ensuremath{(\overset{\shortrightarrow}{\bullet})} (resp. \ensuremath{(\overset{\shortleftarrow}{\bullet})}) given by \emph{left}-(resp. \emph{right}-)multiplication (see below). However, it is \emph{not} sufficient to take (a restricted subset of) \ensuremath{\Varid{d}\to \Varid{e}} as a representative of the set of \ensuremath{\Conid{Bimodule}} homomorphisms from \ensuremath{\Varid{d}} to \ensuremath{\Varid{e}}. 

Instead, we can avoid introducing this new \ensuremath{\Varid{d}}-\ensuremath{\Varid{d}}-bimodule abstraction and 
reuse the existing \ensuremath{\Conid{Module}}-related infrastructure by means of \ensuremath{(\Varid{d},\Varid{d})}-modules.
At the core of this idea is the following semiring construction for pairs.
\indentbegin \begin{hscode}\SaveRestoreHook
\column{B}{@{}>{\hspre}l<{\hspost}@{}}%
\column{3}{@{}>{\hspre}l<{\hspost}@{}}%
\column{5}{@{}>{\hspre}l<{\hspost}@{}}%
\column{11}{@{}>{\hspre}l<{\hspost}@{}}%
\column{22}{@{}>{\hspre}l<{\hspost}@{}}%
\column{31}{@{}>{\hspre}c<{\hspost}@{}}%
\column{31E}{@{}l@{}}%
\column{34}{@{}>{\hspre}l<{\hspost}@{}}%
\column{E}{@{}>{\hspre}l<{\hspost}@{}}%
\>[3]{}\mathbf{instance}\;\Conid{Semiring}\;\Varid{d}\Rightarrow \Conid{Semiring}\;(\Varid{d},\Varid{d})\;\mathbf{where}{}\<[E]%
\\
\>[3]{}\hsindent{2}{}\<[5]%
\>[5]{}\Varid{zero}{}\<[11]%
\>[11]{}\mathrel{=}(\Varid{zero},\Varid{zero}){}\<[E]%
\\
\>[3]{}\hsindent{2}{}\<[5]%
\>[5]{}\Varid{one}{}\<[11]%
\>[11]{}\mathrel{=}(\Varid{one},\Varid{one}){}\<[E]%
\\
\>[3]{}\hsindent{2}{}\<[5]%
\>[5]{}(\Varid{lx},\Varid{rx})\oplus{}\<[22]%
\>[22]{}(\Varid{ly},\Varid{ry}){}\<[31]%
\>[31]{}\mathrel{=}{}\<[31E]%
\>[34]{}(\Varid{lx}\oplus\Varid{ly},\Varid{rx}\oplus\Varid{ry}){}\<[E]%
\\
\>[3]{}\hsindent{2}{}\<[5]%
\>[5]{}(\Varid{lx},\Varid{rx})\otimes{}\<[22]%
\>[22]{}(\Varid{ly},\Varid{ry}){}\<[31]%
\>[31]{}\mathrel{=}{}\<[31E]%
\>[34]{}(\Varid{lx}\otimes\Varid{ly},\colorbox{lightgray}{$\Varid{ry}\otimes\Varid{rx}$}){}\<[E]%
\ColumnHook
\end{hscode}\resethooks
\indentend This is the usual component-wise instance, with one exception: the factors in the multiplication
are flipped for the second component.
\nothingtodo{I don't think we actually use this instance anywhere except to satisfy the \ensuremath{\Conid{Semiring}} constraint on
  the \ensuremath{(\Varid{d},\Varid{d})} module definition below?}
Then \ensuremath{\Varid{d}} is a \ensuremath{(\Varid{d},\Varid{d})} module as follows:
\indentbegin \begin{hscode}\SaveRestoreHook
\column{B}{@{}>{\hspre}l<{\hspost}@{}}%
\column{3}{@{}>{\hspre}l<{\hspost}@{}}%
\column{5}{@{}>{\hspre}l<{\hspost}@{}}%
\column{E}{@{}>{\hspre}l<{\hspost}@{}}%
\>[3]{}\mathbf{instance}\;\Conid{Semiring}\;\Varid{d}\Rightarrow \Conid{Module}\;(\Varid{d},\Varid{d})\;\Varid{d}\;\mathbf{where}{}\<[E]%
\\
\>[3]{}\hsindent{2}{}\<[5]%
\>[5]{}(\Varid{l},\Varid{r})\bullet\Varid{d}\mathrel{=}\Varid{l}\otimes\Varid{d}\otimes\Varid{r}{}\<[E]%
\ColumnHook
\end{hscode}\resethooks
\indentend Here, the first component of the tuple scales from the left and the second
component scales from the right. We can make use of the left and right
injections below to turn a \ensuremath{\Varid{d}} value either into a left or right factor,
using the neutral element \ensuremath{\Varid{one}} for the other factor.
\indentbegin \begin{hscode}\SaveRestoreHook
\column{B}{@{}>{\hspre}l<{\hspost}@{}}%
\column{3}{@{}>{\hspre}l<{\hspost}@{}}%
\column{8}{@{}>{\hspre}l<{\hspost}@{}}%
\column{E}{@{}>{\hspre}l<{\hspost}@{}}%
\>[3]{}\Varid{inl},\Varid{inr}\mathbin{::}\Conid{Semiring}\;\Varid{d}\Rightarrow \Varid{d}\to (\Varid{d},\Varid{d}){}\<[E]%
\\
\>[3]{}\Varid{inl}\;{}\<[8]%
\>[8]{}\Varid{x}\mathrel{=}(\Varid{x},\Varid{one}){}\<[E]%
\\
\>[3]{}\Varid{inr}\;{}\<[8]%
\>[8]{}\Varid{y}\mathrel{=}(\Varid{one},\Varid{y}){}\<[E]%
\ColumnHook
\end{hscode}\resethooks
\indentend More generally, with a \ensuremath{(\Varid{d},\Varid{d})}-module \ensuremath{\Varid{e}} we obtain a variation on the Nagata numbers:
\indentbegin \begin{hscode}\SaveRestoreHook
\column{B}{@{}>{\hspre}l<{\hspost}@{}}%
\column{3}{@{}>{\hspre}l<{\hspost}@{}}%
\column{5}{@{}>{\hspre}l<{\hspost}@{}}%
\column{11}{@{}>{\hspre}c<{\hspost}@{}}%
\column{11E}{@{}l@{}}%
\column{14}{@{}>{\hspre}l<{\hspost}@{}}%
\column{20}{@{}>{\hspre}l<{\hspost}@{}}%
\column{22}{@{}>{\hspre}l<{\hspost}@{}}%
\column{28}{@{}>{\hspre}c<{\hspost}@{}}%
\column{28E}{@{}l@{}}%
\column{31}{@{}>{\hspre}l<{\hspost}@{}}%
\column{48}{@{}>{\hspre}l<{\hspost}@{}}%
\column{E}{@{}>{\hspre}l<{\hspost}@{}}%
\>[3]{}\mathbf{instance}\;(\Conid{Semiring}\;\Varid{d},\Conid{Module}\;\colorbox{lightgray}{$(\Varid{d},\Varid{d})$}\;\Varid{e})\Rightarrow \Conid{Semiring}\;(\Conid{Nagata}\;\Varid{d}\;\Varid{e})\;\mathbf{where}{}\<[E]%
\\
\>[3]{}\hsindent{2}{}\<[5]%
\>[5]{}\Varid{zero}{}\<[11]%
\>[11]{}\mathrel{=}{}\<[11E]%
\>[14]{}\Conid{N}\;\Varid{zero}\;{}\<[22]%
\>[22]{}\Varid{zero}{}\<[E]%
\\
\>[3]{}\hsindent{2}{}\<[5]%
\>[5]{}\Varid{one}{}\<[11]%
\>[11]{}\mathrel{=}{}\<[11E]%
\>[14]{}\Conid{N}\;\Varid{one}\;{}\<[22]%
\>[22]{}\Varid{zero}{}\<[E]%
\\
\>[3]{}\hsindent{2}{}\<[5]%
\>[5]{}\Conid{N}\;\Varid{f}\;\Varid{df}\oplus{}\<[20]%
\>[20]{}\Conid{N}\;\Varid{g}\;\Varid{dg}{}\<[28]%
\>[28]{}\mathrel{=}{}\<[28E]%
\>[31]{}\Conid{N}\;(\Varid{f}\oplus\Varid{g})\;{}\<[48]%
\>[48]{}(\Varid{df}\oplus\Varid{dg}){}\<[E]%
\\
\>[3]{}\hsindent{2}{}\<[5]%
\>[5]{}\Conid{N}\;\Varid{f}\;\Varid{df}\otimes\Conid{N}\;\Varid{g}\;\Varid{dg}{}\<[28]%
\>[28]{}\mathrel{=}{}\<[28E]%
\>[31]{}\Conid{N}\;(\Varid{f}\otimes\Varid{g})\;{}\<[48]%
\>[48]{}((\colorbox{lightgray}{$\Varid{inr}$}\;\Varid{g}\bullet\Varid{df})\oplus(\colorbox{lightgray}{$\Varid{inl}$}\;\Varid{f}\bullet\Varid{dg})){}\<[E]%
\ColumnHook
\end{hscode}\resethooks
\indentend \nothingtodo{
  Unfortunately, the instance head duplicates the existing \ensuremath{\Conid{Semiring}} instance for \ensuremath{\Conid{Nagata}\;\Varid{d}\;\Varid{e}}.
  It's probably fine for a paper. In case you were writing a library you would probably define separate
  modules for commutative and non-commutative AD (duplicating or abstracting over the Nagata
  structure) or go with a bi-module instead. However, similar problems seem to appear for e.g., the \ensuremath{\Conid{Kronecker}} definition of \ensuremath{\Conid{Sparse}}. It also looks like the \ensuremath{\Conid{Module}} instance
  for \ensuremath{(\multimap)} may be flipping the operands to the multiplication.
}
This one localized modification seamlessly integrates non-commutative multiplication
into our AD variants. In particular, following the earlier recipe we now use the \ensuremath{(\Varid{d},\Varid{d})\multimap\Varid{e}} representation for accummulating multiplications, which is that of \ensuremath{(\Varid{d},\Varid{d})\to \Varid{e}}
functions that take two \ensuremath{\Varid{d}} accumulators. The scalar multiplication
of this representation updates both accumulators.
\indentbegin \begin{hscode}\SaveRestoreHook
\column{B}{@{}>{\hspre}l<{\hspost}@{}}%
\column{3}{@{}>{\hspre}l<{\hspost}@{}}%
\column{5}{@{}>{\hspre}l<{\hspost}@{}}%
\column{E}{@{}>{\hspre}l<{\hspost}@{}}%
\>[3]{}\mathbf{instance}\;(\Conid{Semiring}\;\Varid{d},\Conid{Module}\;(\Varid{d},\Varid{d})\;\Varid{e})\Rightarrow \Conid{Module}\;(\Varid{d},\Varid{d})\;((\Varid{d},\Varid{d})\multimap\Varid{e})\;\mathbf{where}{}\<[E]%
\\
\>[3]{}\hsindent{2}{}\<[5]%
\>[5]{}(\Varid{dl'},\Varid{dr'})\bullet\Varid{f}\mathrel{=}\lambda (\Varid{dl},\Varid{dr})\to \Varid{f}\;(\Varid{dl'}\otimes\Varid{dl},\Varid{dr}\otimes\Varid{dr'}){}\<[E]%
\ColumnHook
\end{hscode}\resethooks
\indentend This shows that the first accumulator is for multiplications on the left
and the other for multiplications on the right.

In short, \ensuremath{(\Varid{d},\Varid{d})}-modules support non-commutative multiplication
with minimal upheaval.

\subsection{Overloading Initial Expressions}

Automatic differentiation avoids the intermediate symbolic derivative when
computing precise numeric results.
This is shown by the elimination of \ensuremath{\Varid{eval}} in \Cref{sec:classic}, which was invoked
twice in the naive implementation.
In this subsection we show that, by overloading operators in the initial
expression and thereby avoiding the initial symbolic expression,
we can further eliminate the \ensuremath{\Varid{eval}} call and thus further reduce
verbosity.

In particular, we assume that the initial expression is not given as a value
of type \ensuremath{\Conid{Expr}\;\Varid{v}}, but as a polymorphic function of type \ensuremath{\forall \Varid{d}\hsforall \hsdot{\circ }{.}\Conid{Semiring}\;\Varid{d}\Rightarrow \Varid{d}\to \mathbin{...}\to \Varid{d}\to \Varid{d}}
with as many inputs of type \ensuremath{\Varid{d}} as there are variables \ensuremath{\Varid{v}}.
For example, \ensuremath{\Varid{example}_{2}} from Section~\ref{sec:forward-dense},
can be written as a polymorphic function of arity 2.
\indentbegin \begin{hscode}\SaveRestoreHook
\column{B}{@{}>{\hspre}l<{\hspost}@{}}%
\column{3}{@{}>{\hspre}l<{\hspost}@{}}%
\column{E}{@{}>{\hspre}l<{\hspost}@{}}%
\>[3]{}\Varid{example}_{2}'\mathbin{::}\Conid{Semiring}\;\Varid{d}\Rightarrow \Varid{d}\to \Varid{d}\to \Varid{d}{}\<[E]%
\\
\>[3]{}\Varid{example}_{2}'\;\Varid{x}\;\Varid{y}\mathrel{=}((\Varid{x}\otimes\Varid{y})\oplus\Varid{x})\oplus\Varid{one}{}\<[E]%
\ColumnHook
\end{hscode}\resethooks
\indentend The mechanism behind deriving \ensuremath{\Varid{example}_{2}'} is that of \emph{fold/build fusion}~\cite{fpca/GillLJ93},
applied to \ensuremath{\Varid{eval}\;\Varid{var}\;\Varid{example}_{2}}, where \ensuremath{\Varid{example}_{2}} is an instance of the
builder pattern, building a symbolic expression, and \ensuremath{\Varid{eval}} is a fold. 
Fold/build fusion eliminates the intermediate symbolic structure and directly builds the
result of evaluation.
%
Gibbons and Wu~\citep{gibbons2014folding} show how to exploit such structure systematically (generalising to all algebraic structures defined by Haskell type classes, not only \ensuremath{\Conid{Semiring}}s), by relating the \squote[deep] embedding in terms of symbolic expressions \ensuremath{\Conid{Expr}\;\Varid{v}} with the corresponding \squote[shallow] embedding \ensuremath{\forall \Varid{d}\hsforall \hsdot{\circ }{.}\Conid{Semiring}\;\Varid{d}\Rightarrow (\Varid{v}\to \Varid{d})\to \Varid{d}}. Here, \ensuremath{\Varid{eval}} is one half of an isomorphism pair between these two types, with the inverse given by instantiation at \ensuremath{\Conid{Var}}.

Hence, instead of the verbose calls (left), we
readily invoke \ensuremath{\Varid{example}_{2}'} (right).
\noindent
\begin{minipage}[t]{0.5\textwidth}\indentbegin \begin{hscode}\SaveRestoreHook
\column{B}{@{}>{\hspre}l<{\hspost}@{}}%
\column{3}{@{}>{\hspre}l<{\hspost}@{}}%
\column{6}{@{}>{\hspre}l<{\hspost}@{}}%
\column{E}{@{}>{\hspre}l<{\hspost}@{}}%
\>[3]{}\mathbin{>}{}\<[6]%
\>[6]{}\mathbf{let}\;\{\mskip1.5mu \Varid{var}\;\Varid{X}\mathrel{=}\mathrm{5};\Varid{var}\;\Varid{Y}\mathrel{=}\mathrm{3}\mskip1.5mu\}{}\<[E]%
\\
\>[6]{}\mathbf{in}\;\Varid{eval}\;\Varid{var}\;\Varid{example}_{2}{}\<[E]%
\\
\>[3]{}\mathrm{21}{}\<[E]%
\\
\>[3]{}\mathbin{>}{}\<[6]%
\>[6]{}\mathbf{let}\;\{\mskip1.5mu \Varid{var}\;\Varid{X}\mathrel{=}\mathrm{5};\Varid{var}\;\Varid{Y}\mathrel{=}\mathrm{3}\mskip1.5mu\}{}\<[E]%
\\
\>[6]{}\mathbf{in}\;\Varid{forwardAD}_{Sparse}\;\Varid{var}\;\Varid{example}_{2}{}\<[E]%
\\
\>[3]{}\Varid{N}\;\mathrm{21}\;\{\mskip1.5mu \Varid{X}\;\mapsto\;\mathrm{4};\Varid{Y}\;\mapsto\;\mathrm{5}\mskip1.5mu\}{}\<[E]%
\ColumnHook
\end{hscode}\resethooks
\indentend \end{minipage}%
\begin{minipage}[t]{0.5\textwidth}\indentbegin \begin{hscode}\SaveRestoreHook
\column{B}{@{}>{\hspre}l<{\hspost}@{}}%
\column{3}{@{}>{\hspre}l<{\hspost}@{}}%
\column{16}{@{}>{\hspre}l<{\hspost}@{}}%
\column{E}{@{}>{\hspre}l<{\hspost}@{}}%
\>[3]{}\mathbin{>}\Varid{example}_{2}'\;\mathrm{5}\;\mathrm{3}{}\<[E]%
\\
\>[3]{}\mathrm{21}{}\<[E]%
\\[\blanklineskip]%
\>[3]{}\mathbin{>}\Varid{example}_{2}'\;{}\<[16]%
\>[16]{}(\Varid{N}\;\mathrm{5}\;(\Varid{delta}\;\Varid{X}))\;{}\<[E]%
\\
\>[16]{}(\Varid{N}\;\mathrm{3}\;(\Varid{delta}\;\Varid{Y})){}\<[E]%
\\
\>[3]{}\Varid{N}\;\mathrm{21}\;\{\mskip1.5mu \Varid{X}\;\mapsto\;\mathrm{4};\Varid{Y}\;\mapsto\;\mathrm{5}\mskip1.5mu\}{}\<[E]%
\ColumnHook
\end{hscode}\resethooks
\indentend 
\end{minipage}%

\noindent
Both cases avoid the interpretative overhead of \ensuremath{\Varid{eval}}uating the symbolic expression.
Yet, we can also recover the original symbolic expression.
\indentbegin \begin{hscode}\SaveRestoreHook
\column{B}{@{}>{\hspre}l<{\hspost}@{}}%
\column{3}{@{}>{\hspre}l<{\hspost}@{}}%
\column{E}{@{}>{\hspre}l<{\hspost}@{}}%
\>[3]{}\mathbin{>}\Varid{example}_{2}'\;(\Conid{Var}\;\Varid{X})\;(\Conid{Var}\;\Varid{Y}){}\<[E]%
\\
\>[3]{}\Conid{Plus}\;(\Conid{Plus}\;(\Conid{Times}\;(\Conid{Var}\;\Varid{X})\;(\Conid{Var}\;\Varid{Y}))\;(\Conid{Var}\;\Varid{X}))\;\Conid{One}{}\<[E]%
\ColumnHook
\end{hscode}\resethooks
\indentend 

\subsection{Higher Derivatives}\label{sec:ext:hod}

%

\noindent
Higher derivatives are obtained through repeated differentiation.
For instance, the second-order derivative is the derivative
of the derivative.
The naive approach to compute the second-order derivative at a point is through
repeated evaluation of the symbolic derivative.
\indentbegin \begin{hscode}\SaveRestoreHook
\column{B}{@{}>{\hspre}l<{\hspost}@{}}%
\column{3}{@{}>{\hspre}l<{\hspost}@{}}%
\column{39}{@{}>{\hspre}l<{\hspost}@{}}%
\column{45}{@{}>{\hspre}l<{\hspost}@{}}%
\column{E}{@{}>{\hspre}l<{\hspost}@{}}%
\>[3]{}\Varid{derive}_{\mathit{2^{nd}}}\mathbin{::}(\Conid{Eq}\;\Varid{v},\Conid{Semiring}\;\Varid{d})\Rightarrow (\Varid{v}\to \Varid{d})\to \Conid{Expr}\;\Varid{v}\to \Conid{Dense}\;\Varid{v}\;(\Conid{Dense}\;\Varid{v}\;\Varid{d}){}\<[E]%
\\
\>[3]{}\Varid{derive}_{\mathit{2^{nd}}}\;\Varid{var}\;\Varid{e}\mathrel{=}\lambda \Varid{x}\;\Varid{y}\to \Varid{eval}\;\Varid{var}\;{}\<[39]%
\>[39]{}(\Varid{tan}^{N}\;{}\<[45]%
\>[45]{}(\Varid{forwardAD}_{Dense}\;\Conid{Var}{}\<[E]%
\\
\>[39]{}(\Varid{tan}^{N}\;{}\<[45]%
\>[45]{}(\Varid{forwardAD}_{Dense}\;\Conid{Var}\;\Varid{e})\;\Varid{x}))\;\Varid{y}){}\<[E]%
\ColumnHook
\end{hscode}\resethooks
\indentend 
The Haskell call \ensuremath{\Varid{derive}_{\mathit{2^{nd}}}\;\Varid{var}\;\Varid{e}\;\Varid{x}\;\Varid{y}} corresponds to the mathematical notation
$\left. \frac{\partial}{\partial y} \left( \frac{\partial}{\partial x} e \right) \right\vert_{var}$.
By using naturality of \ensuremath{\Varid{tan}^{N}} and \ensuremath{\Varid{eval}}'s fusion property, we merge
the three traversals of \ensuremath{\Varid{derive}_{\mathit{2^{nd}}}} into a single \ensuremath{\Varid{eval}} (see \ref{app:hod}):
\indentbegin \begin{hscode}\SaveRestoreHook
\column{B}{@{}>{\hspre}l<{\hspost}@{}}%
\column{3}{@{}>{\hspre}l<{\hspost}@{}}%
\column{5}{@{}>{\hspre}l<{\hspost}@{}}%
\column{E}{@{}>{\hspre}l<{\hspost}@{}}%
\>[3]{}\Varid{derive}_{\mathit{2^{nd}}}^{\prime}\;\Varid{var}\;\Varid{e}\mathrel{=}\lambda \Varid{x}\;\Varid{y}\to \Varid{tan}^{N}\;(\Varid{tan}^{N}\;(\Varid{forwardAD}_{Dense}\;\Varid{gen}\;\Varid{e})\;\Varid{x})\;\Varid{y}{}\<[E]%
\\
\>[3]{}\hsindent{2}{}\<[5]%
\>[5]{}\mathbf{where}\;\Varid{gen}\;\Varid{z}\mathrel{=}\Varid{N}\;(\Varid{var}\;\Varid{z})\;(\Varid{delta}\;\Varid{z}){}\<[E]%
\ColumnHook
\end{hscode}\resethooks
\indentend This instance of \ensuremath{\Varid{abstractD}} uses the \ensuremath{(\Varid{d}\ltimes(\Conid{Dense}\;\Varid{v}\;\Varid{d}))}-module
of nested Nagata numbers of type
\ensuremath{(\Varid{d}\ltimes(\Conid{Dense}\;\Varid{v}\;\Varid{d}))\ltimes(\Conid{Dense}\;\Varid{v}\;(\Varid{d}\ltimes(\Conid{Dense}\;\Varid{v}\;\Varid{d})))}.

\begin{figure}
\begin{center}
\begin{forest}
 [,phantom
  [$e$
    [$e$,tier=first
      [$e$,tier=second]
      [$\partialfrac{e}{x}$,tier=second]
    ]
    [$\partialfrac{e}{x}$,tier=first
      [$\partialfrac{e}{x}$,tier=second]
      [$\frac{\partial^2 e}{\partial y \partial x}$,tier=second]
    ]
  ]
  [\ensuremath{\Varid{d}}
    [\ensuremath{\Varid{d}\ltimes(\Conid{Dense}\;\Varid{v}\;\Varid{d})},tier=first
       [\ensuremath{(\Varid{d}\ltimes(\Conid{Dense}\;\Varid{v}\;\Varid{d}))\ltimes(\Conid{Dense}\;\Varid{v}\;(\Varid{d}\ltimes(\Conid{Dense}\;\Varid{v}\;\Varid{d})))},tier=second]
    ]
  ]
 ]
\end{forest}
\end{center}
\caption{$0^{th}$--$2^{nd}$-order derivatives and their types.}\label{fig:2ndorder}
\end{figure}

The nested representation (Figure~\ref{fig:2ndorder}) of the second-order derivative
contains redundancy: the value $\partialfrac{e}{x}$ appears twice.
This redundancy gets worse as the order (and thus the level of nesting) increases. In general, for the $n$th-order
derivative we get $2^n$ values of which only $n + 1$ are distinct.
To avoid this, it is more economical to use a compact representation. 
\indentbegin \begin{hscode}\SaveRestoreHook
\column{B}{@{}>{\hspre}l<{\hspost}@{}}%
\column{3}{@{}>{\hspre}l<{\hspost}@{}}%
\column{E}{@{}>{\hspre}l<{\hspost}@{}}%
\>[3]{}\mathbf{data}\;N_{2^{nd}}\;\Varid{v}\;\Varid{d}\mathrel{=}\Varid{N_{2^{nd}}}\;\Varid{d}\;(\Conid{Dense}\;\Varid{v}\;(\Varid{d}\ltimes(\Conid{Dense}\;\Varid{v}\;\Varid{d}))){}\<[E]%
\ColumnHook
\end{hscode}\resethooks
\indentend 
\noindent
The corresponding forward AD is similar to \ensuremath{\Varid{abstractD}}:
\indentbegin \begin{hscode}\SaveRestoreHook
\column{B}{@{}>{\hspre}l<{\hspost}@{}}%
\column{3}{@{}>{\hspre}l<{\hspost}@{}}%
\column{E}{@{}>{\hspre}l<{\hspost}@{}}%
\>[3]{}\Varid{forwardAD}_{2^{nd}}\mathbin{::}(\Conid{Semiring}\;\Varid{d})\Rightarrow (\Varid{v}\to \Varid{d})\to \Conid{Expr}\;\Varid{v}\to N_{2^{nd}}\;\Varid{v}\;\Varid{d}{}\<[E]%
\\
\>[3]{}\Varid{forwardAD}_{2^{nd}}\;\Varid{var}\mathrel{=}\Varid{eval}\;\Varid{gen}\;\mathbf{where}\;\Varid{gen}\;\Varid{x}\mathrel{=}\Varid{N_{2^{nd}}}\;(\Varid{var}\;\Varid{x})\;(\Varid{delta}\;\Varid{x}){}\<[E]%
\ColumnHook
\end{hscode}\resethooks
\indentend Running this on \ensuremath{\Varid{example}_{3}} gives the evaluation result, first and second
derivative:
\indentbegin \begin{hscode}\SaveRestoreHook
\column{B}{@{}>{\hspre}l<{\hspost}@{}}%
\column{3}{@{}>{\hspre}l<{\hspost}@{}}%
\column{E}{@{}>{\hspre}l<{\hspost}@{}}%
\>[3]{}\mathbin{>}\Varid{forwardAD}_{2^{nd}}\;(\lambda \Conid{X}\to \mathrm{5})\;\Varid{example}_{3}{}\<[E]%
\\
\>[3]{}\Varid{N_{2^{nd}}}\;\mathrm{300}\;(\lambda \Conid{X}\to \Varid{N}\;\{\mskip1.5mu \Varid{pri}^{N}\mathrel{=}\mathrm{170},\Varid{tan}^{N}\mathrel{=}(\lambda \Conid{X}\to \mathrm{64})\mskip1.5mu\}){}\<[E]%
\ColumnHook
\end{hscode}\resethooks
\indentend 
This generalizes straightforwardly to other higher derivatives.


\paragraph{All Higher Partial Derivatives}
In the extreme case where $n = \infty$, the economical representation becomes an infinite stream of all successive
derivatives~\citep{DBLP:conf/icfp/Karczmarczuk98}. This stream of derivatives
has the form \ensuremath{\Varid{x}\mathbin{:<}\Varid{dxs}} where \ensuremath{\Varid{x}} is the value and \ensuremath{\Varid{dxs}} are all higher-order
partial derivatives with respect to all variables \ensuremath{\Varid{v}}.\footnote{It differs
from Hinze's stream representation~\cite{10.1145/1411204.1411232} in the interposition
of \ensuremath{\Conid{Dense}} in the tail.}
\indentbegin \begin{hscode}\SaveRestoreHook
\column{B}{@{}>{\hspre}l<{\hspost}@{}}%
\column{3}{@{}>{\hspre}l<{\hspost}@{}}%
\column{E}{@{}>{\hspre}l<{\hspost}@{}}%
\>[3]{}\mathbf{data}\;\Conid{Stream}\;\Varid{v}\;\Varid{d}\mathrel{=}\Varid{d}\mathbin{:<}\Conid{Dense}\;\Varid{v}\;(\Conid{Stream}\;\Varid{v}\;\Varid{d}){}\<[E]%
\ColumnHook
\end{hscode}\resethooks
\indentend Then, the semiring operations for the infinite stream of successive derivatives are
defined as follows.
\indentbegin \begin{hscode}\SaveRestoreHook
\column{B}{@{}>{\hspre}l<{\hspost}@{}}%
\column{3}{@{}>{\hspre}l<{\hspost}@{}}%
\column{5}{@{}>{\hspre}l<{\hspost}@{}}%
\column{8}{@{}>{\hspre}l<{\hspost}@{}}%
\column{11}{@{}>{\hspre}l<{\hspost}@{}}%
\column{14}{@{}>{\hspre}l<{\hspost}@{}}%
\column{20}{@{}>{\hspre}l<{\hspost}@{}}%
\column{29}{@{}>{\hspre}l<{\hspost}@{}}%
\column{44}{@{}>{\hspre}l<{\hspost}@{}}%
\column{103}{@{}>{\hspre}l<{\hspost}@{}}%
\column{E}{@{}>{\hspre}l<{\hspost}@{}}%
\>[3]{}\mathbf{instance}\;\Conid{Semiring}\;\Varid{d}\Rightarrow \Conid{Semiring}\;(\Conid{Stream}\;\Varid{v}\;\Varid{d})\;\mathbf{where}{}\<[E]%
\\
\>[3]{}\hsindent{2}{}\<[5]%
\>[5]{}\Varid{zero}{}\<[11]%
\>[11]{}\mathrel{=}\Varid{zero}\mathbin{:<}\Varid{zero}{}\<[E]%
\\
\>[3]{}\hsindent{2}{}\<[5]%
\>[5]{}\Varid{one}{}\<[11]%
\>[11]{}\mathrel{=}\Varid{one}\mathbin{:<}\Varid{zero}{}\<[E]%
\\
\>[3]{}\hsindent{2}{}\<[5]%
\>[5]{}(\Varid{x}\mathbin{:<}\Varid{xs}){}\<[20]%
\>[20]{}\oplus{}\<[29]%
\>[29]{}(\Varid{y}\mathbin{:<}\Varid{ys}){}\<[44]%
\>[44]{}\mathrel{=}(\Varid{x}\oplus\Varid{y})\mathbin{:<}(\Varid{xs}\oplus\Varid{ys}){}\<[E]%
\\
\>[3]{}\hsindent{2}{}\<[5]%
\>[5]{}\Varid{xxs}{}\<[20]%
\>[20]{}\otimes{}\<[29]%
\>[29]{}\Varid{yys}{}\<[44]%
\>[44]{}\mathrel{=}(\Varid{x}\otimes\Varid{y})\mathbin{:<}\lambda \Varid{v}\to ((\Varid{xs}\;\Varid{v}\otimes\Varid{yys})\oplus(\Varid{xxs}{}\<[103]%
\>[103]{}\otimes\Varid{ys}\;\Varid{v})){}\<[E]%
\\
\>[5]{}\hsindent{3}{}\<[8]%
\>[8]{}\mathbf{where}\;\Varid{xxs}\mathrel{=}(\Varid{x}\mathbin{:<}\Varid{xs}){}\<[E]%
\\
\>[8]{}\hsindent{6}{}\<[14]%
\>[14]{}\Varid{yys}\mathrel{=}(\Varid{y}\mathbin{:<}\Varid{ys}){}\<[E]%
\ColumnHook
\end{hscode}\resethooks
\indentend The \ensuremath{\Conid{Module}} and \ensuremath{\Conid{Kronecker}} instances follow naturally from this definition.
They allow us to automatically differentiate a symbolic expression and obtain
its stream of derivatives.
\indentbegin \begin{hscode}\SaveRestoreHook
\column{B}{@{}>{\hspre}l<{\hspost}@{}}%
\column{3}{@{}>{\hspre}l<{\hspost}@{}}%
\column{19}{@{}>{\hspre}c<{\hspost}@{}}%
\column{19E}{@{}l@{}}%
\column{22}{@{}>{\hspre}l<{\hspost}@{}}%
\column{E}{@{}>{\hspre}l<{\hspost}@{}}%
\>[3]{}\Varid{forwardAD}_{All}\mathbin{::}(\Conid{Eq}\;\Varid{v},\Conid{Semiring}\;\Varid{d})\Rightarrow (\Varid{v}\to \Varid{d})\to \Conid{Expr}\;\Varid{v}\to \Conid{Stream}\;\Varid{v}\;\Varid{d}{}\<[E]%
\\
\>[3]{}\Varid{forwardAD}_{All}\;\Varid{var}{}\<[19]%
\>[19]{}\mathrel{=}{}\<[19E]%
\>[22]{}\Varid{eval}\;\Varid{gen}\;\mathbf{where}\;\Varid{gen}\;\Varid{y}\mathrel{=}\Varid{var}\;\Varid{y}\mathbin{:<}\Varid{delta}\;\Varid{y}{}\<[E]%
\\[\blanklineskip]%
\>[3]{}\mathbin{>}\Varid{forwardAD}_{All}\;(\lambda \Conid{X}\to (\mathrm{5}\mathbin{::}\Conid{Int}))\;\Varid{example}_{3}{}\<[E]%
\\
\>[3]{}\mathrm{300}\mathbin{:<}(\lambda \Conid{X}\to \mathrm{170}\mathbin{:<}(\lambda \Conid{X}\to \mathrm{64}\mathbin{:<}(\lambda \Conid{X}\to \mathrm{12}\mathbin{:<}(\lambda \Conid{X}\to \mathrm{0}\mathbin{:<}\mathbin{...}{}\<[E]%
\ColumnHook
\end{hscode}\resethooks
\indentend %

\paragraph{Hybrid Automatic Differentiation}
In second-order optimization, it is often preferred to use a hybrid approach,
i.e., combine forward mode and backward mode to get a more efficient algorithm.
For instance, when taking a function's curvature into account, it is necessary
to compute the Hessian\footnote{The Hessian
of a function \(f\) in \(n\) variables \(x_i\) is the \(n\times n\)
square matrix consisting of all second-order derivatives
\(\frac{\partial^2 f}{\partial x_i \partial x_j}\).} applied to a vector.
Pearlmutter~\cite{Pearlmutter94} introduces this mix of the two modes.
In a forward sweep the first derivative is computed, whereupon the second,
reverse sweep computes the second-order derivative at a point.
This can be obtained by exploiting the \ensuremath{\Varid{d}}-module isomorphisms to
work with\indentbegin \begin{hscode}\SaveRestoreHook
\column{B}{@{}>{\hspre}l<{\hspost}@{}}%
\column{3}{@{}>{\hspre}l<{\hspost}@{}}%
\column{E}{@{}>{\hspre}l<{\hspost}@{}}%
\>[3]{}\mathbf{data}\;N_{H}\;\Varid{v}\;\Varid{d}\mathrel{=}\Varid{N_{H}}\;\Varid{d}\;(\Conid{Dense}\;\Varid{v}\;(\Varid{d}\ltimes(\Varid{d}\multimap(\Conid{Cayley}\;(\Conid{Sparse}\;\Varid{v}\;\Varid{d}))))){}\<[E]%
\ColumnHook
\end{hscode}\resethooks
\indentend 
\subsection{Basic Sharing of Subexpressions}
\label{sec:sharing-and-let}

Our approach so far ignores any sharing in the source expression. Consider
for instance \ensuremath{\Varid{example}_{4}}, which denotes the expression $(x + x) \times (x + x)$, but
at the level of Haskell shares the two occurrences of \ensuremath{\Varid{x}\mathbin{+}\Varid{x}} by means of a
\ensuremath{\mathbf{let}} binding.
\indentbegin \begin{hscode}\SaveRestoreHook
\column{B}{@{}>{\hspre}l<{\hspost}@{}}%
\column{3}{@{}>{\hspre}l<{\hspost}@{}}%
\column{E}{@{}>{\hspre}l<{\hspost}@{}}%
\>[3]{}\Varid{example}_{4}\mathbin{::}\Conid{Expr}\;\Conid{X}{}\<[E]%
\\
\>[3]{}\Varid{example}_{4}\mathrel{=}\mathbf{let}\;\Varid{y}\mathrel{=}\Conid{Plus}\;(\Conid{Var}\;\Conid{X})\;(\Conid{Var}\;\Conid{X})\;\mathbf{in}\;\Conid{Times}\;\Varid{y}\;\Varid{y}{}\<[E]%
\ColumnHook
\end{hscode}\resethooks
\indentend 
%

\noindent
The \ensuremath{\Varid{eval}} recursion scheme used by \ensuremath{\Varid{abstractD}} does not see the \ensuremath{\mathbf{let}} binding
and processes both occurrences of $x + x$. In particular, \ensuremath{\Varid{abstractD}\;(\lambda \Conid{X}\to \mathrm{5})\;\Varid{example}_{4}}
reduces to
\ensuremath{(\Varid{N}\;\mathrm{5}\;\Varid{one}\oplus\Varid{N}\;\mathrm{5}\;\Varid{one})\otimes(\Varid{N}\;\mathrm{5}\;\Varid{one}\oplus\Varid{N}\;\mathrm{5}\;\Varid{one})}
which features two \ensuremath{(\oplus)} operations instead of one.
To avoid this recomputation of intermediate results, we extend the symbolic
expression language with a let-construct so that
we can write \ensuremath{\Varid{example}_{4}'} with this new syntax.

\noindent
\begin{minipage}{0.4\textwidth}
\indentbegin \begin{hscode}\SaveRestoreHook
\column{B}{@{}>{\hspre}l<{\hspost}@{}}%
\column{3}{@{}>{\hspre}l<{\hspost}@{}}%
\column{9}{@{}>{\hspre}l<{\hspost}@{}}%
\column{19}{@{}>{\hspre}c<{\hspost}@{}}%
\column{19E}{@{}l@{}}%
\column{22}{@{}>{\hspre}l<{\hspost}@{}}%
\column{30}{@{}>{\hspre}l<{\hspost}@{}}%
\column{E}{@{}>{\hspre}l<{\hspost}@{}}%
\>[3]{}\mathbf{data}\;{}\<[9]%
\>[9]{}\Varid{Expr}\;\Varid{v}{}\<[19]%
\>[19]{}\mathrel{=}{}\<[19E]%
\>[22]{}\mathbin{...}{}\<[E]%
\\
\>[19]{}\mid {}\<[19E]%
\>[22]{}\Varid{Let}\;\Varid{v}\;{}\<[30]%
\>[30]{}(\Varid{Expr}\;\Varid{v})\;{}\<[E]%
\\
\>[30]{}(\Varid{Expr}\;\Varid{v}){}\<[E]%
\ColumnHook
\end{hscode}\resethooks
\indentend \end{minipage}%
\begin{minipage}{0.6\textwidth}
\indentbegin \begin{hscode}\SaveRestoreHook
\column{B}{@{}>{\hspre}l<{\hspost}@{}}%
\column{3}{@{}>{\hspre}l<{\hspost}@{}}%
\column{24}{@{}>{\hspre}l<{\hspost}@{}}%
\column{33}{@{}>{\hspre}l<{\hspost}@{}}%
\column{44}{@{}>{\hspre}l<{\hspost}@{}}%
\column{E}{@{}>{\hspre}l<{\hspost}@{}}%
\>[3]{}\Varid{example}_{4}'\mathbin{::}\Varid{Expr}\;\Varid{XY}{}\<[E]%
\\
\>[3]{}\Varid{example}_{4}'\mathrel{=}\Varid{Let}\;\Varid{Y}\;{}\<[24]%
\>[24]{}(\Varid{Plus}\;{}\<[33]%
\>[33]{}(\Varid{Var}\;\Varid{X})\;{}\<[44]%
\>[44]{}(\Varid{Var}\;\Varid{X}))\;{}\<[E]%
\\
\>[24]{}(\Varid{Times}\;{}\<[33]%
\>[33]{}(\Varid{Var}\;\Varid{Y})\;{}\<[44]%
\>[44]{}(\Varid{Var}\;\Varid{Y})){}\<[E]%
\ColumnHook
\end{hscode}\resethooks
\indentend \end{minipage}

%


\noindent
We adapt the evaluator accordingly with the standard semantics for \ensuremath{\mathbf{let}}-binding.
\indentbegin \begin{hscode}\SaveRestoreHook
\column{B}{@{}>{\hspre}l<{\hspost}@{}}%
\column{3}{@{}>{\hspre}l<{\hspost}@{}}%
\column{22}{@{}>{\hspre}l<{\hspost}@{}}%
\column{26}{@{}>{\hspre}l<{\hspost}@{}}%
\column{31}{@{}>{\hspre}c<{\hspost}@{}}%
\column{31E}{@{}l@{}}%
\column{34}{@{}>{\hspre}l<{\hspost}@{}}%
\column{39}{@{}>{\hspre}l<{\hspost}@{}}%
\column{45}{@{}>{\hspre}l<{\hspost}@{}}%
\column{48}{@{}>{\hspre}l<{\hspost}@{}}%
\column{E}{@{}>{\hspre}l<{\hspost}@{}}%
\>[3]{}\Varid{eval}\mathbin{::}(\Conid{Eq}\;\Varid{v},\Conid{Semiring}\;\Varid{d})\Rightarrow (\Varid{v}\to \Varid{d})\to \Varid{Expr}\;\Varid{v}\to \Varid{d}{}\<[E]%
\\
\>[3]{}\mathbin{...}{}\<[E]%
\\
\>[3]{}\Varid{eval}\;\Varid{gen}\;(\Varid{Let}\;\Varid{y}\;{}\<[22]%
\>[22]{}\Varid{e}_{1}\;{}\<[26]%
\>[26]{}\Varid{e}_{2}){}\<[31]%
\>[31]{}\mathrel{=}{}\<[31E]%
\>[34]{}\mathbf{let}\;{}\<[39]%
\>[39]{}\Varid{d}_{1}{}\<[45]%
\>[45]{}\mathrel{=}{}\<[48]%
\>[48]{}\Varid{eval}\;\Varid{gen}\;\Varid{e}_{1}{}\<[E]%
\\
\>[39]{}\Varid{gen}_{2}{}\<[45]%
\>[45]{}\mathrel{=}\lambda \Varid{x}\to \mathbf{if}\;\Varid{x}= =\Varid{y}\;\mathbf{then}\;\Varid{d}_{1}\;\mathbf{else}\;\Varid{gen}\;\Varid{x}{}\<[E]%
\\
\>[34]{}\mathbf{in}\;{}\<[39]%
\>[39]{}\Varid{eval}\;\Varid{gen}_{2}\;\Varid{e}_{2}{}\<[E]%
\ColumnHook
\end{hscode}\resethooks
\indentend Observe that, because \ensuremath{\Varid{e}_{2}} has the additional \ensuremath{\mathbf{let}}-bound variable in scope, we
evaluate it with the extended generator \ensuremath{\Varid{gen}_{2}}, which maps the free
variables to their given \ensuremath{\Varid{gen}} meaning and the new \ensuremath{\mathbf{let}}-bound variable \ensuremath{\Varid{y}} to the
meaning \ensuremath{\Varid{d}_{1}} of \ensuremath{\Varid{e}_{1}}.
The definition of \ensuremath{\Varid{abstractD}} is as before, now with support for sharing.
%

\paragraph{Sharing for Sparse Maps}
Now, \ensuremath{\Varid{abstractD}\;(\lambda \Conid{X}\to \mathrm{5})\;\Varid{example}_{4}'} reduces to an intermediate form with
only one \ensuremath{(\oplus)} operation,
%
%
and if we commit for the tangent to the sparse module, this reduces further in
a way that nicely preserves the sharing.
\indentbegin \begin{hscode}\SaveRestoreHook
\column{B}{@{}>{\hspre}l<{\hspost}@{}}%
\column{3}{@{}>{\hspre}l<{\hspost}@{}}%
\column{10}{@{}>{\hspre}l<{\hspost}@{}}%
\column{24}{@{}>{\hspre}l<{\hspost}@{}}%
\column{26}{@{}>{\hspre}l<{\hspost}@{}}%
\column{27}{@{}>{\hspre}l<{\hspost}@{}}%
\column{42}{@{}>{\hspre}l<{\hspost}@{}}%
\column{53}{@{}>{\hspre}l<{\hspost}@{}}%
\column{59}{@{}>{\hspre}l<{\hspost}@{}}%
\column{71}{@{}>{\hspre}l<{\hspost}@{}}%
\column{E}{@{}>{\hspre}l<{\hspost}@{}}%
\>[3]{}\mathbf{let}\;\Varid{y}\mathrel{=}\Varid{N}\;\mathrm{5}\;\Varid{one}\oplus\Varid{N}\;\mathrm{5}\;\Varid{one}\;\mathbf{in}\;(\Varid{y}\otimes\Varid{y}){}\<[E]%
\\
\>[3]{}\equiv\;{}\<[10]%
\>[10]{}\mathbf{let}\;\Varid{y}\mathrel{=}\Varid{N}\;\mathrm{5}\;{}\<[26]%
\>[26]{}\{\mskip1.5mu \Conid{X}\;\mapsto\;\mathrm{1}\mskip1.5mu\}{}\<[42]%
\>[42]{}\oplus\Varid{N}\;\mathrm{5}\;\{\mskip1.5mu \Conid{X}\;\mapsto\;\mathrm{1}\mskip1.5mu\}\;{}\<[71]%
\>[71]{}\mathbf{in}\;(\Varid{y}\otimes\Varid{y}){}\<[E]%
\\
\>[3]{}\equiv\;{}\<[10]%
\>[10]{}\mathbf{let}\;\Varid{y}\mathrel{=}\Varid{N}\;\mathrm{10}\;{}\<[26]%
\>[26]{}(\Varid{fromList}\;[\mskip1.5mu (\Conid{X},\mathrm{2})\mskip1.5mu])\;{}\<[71]%
\>[71]{}\mathbf{in}\;(\Varid{y}\otimes\Varid{y}){}\<[E]%
\\
\>[3]{}\equiv\;{}\<[10]%
\>[10]{}\Varid{N}\;(\mathrm{10}\;\times\;\mathrm{10})\;{}\<[27]%
\>[27]{}(\Varid{fmap}\;(\mathrm{10}\;\times)\;\{\mskip1.5mu \Conid{X}\;\mapsto\;\mathrm{2}\mskip1.5mu\}{}\<[59]%
\>[59]{}\oplus\Varid{fmap}\;(\mathrm{10}\;\times)\;\{\mskip1.5mu \Conid{X}\;\mapsto\;\mathrm{2}\mskip1.5mu\}){}\<[E]%
\\
\>[3]{}\equiv\;{}\<[10]%
\>[10]{}\Varid{N}\;\mathrm{100}\;{}\<[24]%
\>[24]{}(\{\mskip1.5mu \Conid{X}\;\mapsto\;\mathrm{20}\mskip1.5mu\}{}\<[53]%
\>[53]{}\oplus\{\mskip1.5mu \Conid{X}\;\mapsto\;\mathrm{20}\mskip1.5mu\}){}\<[E]%
\\
\>[3]{}\equiv\;{}\<[10]%
\>[10]{}\Varid{N}\;\mathrm{100}\;{}\<[24]%
\>[24]{}\{\mskip1.5mu \Conid{X}\;\mapsto\;\mathrm{40}\mskip1.5mu\}{}\<[E]%
\ColumnHook
\end{hscode}\resethooks
\indentend \paragraph{Sharing for Functions}
However, if we use a function-based module representation, such as \ensuremath{\Conid{Dense}}
and our three reverse-mode variants, then the \ensuremath{(\oplus)} is duplicated
again later in the evaluation. This happens because the \ensuremath{(\oplus)} under the
function binder is not reduced until the function is applied. When both
references of the shared function are applied, the \ensuremath{(\oplus)} in its body will be
evaluated each time. The following shows this for dense functions
because it is the simplest of the function-based representations.

\indentbegin \begin{hscode}\SaveRestoreHook
\column{B}{@{}>{\hspre}l<{\hspost}@{}}%
\column{3}{@{}>{\hspre}l<{\hspost}@{}}%
\column{10}{@{}>{\hspre}l<{\hspost}@{}}%
\column{22}{@{}>{\hspre}l<{\hspost}@{}}%
\column{25}{@{}>{\hspre}l<{\hspost}@{}}%
\column{27}{@{}>{\hspre}l<{\hspost}@{}}%
\column{32}{@{}>{\hspre}l<{\hspost}@{}}%
\column{39}{@{}>{\hspre}l<{\hspost}@{}}%
\column{47}{@{}>{\hspre}l<{\hspost}@{}}%
\column{55}{@{}>{\hspre}l<{\hspost}@{}}%
\column{67}{@{}>{\hspre}l<{\hspost}@{}}%
\column{71}{@{}>{\hspre}l<{\hspost}@{}}%
\column{E}{@{}>{\hspre}l<{\hspost}@{}}%
\>[3]{}\mathbf{let}\;\Varid{y}\mathrel{=}\Varid{N}\;\mathrm{5}\;\Varid{one}\oplus\Varid{N}\;\mathrm{5}\;\Varid{one}\;\mathbf{in}\;(\Varid{y}\otimes\Varid{y}){}\<[E]%
\\
\>[3]{}\equiv\;{}\<[10]%
\>[10]{}\mathbf{let}\;\Varid{y}\mathrel{=}\Varid{N}\;\mathrm{5}\;{}\<[25]%
\>[25]{}(\lambda \Conid{X}\to \mathrm{1})\oplus\Varid{N}\;\mathrm{5}\;(\lambda \Conid{X}\to \mathrm{1})\;{}\<[71]%
\>[71]{}\mathbf{in}\;(\Varid{y}\otimes\Varid{y}){}\<[E]%
\\
\>[3]{}\equiv\;{}\<[10]%
\>[10]{}\mathbf{let}\;\Varid{y}\mathrel{=}\Varid{N}\;\mathrm{10}\;{}\<[25]%
\>[25]{}(\lambda \Conid{X'}\to (\lambda \Conid{X}\to \mathrm{1})\;\Conid{X'}\oplus(\lambda \Conid{X}\to \mathrm{1})\;\Conid{X'})\;{}\<[71]%
\>[71]{}\mathbf{in}\;(\Varid{y}\otimes\Varid{y}){}\<[E]%
\\
\>[3]{}\equiv\;{}\<[10]%
\>[10]{}(\Varid{N}\;\mathrm{10}\;{}\<[22]%
\>[22]{}(\lambda \Conid{X'}\to {}\<[32]%
\>[32]{}(\lambda \Conid{X}\to \mathrm{1})\;\Conid{X'}{}\<[47]%
\>[47]{}\oplus{}\<[55]%
\>[55]{}(\lambda \Conid{X}\to \mathrm{1})\;\Conid{X'}))\otimes{}\<[E]%
\\
\>[10]{}(\Varid{N}\;\mathrm{10}\;{}\<[22]%
\>[22]{}(\lambda \Conid{X'}\to {}\<[32]%
\>[32]{}(\lambda \Conid{X}\to \mathrm{1})\;\Conid{X'}{}\<[47]%
\>[47]{}\oplus{}\<[55]%
\>[55]{}(\lambda \Conid{X}\to \mathrm{1})\;\Conid{X'})){}\<[E]%
\\
\>[3]{}\equiv\;{}\<[10]%
\>[10]{}\Varid{N}\;(\mathrm{10}\;\times\;\mathrm{10})\;{}\<[27]%
\>[27]{}(\lambda \Conid{X'''}\to {}\<[39]%
\>[39]{}(\lambda \Conid{X''}\to \mathrm{10}\;\times\;(\lambda \Conid{X'}\to {}\<[67]%
\>[67]{}(\lambda \Conid{X}\to \mathrm{1})\;\Conid{X'}\oplus{}\<[E]%
\\
\>[67]{}(\lambda \Conid{X}\to \mathrm{1})\;\Conid{X'})\;\Conid{X''})\;\Conid{X'''}\mathbin{+}{}\<[E]%
\\
\>[39]{}(\lambda \Conid{X''}\to \mathrm{10}\;\times\;(\lambda \Conid{X'}\to {}\<[67]%
\>[67]{}(\lambda \Conid{X}\to \mathrm{1})\;\Conid{X'}\oplus{}\<[E]%
\\
\>[67]{}(\lambda \Conid{X}\to \mathrm{1})\;\Conid{X'})\;\Conid{X''})\;\Conid{X'''}){}\<[E]%
\ColumnHook
\end{hscode}\resethooks
\indentend Fortunately, we can address the problem of sharing for function-representations with a
custom symbolic rule for \ensuremath{\Varid{Let}} (Proof in \ref{app:let})\footnote{We ignore the fact that some primitive functions may have non-differentiable points.}.
\begin{lemma}[Let Rule]\label{lemma:let}
\begin{equation*}
\frac{\partial (\mathbf{let}\,y\,= \mathit{e}_1\,\mathbf{in}\,\mathit{e}_2)}{\partial x_i}
=
\left.\frac{\partial \mathit{e}_2}{\partial y}\right\vert_{y = \mathit{e}_1} \times \frac{\partial \mathit{e}_1}{\partial x_i} + \left.\frac{\partial \mathit{e}_2}{\partial x_i}\right\vert_{y = \mathit{e}_1}
\end{equation*}
\end{lemma}

\noindent
For instance, the expression \ensuremath{\Varid{example}_{4}'}
has partial derivative:
\begin{eqnarray*}
\partialfrac{\ensuremath{\Varid{example}_{4}'}}{x} & = & \left.\partialfrac{y \times y}{y}\right\vert_{y = x + x} \times \partialfrac{x+x}{x} + \left.\partialfrac{y \times y}{x}\right\vert_{y = x + x} \\
 & = & ((x + x) + (x + x)) \times (1 + 1) + 0
\end{eqnarray*}


\noindent
This formula (\Cref{lemma:let}) is captured in a \ensuremath{\Conid{Letin}} type class with a method \ensuremath{\Varid{letin}},
which in turn is defined in terms of \ensuremath{\Varid{abs}\mathbin{::}\Varid{e}\to \Varid{v}\to \Varid{d}} of the \ensuremath{\Conid{CorrectAD}} type class, that
extracts the \ensuremath{\Varid{v}}-component of a \ensuremath{\Varid{d}}-module \ensuremath{\Varid{e}}.
\indentbegin \begin{hscode}\SaveRestoreHook
\column{B}{@{}>{\hspre}l<{\hspost}@{}}%
\column{3}{@{}>{\hspre}l<{\hspost}@{}}%
\column{5}{@{}>{\hspre}l<{\hspost}@{}}%
\column{E}{@{}>{\hspre}l<{\hspost}@{}}%
\>[3]{}\mathbf{class}\;\Conid{CorrectAD}\;\Varid{v}\;\Varid{d}\;\Varid{e}\Rightarrow \Conid{Letin}\;\Varid{v}\;\Varid{d}\;\Varid{e}\;\mathbf{where}{}\<[E]%
\\
\>[3]{}\hsindent{2}{}\<[5]%
\>[5]{}\Varid{letin}\mathbin{::}\Varid{v}\to \Varid{e}\to \Varid{e}\to \Varid{e}{}\<[E]%
\\
\>[3]{}\hsindent{2}{}\<[5]%
\>[5]{}\Varid{letin}\;\Varid{y}\;\Varid{de}_{1}\;\Varid{de}_{2}\mathrel{=}((\Varid{abs}\;\Varid{de}_{2}\;\Varid{y})\bullet\Varid{de}_{1})\oplus\Varid{de}_{2}{}\<[E]%
\ColumnHook
\end{hscode}\resethooks
\indentend We specialize the \ensuremath{\Varid{eval}}uator used by \ensuremath{\Varid{abstractD}} to use this formula,
also specializing its type \ensuremath{\Varid{d}} to \ensuremath{\Varid{d}\ltimes\Varid{e}}.


\indentbegin \begin{hscode}\SaveRestoreHook
\column{B}{@{}>{\hspre}l<{\hspost}@{}}%
\column{3}{@{}>{\hspre}l<{\hspost}@{}}%
\column{6}{@{}>{\hspre}l<{\hspost}@{}}%
\column{11}{@{}>{\hspre}l<{\hspost}@{}}%
\column{22}{@{}>{\hspre}l<{\hspost}@{}}%
\column{25}{@{}>{\hspre}l<{\hspost}@{}}%
\column{26}{@{}>{\hspre}l<{\hspost}@{}}%
\column{31}{@{}>{\hspre}c<{\hspost}@{}}%
\column{31E}{@{}l@{}}%
\column{E}{@{}>{\hspre}l<{\hspost}@{}}%
\>[3]{}\Varid{eval}\mathbin{::}(\Conid{Eq}\;\Varid{v},\Conid{Letin}\;\Varid{v}\;\Varid{d}\;\Varid{e})\Rightarrow (\Varid{v}\to \colorbox{lightgray}{$\Varid{d}\ltimes\Varid{e}$})\to \Varid{Expr}\;\Varid{v}\to \colorbox{lightgray}{$\Varid{d}\ltimes\Varid{e}$}{}\<[E]%
\\
\>[3]{}\mathbin{...}{}\<[E]%
\\
\>[3]{}\Varid{eval}\;\Varid{gen}\;(\Varid{Let}\;\Varid{y}\;{}\<[22]%
\>[22]{}\Varid{e}_{1}\;{}\<[26]%
\>[26]{}\Varid{e}_{2}){}\<[31]%
\>[31]{}\mathrel{=}{}\<[31E]%
\\
\>[3]{}\hsindent{3}{}\<[6]%
\>[6]{}\mathbf{let}\;{}\<[11]%
\>[11]{}\Varid{N}\;\Varid{f}_{1}\;\Varid{df}_{1}{}\<[22]%
\>[22]{}\mathrel{=}{}\<[25]%
\>[25]{}\Varid{eval}\;\Varid{gen}\;\Varid{e}_{1}{}\<[E]%
\\
\>[11]{}\Varid{gen}_{2}{}\<[22]%
\>[22]{}\mathrel{=}{}\<[25]%
\>[25]{}\lambda \Varid{x}\to \mathbf{if}\;\Varid{x}= =\Varid{y}\;\mathbf{then}\;\Varid{N}\;\Varid{f}_{1}\;(\Varid{delta}\;\Varid{y})\;\mathbf{else}\;\Varid{gen}\;\Varid{x}{}\<[E]%
\\
\>[11]{}\Varid{N}\;\Varid{f}_{2}\;\Varid{df}_{2}{}\<[22]%
\>[22]{}\mathrel{=}{}\<[25]%
\>[25]{}\Varid{eval}\;\Varid{gen}_{2}\;\Varid{e}_{2}{}\<[E]%
\\
\>[3]{}\hsindent{3}{}\<[6]%
\>[6]{}\mathbf{in}\;{}\<[11]%
\>[11]{}\Varid{N}\;\Varid{f}_{2}\;(\Varid{letin}\;\Varid{y}\;\Varid{df}_{1}\;\Varid{df}_{2}){}\<[E]%
\ColumnHook
\end{hscode}\resethooks
\indentend This evaluator uses the standard substitution behavior for the primal, but
defers to the \ensuremath{\Varid{letin}} method of the \ensuremath{\Conid{Letin}} type class for the behavior for
the tangent.
If we inline and specialize \ensuremath{\Varid{letin}} for \ensuremath{\Varid{d}\multimap\Conid{Cayley}\;(\Conid{Sparse}\;\Varid{v}\;\Varid{d})},
we get a tight definition that computes the maps produced by \ensuremath{\Varid{de}_{1}}
and \ensuremath{\Varid{de}_{2}} only once.
\indentbegin \begin{hscode}\SaveRestoreHook
\column{B}{@{}>{\hspre}l<{\hspost}@{}}%
\column{3}{@{}>{\hspre}l<{\hspost}@{}}%
\column{5}{@{}>{\hspre}l<{\hspost}@{}}%
\column{22}{@{}>{\hspre}l<{\hspost}@{}}%
\column{34}{@{}>{\hspre}l<{\hspost}@{}}%
\column{39}{@{}>{\hspre}l<{\hspost}@{}}%
\column{57}{@{}>{\hspre}l<{\hspost}@{}}%
\column{E}{@{}>{\hspre}l<{\hspost}@{}}%
\>[3]{}\mathbf{instance}\;(\Conid{Ord}\;\Varid{v},\Conid{Semiring}\;\Varid{d})\Rightarrow \Conid{Letin}\;\Varid{v}\;\Varid{d}\;(\Varid{d}\multimap\Conid{Cayley}\;(\Conid{Sparse}\;\Varid{v}\;\Varid{d}))\;\mathbf{where}{}\<[E]%
\\
\>[3]{}\hsindent{2}{}\<[5]%
\>[5]{}\Varid{letin}\;\Varid{y}\;\Varid{de}_{1}\;\Varid{de}_{2}{}\<[22]%
\>[22]{}\mathrel{=}\lambda \Varid{n}\;\Varid{m}\to {}\<[34]%
\>[34]{}\mathbf{let}\;{}\<[39]%
\>[39]{}\Varid{m'}\mathrel{=}\Varid{de}_{2}\;\Varid{n}\;\Varid{m}\;\mathbf{in}\;{}\<[57]%
\>[57]{}\Varid{de}_{1}\;(\Varid{m'}\mathbin{!}\Varid{y})\;\Varid{m'}{}\<[E]%
\ColumnHook
\end{hscode}\resethooks
\indentend With the above definition, the \ensuremath{\mathbf{let}}-sharing is properly preserved.








\paragraph{Complications}
In order to get the main points across, we have ignored two complicating
aspects above.

Firstly, the way the \ensuremath{\Varid{Let}} case of \ensuremath{\Varid{eval}} extends the generator \ensuremath{\Varid{gen}} results
in a lookup time that is linear in the number ($L$) of nested
\ensuremath{\mathbf{let}}-bindings. A more efficient datastructure (a map or array) can be used
instead to improve the lookup time complexity to \bigO{\log L}\ or
\bigO{1}.

Secondly, we have implicitly assumed that \ensuremath{\mathbf{let}}-bound variables do not shadow
other variables, be they free variables or other \ensuremath{\mathbf{let}}-bound variables. To
support shadowing, the partial derivative for the shadowed variable \ensuremath{\Varid{y}}---or at
least, as much of it as has been accumulated so far in \ensuremath{\Varid{m}}---should be saved
before running \ensuremath{\Varid{dx}_{2}} and restored afterwards. Regardless, other than for
debugging purposes, one may want to purge the partial derivative for the
\ensuremath{\mathbf{let}}-bound variable \ensuremath{\Varid{y}} from \ensuremath{\Varid{m'}} as it is no longer relevant. Both together
are accomplished as follows.
\indentbegin \begin{hscode}\SaveRestoreHook
\column{B}{@{}>{\hspre}l<{\hspost}@{}}%
\column{3}{@{}>{\hspre}l<{\hspost}@{}}%
\column{10}{@{}>{\hspre}l<{\hspost}@{}}%
\column{21}{@{}>{\hspre}l<{\hspost}@{}}%
\column{33}{@{}>{\hspre}l<{\hspost}@{}}%
\column{38}{@{}>{\hspre}l<{\hspost}@{}}%
\column{43}{@{}>{\hspre}l<{\hspost}@{}}%
\column{E}{@{}>{\hspre}l<{\hspost}@{}}%
\>[3]{}\Varid{letin}\;{}\<[10]%
\>[10]{}\Varid{y}\;\Varid{de}_{1}\;\Varid{de}_{2}{}\<[21]%
\>[21]{}\mathrel{=}\lambda \Varid{n}\;\Varid{m}\to {}\<[33]%
\>[33]{}\mathbf{let}\;{}\<[38]%
\>[38]{}\Varid{old}{}\<[43]%
\>[43]{}\mathrel{=}\Varid{lookup}\;\Varid{y}\;\Varid{m}{}\<[E]%
\\
\>[38]{}\Varid{m'}{}\<[43]%
\>[43]{}\mathrel{=}\Varid{de}_{2}\;\Varid{n}\;\Varid{m}{}\<[E]%
\\
\>[33]{}\mathbf{in}\;{}\<[38]%
\>[38]{}\Varid{de}_{1}\;(\Varid{m'}\mathbin{!}\Varid{y})\;(\Varid{update}\;(\Varid{const}\;\Varid{old})\;\Varid{y}\;\Varid{m'}){}\<[E]%
\ColumnHook
\end{hscode}\resethooks
\indentend 
%
%
%

\section{Related Work}\label{sec:related}

There is a vast amount of literature on automatic differentiation since the
idea was first presented by Wengert~\cite{Wengert64}.

\paragraph{State-of-the-Art AD Algorithms}
Much of the effort in recent years has been devoted to establishing the
correctness of different AD variants, and to extending the supported language
features. Most works focus on the operational semantics of formal calculi they
define, and reason about the correctness of their AD approaches in operational
terms.
One of the most comprehensive and accessible
of these, in our view, is the recent work of
Krawiec et al.~\cite{pacmpl/KrawiecJKEEF22}. It too goes through several steps
of refinement, from a standard forward mode AD to an efficient reverse mode
variant. While we present several purely functional map-based variants, their
main focus is the imperative version, which uses a defunctionalized form of our \ensuremath{\Varid{d}\multimap\Conid{STCayley}\;\Varid{v}\;\Varid{d}}
representation  that resembles the so-called trace or Wengert list~\cite{Wengert64} used in many AD implementations.
They cover several additional language features, which we believe can also be
incorporated into our approach, such as support for higher-order
functions\footnote{Observe that features like higher-order functions are
parametric with respect to the semi-ring type and can thus be used as is.  },
sum and product types.

A virtue of their paper is in identifying criteria for a (successful!) approach to AD:
that it should demonstrate efficient reverse-mode
AD; that it be higher-order; that it be asymptotically efficient; and that it
be provably correct.
The approach we have presented here, thanks to our careful algebraic treatment,
ensures that (functional) correctness is more or less
immediate, via our use of Kronecker type-class isomorphisms; efficiency
relative to the computation of the primal follows from our use of a single
abstract computation \ensuremath{\Varid{abstractD}}, based on \ensuremath{\Varid{eval}}; and our ability to capture
all the various modes of AD may be seen as arising by careful instantiation of
Nagata's fundamental construction, which nevertheless may be summarised in
completely elementary terms by the \ensuremath{\Conid{Semiring}} typeclass instance given in
Section~\ref{sec:nagata}.

The authors of ~\cite{pacmpl/KrawiecJKEEF22}
give only informal arguments for the relation between different variants,
but the correctness proof of their final version constitutes a non-trivial
effort involving a Kripke logical relation. This proof approach is inspired by
the earlier work of Huot et al.~\cite{fossacs/HuotSV20,lmcs/HuotSV22} on a
logical-relation-based correctness proof of a basic forward mode algorithm for
a higher-order language.

Another recent work in a similar vein is that of Smeding and
V\'ak\'ar~\cite{popl2023}, who derive several variants of reverse-mode AD.
They use ``well-known program transformations'', reasoning mostly in
algorithmic terms.

\paragraph{AD and Functional Programming}
Karczmarczuk~\cite{DBLP:conf/icfp/Karczmarczuk98} shows how to use type classes
to overload arithmetic operations
to obtain forward-mode AD with dual numbers, and
generalizes that to an infinite stream of higher derivatives.

Elliott takes a more algebraic approach based on program derivation.
First, he explores a derivation of forward-mode AD,
focusing on the chain rule as the guiding principle, and generalizing from
scalars to vector spaces \cite{icfp/Elliott09}. Next,
he shifts his view to derivatives as linear maps and the categorical
structure of differentiation \cite{DBLP:journals/pacmpl/Elliott18}. More recently, Elliott
has also explored the application of automatic differentiation to
languages \cite{pacmpl/Elliott21}.

Elsman et al.~\cite{corr/abs-2207-00847} follow Elliott's lead, with their
functional analysis for differentiation based on linear functions. They show
that symbolic and forward-mode derivatives follow from the same rules, but
reverse-mode essentially computes the adjoint of a derivative.
%
%
Adjoints arise by taking a curried form of \ensuremath{\Varid{d}}-valued inner product structure on vector spaces \ensuremath{\Varid{e}}, giving rise to a duality between \ensuremath{\Varid{e}} and the linear function space \ensuremath{\Varid{e}\multimap\Varid{d}}. Our more general setting does not consider such normed/inner product structure on our \ensuremath{\Varid{d}}-modules \ensuremath{\Varid{e}}, but we expect that such richer structure may be smoothly incorporated into a general picture, and we leave this as a topic for future work.

Wang et al.~\cite{pacmpl/WangZDWER19} show that the two passes of reverse-mode
AD can be conveniently implemented using delimited control in an impure
functional setting like Scala; they provide no formal argument for its
correctness.

Kmett's \texttt{ad} library~\cite{kmett} is a Haskell implementation based on
Pearlmutter and Siskind's work~\cite{DBLP:journals/toplas/PearlmutterS08},
which explained how to incorporate AD as a first-class function in a functional
language. The library exploits Haskell's abstraction and overloading mechanisms
to support different AD modes.

Oleksandr et al. \cite{manzyuk_pearlmutter_radul_rush_siskind_2019} discuss 
the challenges in automatic differentiation of higher-order functions, for example
functions in curried form. 

\paragraph{Nagata Numbers}
As far as we know, our use of Nagata's idealization of a module (``Nagata
numbers'') in the context of AD is novel. Most works in the AD literature use
the standard homogenous dual numbers. Some, like those of Krawiec et
al.~\cite{pacmpl/KrawiecJKEEF22} and Smeding and V\'ak\'ar~\cite{popl2023}, do feature
various heterogeneous dual numbers, but they do not observe that these are
instances of a general structure.
Nagata originally defined his idealization of a module over a ring.
Our weakening to consider only \emph{semi}ring structure is just cosmetic, and
the extension of our methods to consider a (commutative) \emph{ring} \ensuremath{\Varid{d}} of
coefficients is straightforward, and we appeal to it without comment in
Section~\ref{sec:extensions}. See also~\cite{anderson-winder-2009:JCA} for a
comprehensive survey.

\paragraph{Leibniz vs.~Newton}
One way to understand our contribution, in contrast to those of Elliott and others, is that while they take a \squote[Newtonian] view of AD as an algebraic operation on spaces of (abstract) \emph{functions}, we take the \squote[Leibnizian] approach of understanding both AD and \emph{symbolic} differentiation as an algebraic operation on a language \ensuremath{\Conid{Expr}\;\Varid{v}} of (concrete) symbolic \emph{expressions}.
This in particular contrasts with Baydin et al.'s assertion ~\cite[Section 2.2, under {\dquote[What AD Is Not]}]{JMLR:v18:17-468}
that AD is \dquote[not symbolic differentiation], but echoes their observation, following Griewank~\cite{Griewank2003:Acta}, that it has a \dquote[two-sided nature that is partly symbolic and partly numerical].

\section{Conclusion}\label{sec:conclusion}

Having identified three algebraic abstractions, we can write symbolic
differentiation, forward-mode and reverse-mode AD as different instances of one
and the same abstract algorithm.  A significant contribution is the observation
that we can generalize the well-known dual numbers into the Nagata idealization 
abstraction \ensuremath{\Varid{d}\ltimes\Varid{e}} for any \ensuremath{\Varid{d}}-module \ensuremath{\Varid{e}}, which allows the \ensuremath{\Varid{e}} parameter to vary
independently of the semiring \ensuremath{\Varid{d}} of coefficients, and that we can use abstract
Kronecker deltas to represent partial derivatives.
Furthermore, guided by structure-preserving maps in the form of Kronecker isomorphisms, we have taken successive steps in optimising the representation, in a correct-by-construction manner, from basic symbolic differentiation via forward-mode AD to the (typically) much more efficient reverse-mode AD.

Future work includes further exploring matrix differentiation with this framework. 
Square matrices form a (non-commutative) semiring but matrices of arbitrary dimensions
form an extra challenge, in particular when reflecting these dimensions in the types. 


\section{Acknowledgements}\label{sec:acks}

This work has been funded by KU Leuven's BOF grant number C14/20/079,
and by the Flemish Fund for Scientific Research (FWO) grant number G0A9423N.
The third author would like to thank Bernhard Rybo{\l}owicz for background insight into Nagata's construction. 

\bibliographystyle{elsarticle-num}
\bibliography{bibliography}

\label{lastpage01}

\appendix

\section{Derivation of the Second-Order Derivative}
\label{app:hod}

This appendix expands on Section~\ref{sec:ext:hod}'s claim that
we can derive from the specification
\indentbegin \begin{hscode}\SaveRestoreHook
\column{B}{@{}>{\hspre}l<{\hspost}@{}}%
\column{3}{@{}>{\hspre}l<{\hspost}@{}}%
\column{35}{@{}>{\hspre}l<{\hspost}@{}}%
\column{41}{@{}>{\hspre}l<{\hspost}@{}}%
\column{E}{@{}>{\hspre}l<{\hspost}@{}}%
\>[3]{}\Varid{derive}_{\mathit{2^{nd}}}\mathbin{::}(\Conid{Eq}\;\Varid{v},\Conid{Semiring}\;\Varid{d})\Rightarrow (\Varid{v}\to \Varid{d})\to \Conid{Expr}\;\Varid{v}\to \Conid{Dense}\;\Varid{v}\;(\Conid{Dense}\;\Varid{v}\;\Varid{d}){}\<[E]%
\\
\>[3]{}\Varid{derive}_{\mathit{2^{nd}}}\;\Varid{var}\;\Varid{e}\;\Varid{y}\;\Varid{x}\mathrel{=}\Varid{eval}\;\Varid{var}\;{}\<[35]%
\>[35]{}(\Varid{tan}^{N}\;{}\<[41]%
\>[41]{}(\Varid{forwardAD}_{Dense}\;\Conid{Var}{}\<[E]%
\\
\>[35]{}(\Varid{tan}^{N}\;{}\<[41]%
\>[41]{}(\Varid{forwardAD}_{Dense}\;\Conid{Var}\;\Varid{e})\;\Varid{y}))\;\Varid{x}){}\<[E]%
\ColumnHook
\end{hscode}\resethooks
\indentend that consists of three successive appeals to \ensuremath{\Varid{eval}}, an equivalent
version that is an instance of \ensuremath{\Varid{abstractD}} (which consists of
a single \ensuremath{\Varid{eval}}):
\indentbegin \begin{hscode}\SaveRestoreHook
\column{B}{@{}>{\hspre}l<{\hspost}@{}}%
\column{3}{@{}>{\hspre}l<{\hspost}@{}}%
\column{5}{@{}>{\hspre}l<{\hspost}@{}}%
\column{E}{@{}>{\hspre}l<{\hspost}@{}}%
\>[3]{}\Varid{derive}_{\mathit{2^{nd}}}^{\prime}\;\Varid{var}\;\Varid{x}\;\Varid{y}\;\Varid{e}\mathrel{=}\Varid{tan}^{N}\;(\Varid{tan}^{N}\;(\Varid{abstractD}\;\Varid{gen}\;\Varid{e})\;\Varid{x})\;\Varid{y}{}\<[E]%
\\
\>[3]{}\hsindent{2}{}\<[5]%
\>[5]{}\mathbf{where}\;\Varid{gen}\;\Varid{z}\mathrel{=}\Varid{N}\;(\Varid{var}\;\Varid{z})\;(\Varid{delta}\;\Varid{z}){}\<[E]%
\ColumnHook
\end{hscode}\resethooks
\indentend We start the derivation from a point-free formulation of \ensuremath{\Varid{derive}_{\mathit{2^{nd}}}}:
\indentbegin \begin{hscode}\SaveRestoreHook
\column{B}{@{}>{\hspre}l<{\hspost}@{}}%
\column{3}{@{}>{\hspre}l<{\hspost}@{}}%
\column{5}{@{}>{\hspre}l<{\hspost}@{}}%
\column{8}{@{}>{\hspre}l<{\hspost}@{}}%
\column{11}{@{}>{\hspre}l<{\hspost}@{}}%
\column{15}{@{}>{\hspre}l<{\hspost}@{}}%
\column{E}{@{}>{\hspre}l<{\hspost}@{}}%
\>[5]{}\lambda \Varid{y}\;\Varid{x}\to {}\<[15]%
\>[15]{}\Varid{eval}\;\Varid{var}\hsdot{\circ }{.}(\mathbin{\$}\Varid{x})\hsdot{\circ }{.}\Varid{tan}^{N}\hsdot{\circ }{.}\Varid{forwardAD}_{Dense}\;\Conid{Var}{}\<[E]%
\\
\>[15]{}\hsdot{\circ }{.}(\mathbin{\$}\Varid{y})\hsdot{\circ }{.}\Varid{tan}^{N}\hsdot{\circ }{.}\Varid{forwardAD}_{Dense}\;\Conid{Var}{}\<[E]%
\\
\>[3]{}\mathrel{=}\mbox{\commentbegin  naturality of function application  \commentend}{}\<[E]%
\\
\>[3]{}\hsindent{2}{}\<[5]%
\>[5]{}\lambda \Varid{y}\;\Varid{x}\to {}\<[15]%
\>[15]{}(\mathbin{\$}\Varid{x})\hsdot{\circ }{.}\Varid{fmap}\;(\Varid{eval}\;\Varid{var})\hsdot{\circ }{.}\Varid{tan}^{N}\hsdot{\circ }{.}\Varid{forwardAD}_{Dense}\;\Conid{Var}{}\<[E]%
\\
\>[15]{}\hsdot{\circ }{.}(\mathbin{\$}\Varid{y})\hsdot{\circ }{.}\Varid{tan}^{N}\hsdot{\circ }{.}\Varid{forwardAD}_{Dense}\;\Conid{Var}{}\<[E]%
\\
\>[3]{}\mathrel{=}\mbox{\commentbegin  $\eta$-reduction  \commentend}{}\<[E]%
\\
\>[3]{}\hsindent{2}{}\<[5]%
\>[5]{}\lambda \Varid{y}\to {}\<[15]%
\>[15]{}\Varid{fmap}\;(\Varid{eval}\;\Varid{var})\hsdot{\circ }{.}\Varid{tan}^{N}\hsdot{\circ }{.}\Varid{forwardAD}_{Dense}\;\Conid{Var}{}\<[E]%
\\
\>[15]{}\hsdot{\circ }{.}(\mathbin{\$}\Varid{y})\hsdot{\circ }{.}\Varid{tan}^{N}\hsdot{\circ }{.}\Varid{forwardAD}_{Dense}\;\Conid{Var}{}\<[E]%
\\
\>[3]{}\mathrel{=}\mbox{\commentbegin  naturality of function application  \commentend}{}\<[E]%
\\
\>[3]{}\hsindent{2}{}\<[5]%
\>[5]{}\lambda \Varid{y}\to {}\<[15]%
\>[15]{}(\mathbin{\$}\Varid{y})\hsdot{\circ }{.}\Varid{fmap}\;(\Varid{fmap}\;(\Varid{eval}\;\Varid{var})\hsdot{\circ }{.}\Varid{tan}^{N}\hsdot{\circ }{.}\Varid{forwardAD}_{Dense}\;\Conid{Var}){}\<[E]%
\\
\>[15]{}\hsdot{\circ }{.}\Varid{tan}^{N}\hsdot{\circ }{.}\Varid{forwardAD}_{Dense}\;\Conid{Var}{}\<[E]%
\\
\>[3]{}\mathrel{=}\mbox{\commentbegin  $\eta$-reduction  \commentend}{}\<[E]%
\\
\>[3]{}\hsindent{2}{}\<[5]%
\>[5]{}\Varid{fmap}\;{}\<[11]%
\>[11]{}(\Varid{fmap}\;(\Varid{eval}\;\Varid{var})\hsdot{\circ }{.}\Varid{tan}^{N}\hsdot{\circ }{.}\Varid{forwardAD}_{Dense}\;\Conid{Var}){}\<[E]%
\\
\>[11]{}\hsdot{\circ }{.}\Varid{tan}^{N}\hsdot{\circ }{.}\Varid{forwardAD}_{Dense}\;\Conid{Var}{}\<[E]%
\\
\>[3]{}\mathrel{=}\mbox{\commentbegin  naturality of \ensuremath{\Varid{tan}^{N}}; see equation (\ref{eq:binaturality}) below  \commentend}{}\<[E]%
\\
\>[3]{}\hsindent{2}{}\<[5]%
\>[5]{}\Varid{fmap}\;{}\<[11]%
\>[11]{}(\Varid{tan}^{N}\hsdot{\circ }{.}\Varid{bimap}\;(\Varid{eval}\;\Varid{var})\;(\Varid{fmap}\;(\Varid{eval}\;\Varid{var}))\hsdot{\circ }{.}\Varid{forwardAD}_{Dense}\;\Conid{Var}){}\<[E]%
\\
\>[11]{}\hsdot{\circ }{.}\Varid{tan}^{N}\hsdot{\circ }{.}\Varid{forwardAD}_{Dense}\;\Conid{Var}{}\<[E]%
\\
\>[3]{}\mathrel{=}\mbox{\commentbegin  \ensuremath{\Varid{eval}} fusion  \commentend}{}\<[E]%
\\
\>[3]{}\hsindent{2}{}\<[5]%
\>[5]{}\Varid{fmap}\;(\Varid{tan}^{N}\hsdot{\circ }{.}\Varid{eval}\;\Varid{gen'})\hsdot{\circ }{.}\Varid{tan}^{N}\hsdot{\circ }{.}\Varid{forwardAD}_{Dense}\;\Conid{Var}{}\<[E]%
\\
\>[5]{}\hsindent{3}{}\<[8]%
\>[8]{}\mathbf{where}\;{}\<[15]%
\>[15]{}\Varid{gen'}\;\Varid{z}\mathrel{=}\Varid{N}\;(\Varid{var}\;\Varid{z})\;(\Varid{delta}\;\Varid{z}){}\<[E]%
\\
\>[3]{}\mathrel{=}\mbox{\commentbegin  \ensuremath{\Varid{fmap}} fission + naturality of \ensuremath{\Varid{tan}^{N}}  \commentend}{}\<[E]%
\\
\>[3]{}\hsindent{2}{}\<[5]%
\>[5]{}\Varid{fmap}\;\Varid{tan}^{N}\hsdot{\circ }{.}\Varid{tan}^{N}\hsdot{\circ }{.}\Varid{bimap}\;(\Varid{eval}\;\Varid{gen'})\;(\Varid{fmap}\;(\Varid{eval}\;\Varid{gen'})){}\<[E]%
\\
\>[3]{}\hsindent{2}{}\<[5]%
\>[5]{}\hsdot{\circ }{.}\Varid{forwardAD}_{Dense}\;\Conid{Var}{}\<[E]%
\\
\>[5]{}\hsindent{3}{}\<[8]%
\>[8]{}\mathbf{where}\;{}\<[15]%
\>[15]{}\Varid{gen'}\;\Varid{z}\mathrel{=}\Varid{N}\;(\Varid{var}\;\Varid{z})\;(\Varid{delta}\;\Varid{z}){}\<[E]%
\\
\>[3]{}\mathrel{=}\mbox{\commentbegin  \ensuremath{\Varid{eval}} fusion  \commentend}{}\<[E]%
\\
\>[3]{}\hsindent{2}{}\<[5]%
\>[5]{}\Varid{fmap}\;\Varid{tan}^{N}\hsdot{\circ }{.}\Varid{tan}^{N}\hsdot{\circ }{.}\Varid{eval}\;\Varid{gen''}{}\<[E]%
\\
\>[5]{}\hsindent{3}{}\<[8]%
\>[8]{}\mathbf{where}\;{}\<[15]%
\>[15]{}\Varid{gen''}\;\Varid{z}\mathrel{=}\Varid{N}\;(\Varid{N}\;(\Varid{var}\;\Varid{z})\;(\Varid{delta}\;\Varid{z}))\;(\Varid{delta}\;\Varid{z}){}\<[E]%
\\
\>[3]{}\mathrel{=}\mbox{\commentbegin  \ensuremath{\Varid{abstractD}} definition  \commentend}{}\<[E]%
\\
\>[3]{}\hsindent{2}{}\<[5]%
\>[5]{}\Varid{fmap}\;\Varid{tan}^{N}\hsdot{\circ }{.}\Varid{tan}^{N}\hsdot{\circ }{.}\Varid{abstractD}{}\<[E]%
\\
\>[5]{}\hsindent{3}{}\<[8]%
\>[8]{}\mathbf{where}\;{}\<[15]%
\>[15]{}\Varid{gen}\;\Varid{z}\mathrel{=}\Varid{N}\;(\Varid{var}\;\Varid{z})\;(\Varid{delta}\;\Varid{z}){}\<[E]%
\ColumnHook
\end{hscode}\resethooks
\indentend The steps involving the naturality of \ensuremath{\Varid{tan}^{N}} do not concern the \ensuremath{\Conid{Functor}} instance of Nagata numbers,
but their \ensuremath{\Conid{Bifunctor}} instance, which allows to vary both components.
\indentbegin \begin{hscode}\SaveRestoreHook
\column{B}{@{}>{\hspre}l<{\hspost}@{}}%
\column{3}{@{}>{\hspre}l<{\hspost}@{}}%
\column{5}{@{}>{\hspre}l<{\hspost}@{}}%
\column{E}{@{}>{\hspre}l<{\hspost}@{}}%
\>[3]{}\mathbf{instance}\;\Conid{Bifunctor}\;(\cdot \ltimes\cdot )\;\mathbf{where}{}\<[E]%
\\
\>[3]{}\hsindent{2}{}\<[5]%
\>[5]{}\Varid{bimap}\;\Varid{f}\;\Varid{g}\;(\Varid{N}\;\Varid{d}\;\Varid{e})\mathrel{=}\Varid{N}\;(\Varid{f}\;\Varid{d})\;(\Varid{g}\;\Varid{d}){}\<[E]%
\ColumnHook
\end{hscode}\resethooks
\indentend The general naturality property of \ensuremath{\Varid{tan}^{N}\mathbin{::}\forall \Varid{d}\hsforall \;\Varid{e}\hsdot{\circ }{.}\Varid{d}\ltimes\Varid{e}\to \Varid{e}} for this \ensuremath{\Conid{Bifunctor}} instance is:
\begin{equation*}
\ensuremath{\Varid{g}\hsdot{\circ }{.}\Varid{tan}^{N}} = \ensuremath{\Varid{tan}^{N}\hsdot{\circ }{.}\Varid{bimap}\;\Varid{f}\;\Varid{g}}
\end{equation*}
However, for the specialized type \ensuremath{\Varid{tan}^{N}\mathbin{::}\forall \Varid{d}\hsforall \hsdot{\circ }{.}\Varid{d}\ltimes(\Conid{Dense}\;\Varid{v}\;\Varid{d})\to \Conid{Dense}\;\Varid{v}\;\Varid{d}} where
the two components are not independent, we also have this specialized naturality property:
\begin{equation}
\ensuremath{\Varid{fmap}\;\Varid{f}\hsdot{\circ }{.}\Varid{tan}^{N}} = \ensuremath{\Varid{tan}^{N}\hsdot{\circ }{.}\Varid{bimap}\;\Varid{f}\;(\Varid{fmap}\;\Varid{f})} \label{eq:binaturality}
\end{equation}


\section{Proof of \Cref{lemma:let}}\label{app:let}


\begin{proof}
We prove this rule by means of the chain rule for multi-variate functions: 
\begin{equation*}
\frac{\partial f(e_1,\ldots,e_n)}{\partial x} =
\sum_{i=1}^n \left(
\left.\frac{\partial f(x_1,\ldots,x_n)}{\partial x_i}\right\vert_{x_1 = e_1,\ldots,x_n = e_n} \times \frac{\partial e_i}{\partial x}
\right)
\end{equation*}

\noindent
Define $f$ as
\ensuremath{\Varid{f}\;(\Varid{y},\Varid{x}_{1},\mathbin{...},\Varid{x}_n)\mathrel{=}\Varid{e}_{2}}
where $x_1,\ldots,x_n$ comprise all
the free variables of \ensuremath{\Varid{e}_{1}} and \ensuremath{\Varid{e}_{2}}. Then we have that
$f(\ensuremath{\Varid{e}_{1}},x_1,\ldots,x_n) = \ensuremath{\mathbf{let}\;\Varid{y}\mathrel{=}\Varid{e}_{1}\;\mathbf{in}\;\Varid{e}_{2}}$.

Using this in combination with the chain rule yields:
%
\begin{eqnarray*}
\frac{\partial (\mathbf{let}\,y\,= \mathit{e}_1\,\mathbf{in}\,\mathit{e}_2)}{\partial x_i}
& = & \partialfrac{\ensuremath{\Varid{f}\;(\Varid{e}_{1},\Varid{x}_{1},\mathbin{...},\Varid{x}_n)}}{x_i} \\
& = & \left.\frac{\partial f(y,x_1,\ldots,x_n)}{\partial y}\right\vert_{y=\mathit{e}_1,x_1 = x_1,\ldots,x_n = x_n} \times \frac{\partial \mathit{e}_1}{\partial x_i} \\
& & +\sum_{j=1}^n \left(
\left.\frac{\partial f(y,x_1,\ldots,x_n)}{\partial x_j}\right\vert_{y=\mathit{e}_1,x_1 = x_1,\ldots,x_n = x_n} \times \frac{\partial x_j}{\partial x_i}
\right)
\\
 & = & \left.\frac{\partial \mathit{e}_2}{\partial y}\right\vert_{y = \mathit{e}_1} \times \frac{\partial \mathit{e}_1}{\partial x_i} + \left.\frac{\partial \mathit{e}_2}{\partial x_i}\right\vert_{y = \mathit{e}_1}
\end{eqnarray*}
\end{proof}

\section{Haskell Libraries and Functionality}
\label{app:functions}

This section briefly introduces Haskell's type class feature 
and summarizes the key library functions that we use.

\subsection{Type Classes}

We use Haskell's type classes \cite{WadlerBlott89} to model the interface of
algebraic structures, such as the semirings:\indentbegin \begin{hscode}\SaveRestoreHook
\column{B}{@{}>{\hspre}l<{\hspost}@{}}%
\column{3}{@{}>{\hspre}l<{\hspost}@{}}%
\column{5}{@{}>{\hspre}l<{\hspost}@{}}%
\column{11}{@{}>{\hspre}l<{\hspost}@{}}%
\column{E}{@{}>{\hspre}l<{\hspost}@{}}%
\>[3]{}\mathbf{class}\;\Conid{Semiring}\;\Varid{d}\;\mathbf{where}{}\<[E]%
\\
\>[3]{}\hsindent{2}{}\<[5]%
\>[5]{}\Varid{zero}{}\<[11]%
\>[11]{}\mathbin{::}\Varid{d}{}\<[E]%
\\
\>[3]{}\hsindent{2}{}\<[5]%
\>[5]{}\Varid{one}{}\<[11]%
\>[11]{}\mathbin{::}\Varid{d}{}\<[E]%
\\
\>[3]{}\hsindent{2}{}\<[5]%
\>[5]{}(\oplus){}\<[11]%
\>[11]{}\mathbin{::}\Varid{d}\to \Varid{d}\to \Varid{d}{}\<[E]%
\\
\>[3]{}\hsindent{2}{}\<[5]%
\>[5]{}(\otimes)\mathbin{::}\Varid{d}\to \Varid{d}\to \Varid{d}{}\<[E]%
\ColumnHook
\end{hscode}\resethooks
\indentend Under the hood, the Haskell compiler creates a datatype, called a \emph{dictionary} type, that contains all the methods of the type class.\indentbegin \begin{hscode}\SaveRestoreHook
\column{B}{@{}>{\hspre}l<{\hspost}@{}}%
\column{3}{@{}>{\hspre}l<{\hspost}@{}}%
\column{29}{@{}>{\hspre}l<{\hspost}@{}}%
\column{E}{@{}>{\hspre}l<{\hspost}@{}}%
\>[3]{}\mathbf{data}\;\Conid{SemiringDict}\;\Varid{d}\mathrel{=}\Conid{SD}\;{}\<[29]%
\>[29]{}\{\mskip1.5mu \Varid{zero}\mathbin{::}\Varid{d}{}\<[E]%
\\
\>[29]{},\Varid{one}\mathbin{::}\Varid{d}{}\<[E]%
\\
\>[29]{},(\oplus)\mathbin{::}\Varid{d}\to \Varid{d}\to \Varid{d}{}\<[E]%
\\
\>[29]{},(\otimes)\mathbin{::}\Varid{d}\to \Varid{d}\to \Varid{d}\mskip1.5mu\}{}\<[E]%
\ColumnHook
\end{hscode}\resethooks
\indentend A type class instance provides the implementation of the interface for a specific type. For example, \indentbegin \begin{hscode}\SaveRestoreHook
\column{B}{@{}>{\hspre}l<{\hspost}@{}}%
\column{3}{@{}>{\hspre}l<{\hspost}@{}}%
\column{5}{@{}>{\hspre}l<{\hspost}@{}}%
\column{12}{@{}>{\hspre}c<{\hspost}@{}}%
\column{12E}{@{}l@{}}%
\column{15}{@{}>{\hspre}l<{\hspost}@{}}%
\column{E}{@{}>{\hspre}l<{\hspost}@{}}%
\>[3]{}\mathbf{instance}\;\Conid{Semiring}\;(\Conid{Expr}\;\Varid{v})\;\mathbf{where}{}\<[E]%
\\
\>[3]{}\hsindent{2}{}\<[5]%
\>[5]{}\Varid{zero}{}\<[12]%
\>[12]{}\mathrel{=}{}\<[12E]%
\>[15]{}\Conid{Zero}{}\<[E]%
\\
\>[3]{}\hsindent{2}{}\<[5]%
\>[5]{}\Varid{one}{}\<[12]%
\>[12]{}\mathrel{=}{}\<[12E]%
\>[15]{}\Conid{One}{}\<[E]%
\\
\>[3]{}\hsindent{2}{}\<[5]%
\>[5]{}(\oplus){}\<[12]%
\>[12]{}\mathrel{=}{}\<[12E]%
\>[15]{}\Conid{Plus}{}\<[E]%
\\
\>[3]{}\hsindent{2}{}\<[5]%
\>[5]{}(\otimes){}\<[12]%
\>[12]{}\mathrel{=}{}\<[12E]%
\>[15]{}\Conid{Times}{}\<[E]%
\ColumnHook
\end{hscode}\resethooks
\indentend Under the hood, an instance gives rise to a dictionary value that contains these implementations.\indentbegin \begin{hscode}\SaveRestoreHook
\column{B}{@{}>{\hspre}l<{\hspost}@{}}%
\column{3}{@{}>{\hspre}l<{\hspost}@{}}%
\column{26}{@{}>{\hspre}l<{\hspost}@{}}%
\column{35}{@{}>{\hspre}l<{\hspost}@{}}%
\column{E}{@{}>{\hspre}l<{\hspost}@{}}%
\>[3]{}\Varid{semiringExprDict}\mathbin{::}\Conid{SemiringDict}\;(\Conid{Expr}\;\Varid{v}){}\<[E]%
\\
\>[3]{}\Varid{semiringExprDict}\mathrel{=}\Conid{SD}\;{}\<[26]%
\>[26]{}\{\mskip1.5mu \Varid{zero}{}\<[35]%
\>[35]{}\mathrel{=}\Conid{Zero}{}\<[E]%
\\
\>[26]{},\Varid{one}{}\<[35]%
\>[35]{}\mathrel{=}\Conid{One}{}\<[E]%
\\
\>[26]{},(\oplus){}\<[35]%
\>[35]{}\mathrel{=}\Conid{Plus}{}\<[E]%
\\
\>[26]{},(\otimes){}\<[35]%
\>[35]{}\mathrel{=}\Conid{Times}\mskip1.5mu\}{}\<[E]%
\ColumnHook
\end{hscode}\resethooks
\indentend We can use the type class methods at a specific type for which there is an instance. E.g.,\indentbegin \begin{hscode}\SaveRestoreHook
\column{B}{@{}>{\hspre}l<{\hspost}@{}}%
\column{3}{@{}>{\hspre}l<{\hspost}@{}}%
\column{E}{@{}>{\hspre}l<{\hspost}@{}}%
\>[3]{}\Varid{expr}\mathbin{::}\Conid{Expr}\;\Varid{v}{}\<[E]%
\\
\>[3]{}\Varid{expr}\mathrel{=}\Varid{one}\oplus\Varid{one}{}\<[E]%
\ColumnHook
\end{hscode}\resethooks
\indentend The compiler turns this into code that explicitly extracts the implementations from the appropriate dictionary:
\indentbegin \begin{hscode}\SaveRestoreHook
\column{B}{@{}>{\hspre}l<{\hspost}@{}}%
\column{3}{@{}>{\hspre}l<{\hspost}@{}}%
\column{E}{@{}>{\hspre}l<{\hspost}@{}}%
\>[3]{}\Varid{expr}\mathbin{::}\Conid{Expr}\;\Varid{v}{}\<[E]%
\\
\>[3]{}\Varid{expr}\mathrel{=}\Varid{d}.(\oplus)\;\Varid{d}.\Varid{one}\;\Varid{d}.\Varid{one}\;\mathbf{where}\;\Varid{d}\mathrel{=}\Varid{semiringExprDict}{}\<[E]%
\ColumnHook
\end{hscode}\resethooks
\indentend The selection of the appropriate dictionary is known as \emph{type class resolution} and directed by the type at which the methods are used. Because every type
can have at most one instance of a type class, there is always at most one appropriate dictionary.

We can also abstract over the particular type used with a type variable \ensuremath{\Varid{d}}. Then the type variable
is subject to the type class constraint \ensuremath{\Conid{Semiring}\;\Varid{d}}. This means that \ensuremath{\Varid{d}} can be freely chosen as long
as it has a \ensuremath{\Conid{Semiring}} instance. This is known as constrained polymorphism.\indentbegin \begin{hscode}\SaveRestoreHook
\column{B}{@{}>{\hspre}l<{\hspost}@{}}%
\column{3}{@{}>{\hspre}l<{\hspost}@{}}%
\column{E}{@{}>{\hspre}l<{\hspost}@{}}%
\>[3]{}\Varid{expr}\mathbin{::}\Conid{Semiring}\;\Varid{d}\Rightarrow \Varid{d}{}\<[E]%
\\
\>[3]{}\Varid{expr}\mathrel{=}\Varid{one}\oplus\Varid{one}{}\<[E]%
\ColumnHook
\end{hscode}\resethooks
\indentend Under the hood, the type class constraint is transformed into an explicit dictionary parameter.
\indentbegin \begin{hscode}\SaveRestoreHook
\column{B}{@{}>{\hspre}l<{\hspost}@{}}%
\column{3}{@{}>{\hspre}l<{\hspost}@{}}%
\column{E}{@{}>{\hspre}l<{\hspost}@{}}%
\>[3]{}\Varid{expr}\mathbin{::}\Conid{SemiringDict}\;\Varid{d}\to \Varid{d}{}\<[E]%
\\
\>[3]{}\Varid{expr}\;\Varid{d}\mathrel{=}\Varid{d}.(\oplus)\;\Varid{d}.\Varid{one}\;\Varid{d}.\Varid{one}{}\<[E]%
\ColumnHook
\end{hscode}\resethooks
\indentend 
When using such a constraint polymorphic definition at a specific type, the
Haskell compiler implicitly provides the appropriate dictionary parameter. For example,
it turns \ensuremath{\Varid{expr}\mathbin{::}\Conid{Expr}\;\Varid{v}} into \ensuremath{\Varid{expr}\;\Varid{semiringExprDict}\mathbin{::}\Conid{Expr}\;\Varid{v}}.

In multi-parameter type classes, i.e., type classes with multiple parameters in their signature (e.g., \ensuremath{\Conid{Module}\;\Varid{d}\;\Varid{e}}), 
we can declare \emph{functional dependencies} between parameters to aid type class resolution.
A functional dependency indicates that one type parameter is uniquely determined by the other(s).
For example, our \ensuremath{\Conid{Module}} class uses such a functional dependency:\indentbegin \begin{hscode}\SaveRestoreHook
\column{B}{@{}>{\hspre}l<{\hspost}@{}}%
\column{3}{@{}>{\hspre}l<{\hspost}@{}}%
\column{E}{@{}>{\hspre}l<{\hspost}@{}}%
\>[3]{}\mathbf{class}\;(\Conid{Semiring}\;\Varid{d},\Conid{Monoid}\;\Varid{e})\Rightarrow \Conid{Module}\;\Varid{d}\;\Varid{e}\mid \Varid{e}\to \Varid{d}\;\mathbf{where}\mathbin{...}{}\<[E]%
\ColumnHook
\end{hscode}\resethooks
\indentend Here the semiring type \ensuremath{\Varid{d}} can be determined from type \ensuremath{\Varid{e}}.

For a more thorough introduction to type classes and Haskell in
general, we refer to Bird \cite{bird_2014} and Hutton \cite{Hutton16}.

\subsection{Maps}

\noindent
The functions below are from Haskell's \ensuremath{\Conid{\Conid{Data}.Map}} 
library\footnote{\href{https://hackage.haskell.org/package/containers-0.4.0.0/docs/Data-Map.html}{https://hackage.haskell.org/package/containers-0.4.0.0/docs/Data-Map.html}}, 
copied with their type signature and explanation.

\begin{itemize}
\item
\ensuremath{(\mathbin{!})\mathbin{::}\Conid{Ord}\;\Varid{k}\Rightarrow \Conid{Map}\;\Varid{k}\;\Varid{a}\to \Varid{k}\to \Varid{a}}\\
\bigO{\log n}. Find the value at a key. Calls \ensuremath{\Varid{error}} when the element can not be found.
\item
\ensuremath{\Varid{findWithDefault}\mathbin{::}\Conid{Ord}\;\Varid{k}\Rightarrow \Varid{a}\to \Varid{k}\to \Conid{Map}\;\Varid{k}\;\Varid{a}\to \Varid{a}}\\
\bigO{\log n}. The expression \ensuremath{\Varid{findWithDefault}\;\Varid{def}\;\Varid{k}\;\Varid{map}} returns the value at key \ensuremath{\Varid{k}} or returns default value \ensuremath{\Varid{def}} when the key is not in the \ensuremath{\Varid{map}}.
\item
\ensuremath{\Varid{fromList}\mathbin{::}\Conid{Ord}\;\Varid{k}\Rightarrow [\mskip1.5mu (\Varid{k},\Varid{a})\mskip1.5mu]\to \Conid{Map}\;\Varid{k}\;\Varid{a}}\\
\bigO{n \times \log n}. Build a map from a list of key/value pairs. 
If the list contains more than one value for the same key, the last value for the key is retained.
\item
\ensuremath{\Varid{unionWith}\mathbin{::}\Conid{Ord}\;\Varid{k}\Rightarrow (\Varid{a}\to \Varid{a}\to \Varid{a})\to \Conid{Map}\;\Varid{k}\;\Varid{a}\to \Conid{Map}\;\Varid{k}\;\Varid{a}\to \Conid{Map}\;\Varid{k}\;\Varid{a}}\\
\bigO{n+m}. Union with a combining function. The implementation uses the efficient hedge-union algorithm.
\item
\ensuremath{\Varid{insertWith}\mathbin{::}\Conid{Ord}\;\Varid{k}\Rightarrow (\Varid{a}\to \Varid{a}\to \Varid{a})\to \Varid{k}\to \Varid{a}\to \Conid{Map}\;\Varid{k}\;\Varid{a}\to \Conid{Map}\;\Varid{k}\;\Varid{a}}\\
\bigO{\log n}. Insert with a function, combining new value and old value. \\
\ensuremath{\Varid{insertWith}\;\Varid{f}\;\Varid{key}\;\Varid{value}\;\Varid{mp}} will insert the pair \ensuremath{(\Varid{key},\Varid{value})} into \ensuremath{\Varid{mp}} if \ensuremath{\Varid{key}} does not exist in the map. 
If the key does exist, the function will insert the pair \ensuremath{(\Varid{key},\Varid{f}\;\Varid{new\char95 value}\;\Varid{old\char95 value})}.
\end{itemize}

\subsection{Arrays}

\noindent
The functions below are from Haskell's \ensuremath{\Conid{\Conid{Data}.\Conid{Array}.MArray}} 
library\footnote{\href{https://hackage.haskell.org/package/array-0.5.1.1/candidate/docs/Data-Array-MArray.html}{https://hackage.haskell.org/package/array-0.5.1.1/candidate/docs/Data-Array-MArray.html}}, 
copied with their type signature and explanation.

\begin{itemize}
\item
\ensuremath{\Varid{newArray}\mathbin{::}\Conid{Ix}\;\Varid{i}\Rightarrow (\Varid{i},\Varid{i})\to \Varid{e}\to \Varid{m}\;(\Varid{a}\;\Varid{i}\;\Varid{e})}\\
Builds a new array, with every element initialised to the supplied value.
\item
\ensuremath{\Varid{readArray}\mathbin{::}(\Conid{MArray}\;\Varid{a}\;\Varid{e}\;\Varid{m},\Conid{Ix}\;\Varid{i})\Rightarrow \Varid{a}\;\Varid{i}\;\Varid{e}\to \Varid{i}\to \Varid{m}\;\Varid{e}}\\
Read an element from a mutable array
\item
\ensuremath{\Varid{writeArray}\mathbin{::}(\Conid{MArray}\;\Varid{a}\;\Varid{e}\;\Varid{m},\Conid{Ix}\;\Varid{i})\Rightarrow \Varid{a}\;\Varid{i}\;\Varid{e}\to \Varid{i}\to \Varid{e}\to \Varid{m}\;()}\\
Write an element in a mutable array
\item
\ensuremath{\Varid{getAssocs}\mathbin{::}(\Conid{MArray}\;\Varid{a}\;\Varid{e}\;\Varid{m},\Conid{Ix}\;\Varid{i})\Rightarrow \Varid{a}\;\Varid{i}\;\Varid{e}\to \Varid{m}\;[\mskip1.5mu (\Varid{i},\Varid{e})\mskip1.5mu]}\\
Return a list of all the associations of a mutable array, in index order.
\end{itemize}

\noindent
Below, we document the \ensuremath{\Conid{ST}} monad from Haskell's \ensuremath{\Conid{\Conid{Control}.\Conid{Monad}.ST}} 
library\footnote{\href{https://hackage.haskell.org/package/base-4.18.0.0/docs/Control-Monad-ST.html}
{https://hackage.haskell.org/package/base-4.18.0.0/docs/Control-Monad-ST.html}}.

\begin{itemize}
\item
\ensuremath{\mathbf{data}\;\Conid{ST}\;\Varid{s}\;\Varid{a}}\\
The ST monad allows for destructive updates, but is escapable. 
A computation of type \ensuremath{\Conid{ST}\;\Varid{s}\;\Varid{a}} returns a value of type \ensuremath{\Varid{a}}, and execute in ``thread'' \ensuremath{\Varid{s}}. 
It serves to keep the internal states of different invocations of runST separate from each other.
\item
\ensuremath{\Varid{runST}\mathbin{::}(\forall \Varid{s}\hsforall \hsdot{\circ }{.}\Conid{ST}\;\Varid{s}\;\Varid{a})\to \Varid{a}}\\
Return the value computed by a state thread. 
The \ensuremath{\forall \cdot \hsforall } ensures that the internal state used by the \ensuremath{\Conid{ST}} computation is inaccessible to the rest of the program.
\end{itemize}

\noindent
Below, we document the \ensuremath{\Conid{STArray}}'s from Haskell's \ensuremath{\Conid{\Conid{Data}.\Conid{Array}.ST}} 
library\footnote{\href{https://hackage.haskell.org/package/array-0.5.5.0/docs/Data-Array-ST.html}
{https://hackage.haskell.org/package/array-0.5.5.0/docs/Data-Array-ST.html}}.

\begin{itemize}
\item
\ensuremath{\mathbf{data}\;\Conid{STArray}\;\Varid{s}\;\Varid{i}\;\Varid{e}}\\
Mutable, boxed, non-strict arrays in the \ensuremath{\Conid{ST}} monad. 
\ensuremath{\Varid{s}} is the state variable argument for the \ensuremath{\Conid{ST}} type,
\ensuremath{\Varid{i}} is the index type of the array (should be an instance of \ensuremath{\Conid{Ix}}), and
\ensuremath{\Varid{e}} is the element type of the array.
\end{itemize}

\subsection{Other}

\noindent
Below, we document the \ensuremath{\Varid{foldMap}} function from Haskell's \ensuremath{\Conid{\Conid{Data}.Foldable}} 
library\footnote{\href{https://hackage.haskell.org/package/base-4.18.0.0/docs/Data-Foldable.html}
{https://hackage.haskell.org/package/base-4.18.0.0/docs/Data-Foldable.html}}.

\begin{itemize}
\item
\ensuremath{\Varid{foldMap}\mathbin{::}\Conid{Monoid}\;\Varid{m}\Rightarrow (\Varid{a}\to \Varid{m})\to \Varid{t}\;\Varid{a}\to \Varid{m}}\\
Map each element of the structure into a monoid, and combine the results with \ensuremath{(\oplus)}. 
This fold is right-associative and lazy in the accumulator. 
\end{itemize}

\noindent
Below, we document the \ensuremath{\Conid{ReaderT}} monad transformer taken from Haskell's \ensuremath{\Conid{\Conid{Control}.\Conid{Monad}.\Conid{Trans}.Reader}} 
library\footnote{\href{https://hackage.haskell.org/package/transformers-0.6.1.0/docs/Control-Monad-Trans-Reader.html}
{https://hackage.haskell.org/package/transformers-0.6.1.0/docs/Control-Monad-Trans-Reader.html}}.

\begin{itemize}
\item
\ensuremath{\mathbf{newtype}\;\Conid{ReaderT}\;\Varid{r}\;\Varid{m}\;\Varid{a}\mathrel{=}\Conid{ReaderT}\;\{\mskip1.5mu \Varid{runReaderT}\mathbin{::}\Varid{r}\to \Varid{m}\;\Varid{a}\mskip1.5mu\}}\\
The reader monad transformer, which adds a read-only environment to the given monad \ensuremath{\Varid{m}}.
The \ensuremath{\Varid{return}} function ignores the environment, while \ensuremath{(\bind )} passes the inherited environment to both subcomputations.
\end{itemize}

\end{document}